\documentclass[aps,prd,reprint,onecolumn,superscriptaddress,nofootinbib,10pt]{revtex4-2}
\usepackage{graphicx}
\usepackage{amsmath}
\usepackage{amssymb}
\usepackage{multirow}
\usepackage{siunitx}
\usepackage{bm}
\usepackage{xcolor}
\usepackage{booktabs}
\usepackage{array}
\usepackage{setspace}
\usepackage{geometry}
\usepackage{rotating}
\usepackage{subcaption}
\RequirePackage{mathptmx}
\RequirePackage{flushend}
\usepackage{tabularx}
\usepackage{array}
\usepackage{lipsum}
\usepackage[utf8]{inputenc}
\usepackage{multirow}
\usepackage[colorlinks,allcolors=blue,pdftex]{hyperref}
\usepackage{cleveref}
\usepackage{anyfontsize}
\usepackage{float}
\usepackage[section]{placeins}
\usepackage{caption}
 	
\begin{document}
	\title{Models for differential cross section in neutron-proton scattering and their implications}
	
	\author{Muhammad Saad Ashraf}

	\author{Nosheen Akbar}
	
	\affiliation{Department of Physics, COMSATS University Islamabad, Lahore Campus, Lahore, Pakistan.}
	
	\email{msaadashraf99@gmail.com}
	
	\email{nosheenakbar@cuilahore.edu.pk (corresponding author)}
	\begin{abstract}
		\textbf{\centering Abstract}
		A few analytic exponential models of the elastic differential cross section ($\frac{d\sigma}{dt}$) are proposed to analyze elastic neutron-proton ($np$) scattering. The models incorporate energy-dependent exponential slopes, power-law prefactors, and localized Gaussian modifications. Sub-leading exponential terms representing charge conjugation and isospin roles increase the applicability and quality of fit of the models to describe the differential cross section data of $np$, $n\bar{p}$, $pp$, and $p\bar{p}$ elastic scattering. The models describe the characteristic features of all the elastic scattering data such as the dip-bump structure, shrinkage of the forward peak, and controlled curvature around the diffractive dip. Parameter values of the models are found by fitting the $np$ experimental data in an energy range of $\sqrt{s} = 3.363$ to $26.019$ GeV and a momentum transfer range of $0.065 \le \vert{}t\vert{} \le 5.341 \text{ GeV}^2$. Employing the proposed models, total, elastic, and inelastic cross sections, slope parameters, interaction radii, and cross-section ratios ($\sigma_{\text{el}}/\sigma_{\text{tot}}$ and $\sigma_{\text{inel}}/\sigma_{\text{tot}}$) for $np$ scattering are calculated. A multi-stage error analysis is performed to evaluate the error in the calculations of the models. Implications of the results obtained by the models for nucleon structure and dynamics are discussed. The results show that the proposed models can be used for effective description of the $np$ elastic differential cross section and for accurate estimation of the observables of elastic hadronic scattering. The findings of this study emphasize the significance of combining theoretical and phenomenological approaches, providing valuable information for investigations of future experimental data on the differential cross section from various colliders.
		\singlespacing
	\end{abstract}
	\maketitle
	\section{Introduction}
	\label{sec:I}
	For several decades, elastic scattering of hadrons ($h_1 h_2\rightarrow h_1 h_2$) has served as a foundational process in particle physics to improve our understanding of the strong interaction at high energies. In an elastic collision process, the interacting particles emerge in the same internal states as they entered, with their quantum numbers conserved, and without any production of new particles. Elastic hadronic scattering is governed by strong interactions and should be explained by Quantum Chromodynamics (QCD). However, our understanding of the elastic processes is still at an early stage of development due to the absence of rigorous solutions of QCD equations from first principles \cite{Brodsky1973,Matveev1973,Lepage1980}. The study of the non-perturbative regime of QCD involves quark-gluon dynamics. It becomes quite difficult to perform the theoretical analysis for the forward scattering due to the nonperturbative nature of the underlying partonic dynamics. This necessitates phenomenological models and effective approaches to be developed to effectively describe the experimental data of the elastic scattering \cite{Donnachie:2002en,Collins:1977jy}. The existence of isospin symmetry \cite{Yan:2021fwv} between proton and neutron allows us to probe the isospin dependence in the elastic $np$ scattering ($np \rightarrow np$) \cite{Barone:2002cv}, and to test the universality of Pomeron or Regge descriptions beyond purely proton-based systems \cite{Watanabe2023}. Charge conjugation allows to probe ($np$) and $n\bar{p}$ systems, and $pp$ and $p\bar{p}$ in the elastic scattering \cite{Barone:2002cv,Collins:1977jy}. The study of the elastic scattering phenomena allows direct access to the spatial and dynamical structure of the particles through the differential cross section $\frac{d\sigma}{d|t|}$. We can deepen our understanding of hadron structure and collective effects like gluon saturation by precisely parameterizing elastic scattering data, which anchors both theoretical and experimental high-energy physics. The modeling of the differential cross section has been performed in various models to describe experimental data and establish potential linkages with existing theoretical frameworks. The modeling of $\frac{d\sigma}{d|t|}$ is important for describing the experimental data of TOTEM (Total, Elastic and Diffractive Cross-Section Measurement) and ALFA (Absolute Luminosity for ATLAS) at the Large Hadron Collider (LHC) and Intersecting Storage Rings (ISR) at CERN. It also serves as a critical input for cosmic ray physics, hadronic Monte Carlo generators, and the investigations in theoretical frameworks such as eikonal models, Regge theory, and QCD-driven saturation approaches \cite{MSAshrafEPJ140}.
	
	In this study, we primarily focus on analyzing elastic $np$ scattering, which is an important frontier in high-energy particle physics. In the following, we mention highlights of key experimental measurements of potential phenomenological importance, and theoretical frameworks for elastic $np$ scattering across various kinematic conditions. 
	\subsection{Experimental and Theoretical Contexts}
	Over several decades, investigations in high-energy physics have been centered on the study of hadronic interactions across both the diffractive and transition regimes of QCD. Many of these investigations were carried out through high-precision measurements of $np$ elastic scattering. Among the most significant characteristics of the differential cross section that are observed are the diffractive peak, shoulder, and dip structure. The measurements of the differential cross sections $d\sigma/dt$ performed by early experiments \cite{Gibbard1, Engler1, Bohmer1975, Mperl1970, Miller1971, Mischke1969}, spanned a wide range of momenta. These measurements particularly focused on the small-angle-diffraction peak and the backward charge exchange region, where distinct components of the scattering amplitude can be probed. 
	
	Measurements carried out at Brookhaven National Laboratory (BNL) \cite{Gibbard1} covering $5 \leq p_{\text{lab}} \leq 30~\mathrm{GeV}/c$ and $0.2 \lesssim -t \lesssim 1.2~(\mathrm{GeV}/c)^2$ are prominent which showed that at small $|t|$, $np$ cross sections are very similar to those of $pp$ scattering. This gives an indication of the fact that diffractive Pomeron exchange dominates the forward region of the elastic scattering. Subsequent experiments were performed at CERN Proton Synchrotron (CERN-PS) \cite{Engler1} and IHEP Serpukhov \cite{Bohmer1975} up to $p_{\text{lab}} \approx 70~\mathrm{GeV}/c$ and $0.2 < -t < 2.8~(\mathrm{GeV}/c)^2$. These measurements were performed beyond the diffraction cone, and a smooth, monotonic decrease in $d\sigma/dt$ is confirmed with increasing $|t|$. These results also show consistency with earlier large-angle data \cite{Mperl1970} in the range $2 < p_{\text{lab}} < 7~\mathrm{GeV}/c$. High-statistics measurements at ANL Zero Gradient Synchrotron (ZGS) are carried out in the interval $3 \leq p_{\text{lab}} \leq 12~\mathrm{GeV}/c$, providing almost complete angular distributions. These results offer rigorous constraints on spin-dependent components of the scattering amplitude. In parallel, a smooth exponential behavior is revealed in the studies of $pp$ scattering \cite{Allaby1968,Kammerud1971,Ankenbrandt1968}. 
	
	At high energies, ISR and FNAL \cite{Akerlof, Barbiellini} observed a shoulder that is found to be evolved into a pronounced diffractive dip at $|t| \approx 1.4~(\mathrm{GeV}/c)^2$. The studies in refs. \cite{Carrigan,Akerlof1967,Allaby1968,Kammerud1971}, involve the analyses in the center-of-mass frame, and further indicate $\sigma \propto s^{-n}$ for $s \gtrsim 8~\mathrm{GeV}^2$ along with a steepening of the forward slope for $|t| < 0.2~(\mathrm{GeV}/c)^2$ with a power law behavior at $90^\circ$, which indicates the onset of a large-angle scattering regime for $|t| \gtrsim 2.6~(\mathrm{GeV}/c)^2$, and signals a transition toward perturbative QCD dynamics. In ref.~\cite{RefDataset1}, comprehensive datasets by ANL-ZGS that span $4.5$–$12.5~\mathrm{GeV}/c$ and $0.14 < -t \lesssim 19~(\mathrm{GeV}/c)^2$, enable to perform detailed comparisons with $pp$ data and also show good agreement with Regge phenomenology with a sensitivity to charge symmetry breaking and isovector exchanges. In ref. \cite{RefDataset2}, the universality of forward diffraction that is dominated by Pomeron exchange, is confirmed by precision measurements over $0.06 < |t| < 3~(\mathrm{GeV}/c)^2$. It is revealed that subtle deviations occur at intermediate $|t|$ that are attributable to $C$-odd (Odderon-like) and isospin-dependent contributions \cite{Barone:2002cv,MARTYNOV2018414}. 
	
	In the ref. \cite{RefDataset3}, measurements at BNL-Alternating Gradient Synchrotron (BNL-AGS) further show a qualitative similarity between $np$ and $pp$ differential cross sections, but these are not considered identical. In the cross sections measured in \cite{RefDataset3}, characteristic diffraction peak shrinkage with increasing momentum is found. These cross sections remain consistent with expectations related to Regge trajectory~\cite{Collins:1977jy}. In these cross sections, pronounced oscillatory structures are still lacking, which is a finding indicative of a smoother transition in the $np$ scattering. Other important measurements are those carried out as extensions to higher energies ($70$–$400~\mathrm{GeV}/c$) in ref.~\cite{RefDataset4}. In these measurements, empirical observations such as persistence of the shrinkage of the diffraction cone and logarithmic slopes are confirmed to be consistent with $pp$ parameterizations over $0.15 < -t < 3.6~(\mathrm{GeV}/c)^2$. Alongside this confirmation, the emergence of a diffractive dip near $-t \approx 1.4~(\mathrm{GeV}/c)^2$ for $p_{\text{lab}} \gtrsim 200~\mathrm{GeV}/c$ is confirmed. It is noted that the data favor the Pomeron–Reggeon interference, while being inadequate to provide any conclusive distinguishing feature from Reggeon–Reggeon effects. It is importantly notable that in the intermediate region $0.7 \lesssim |t| \lesssim 1.3~(\mathrm{GeV}/c)^2$ for momenta below $\sim 280~\mathrm{GeV}/c$, $np$ cross sections exceed those of $pp$. A persistent challenge is presented by these empirical findings to conventional Regge-based theoretical descriptions.
	
	The differential cross section for $np$ elastic scattering serves as a key physical quantity that probes the spatial and dynamical structure of strong interactions. It is a sensitive quantity for the investigation of non-perturbative QCD dynamics and transition towards perturbative regimes \cite{Donnachie:2002en}. The evolution of $\frac{d\sigma}{dt}$ with $\sqrt{s}$ and $t$ is important in elastic scattering phenomena, as continuous restructuring of the relevant degrees of freedom in the hadron-hadron system is encoded in it. When the elastic $np$ scattering occurs at low energies ($\sqrt{s} \lesssim 2.5~\mathrm{GeV}$, corresponding to $p_{\text{lab}} \lesssim 2-3~\mathrm{GeV}/c$) and small momentum transfer ($|t| \lesssim 0.2~\mathrm{GeV}^2$), the scattering amplitude is governed by meson exchanges and their dynamics, occurring within an effective field-theoretic description of the nuclear force. In this scenario, long-range one-pion exchange and contributions from intermediate-range heavier vector mesons ($\rho$, $\omega$) also dominate the interaction \cite{Geramb1998}. Partial wave analyses are effective in this regime as strong angular-momentum channels can be revealed, which produce pronounced angular oscillations and near-threshold enhancements linked to the deuteron bound state, large S-wave scattering lengths, and spin-isospin correlations ~\cite{Arndt2007}. 
	
	A view of the structure of the interacting particles reveals that the proton and neutron behave as composite objects that are extended with spatial distributions governed by meson clouds and confined valence quarks. These spatial distributions probe distance scales of order $(1~\mathrm{fm})$, where confinement and chiral dynamics dominate. The elastic scattering in the intermediate region ($2.5 \lesssim \sqrt{s} \lesssim 6~\mathrm{GeV}$, $p_{\text{lab}} \sim 3-10~\mathrm{GeV}/c$) is characterized by an evident transition toward Reggeized dynamics. At small $|t| \lesssim 0.5~\mathrm{GeV}^2$, a diffractive cone with approximately exponential behavior $d\sigma/dt \sim e^{Bt}$ is exhibited by the differential cross section, which is reflective of absorptive corrections and $t$-channel Regge pole exchange. The intermediate momentum-transfer region ($0.5 \lesssim |t| \lesssim 1.5~\mathrm{GeV}^2$) is marked by a characteristic dip. It occurs due to interference between even- and odd-under-crossing contributions, notably the Pomeron and C-odd trajectories~\cite{DONNACHIE1992227,Cudell2002}. It gives a signal of several key features, including the increasing role of multi-gluon exchanges and non-trivial analyticity properties of the scattering amplitude. The effective description undergoes a shift from purely mesonic degrees of freedom toward correlated quark-gluon dynamics; thereby the nucleon reveals substructure beyond its meson cloud. 
	
	At high energies ($\sqrt{s} \gtrsim 6~\mathrm{GeV}$, extending to ISR and collider scales with $p_{\text{lab}} \gtrsim 10~\mathrm{GeV}/c$), diffractive processes dominate the elastic scattering. These processes are associated with Pomeron exchange, which leads to a pronounced forward peak and a steep exponential fall-off at small $|t| \lesssim 0.3~\mathrm{GeV}^2$. It is found to be consistent with an interaction radius that has energy dependence and geometrical scaling behavior. At large momentum transfers ($|t| \gtrsim 2-3~\mathrm{GeV}^2$), the differential cross section shows a power-law tail, $d\sigma/dt \sim |t|^{-n}$. The high-$|t|$ behavior remains in accordance with quark counting rules and predictions related to perturbative QCD. It reflects that the scattering phenomenon is dominated by hard scattering that occurs between nearly point-like partons and is mediated by gluon exchange~\cite{Block2005}. This asymptotic regime resolves the nucleon into its quarks and gluon constituents, and the asymptotic freedom governs the scattering dynamics at short distances ($\lesssim 0.1~\mathrm{fm}$). Thus, a continuous and theoretically coherent evolution is exhibited by elastic $np$ scattering across increasing $\sqrt{s}$ and $|t|$, spanning from meson-exchange dominance and collective hadronic structure to Reggeized gluonic exchanges and ultimately to partonic scattering.
	
	Elastic nucleon–nucleon scattering, including $np$, $pp$, and $p\bar{p}$ processes, has been extensively investigated by researchers using various models that are usually developed in relation to Regge theory. In such models, amplitude can be decomposed into contributions from Pomeron ($P$), secondary Reggeon ($R$), and Odderon ($O$) exchanges with structured $t$-dependence introduced via form factors, signature factors and an explicit energy scaling $\sim s^{\alpha(t)}$ \cite{Donnachie:2002en,Saleem1985,Jenkovszky2011}. Early Regge models and their Regge-cut and Regge-eikonal extensions \cite{Collins:1977jy,Saleem1979,Rarita1968,Yhara1962,VBarger1966,Irving1977,Godizov2022} describe total and differential cross sections and their asymptotic behavior \cite{DONNACHIE1992227,Cudell2002,Block2005.1}. In recent years, analogous decompositions have also emerged in holographic QCD involving Reggeized glueball and vector meson exchanges \cite{Watanabe2023}. Several complementary approaches exist that include impact-parameter and eikonal formalism, such as the Bourrely-Soffer-Wu (BSW) model with Fermi–Dirac-type opaqueness in $b$-space \cite{BSW2014}. Other notable models are Regge–eikonal models with multiple exchange components \cite{Godizov2022}, and Glauber multiple scattering theory \cite{Glauber1959,Mehndiratta2017}. Optical and microscopic potential models \cite{Hodgson1963,Ogawa2019}, geometrical and Chou-Yang models \cite{Zbigniew2016,ChouYang1968} are some widely used approaches to study the scattering of hadrons. Other significant approaches aim to provide a description of scattering based on links with QCD and constituent quarks of hadrons. These are QCD-inspired and constituent quark or multiple diffraction models \cite{Islam2017,Wakaizumi1978}. Dispersion relation approaches \cite{Bronzan1974}, and dual or Veneziano-type amplitudes \cite{Veneziano1968}, as well as empirical parameterizations such as the Phillips–Barger (PB) model and its extensions \cite{PHILLIPS1973412,Goncalves2019}, also aim to describe the experimental data. These models may reproduce specific features such as forward diffraction cones, Regge scaling, or unitarity constraints, and may also provide insights into nucleon structure through form factors and spatial distributions at large $|t|$ \cite{Islam2017,Donnachie:2002en}. However, some limitations also appear within these frameworks; for instance, some Regge-based models rely on asymptotic assumptions and do not effectively constrain the intermediate-$t$ and dip–bump regions. Glauber and optical approaches are reliable mainly at small momentum transfer. Eikonal and QCD-inspired models often involve many parameters. Microscopic approaches are often computationally intensive and model-dependent. Modern high-precision measurements at LHC, specifically from the TOTEM and ATLAS collaborations, particularly of dip-bump structures and possible Odderon effects, highlight parameter degeneracies and limited extrapolation control inherent in these approaches \cite{Goncalves2019,Godizov2022,Rasmussen2018}. Furthermore, the inclusion of $C$-odd exchanges (such as the Odderon) that are required to explain differences between particle-particle and particle-antiparticle scattering introduces additional complexity when attempting a description of $np$, $n\bar{p}$, $pp$, and $p\bar{p}$ data in an effective parameterization of the differential cross section. It is emphasized that progress in theoretical and experimental high-energy physics can be driven by performing a global comparison of various models, parametrizations, and methods. The present methodology and results are intended to advance research in the domain of inverse problems in high-energy hadronic scattering.
	
	In this study, we present purely phenomenological models for the differential cross section as functions of Mandelstam variables s and t, based on empirical observations to describe the experimental data of the elastic $np$ scattering. The methodical approach towards modeling the elastic differential cross section $\frac{d\sigma}{d|t|}$ is described in Section \ref{sec:II}. The results of the data fitting, model comparisons, statistical and error analysis, and implications of the models are written in Section \ref{sec:III}. The calculated results of total, inelastic, and elastic cross sections, slope parameters, interaction radii, and other quantities by our models are also mentioned in this section. Sections \ref{sec:IV} and \ref{sec:V} give the conclusion and recommendations of this study, respectively.
	\section{Methodology}
	\label{sec:II}
	The five distinct composite exponential models, based on their ability to describe the key features of the differential cross-section data, including the forward peak, diffractive dip, and large-${\mid t\mid}$ behavior, are constructed to fit four datasets of the experimental data of elastic ${(np)}$ differential cross-section. Dataset 1 is experimental measurements of the differential cross section made at the ANL-ZGS for 8 different momentum ranges between 4.5 and 12.5 GeV/c \cite{RefDataset1}. Dataset 2 comprises measurements of the differential cross section for incident momenta between 10 and 24 GeV/c using wire spark chambers performed at the CERN-PS \cite{RefDataset2}. Dataset 3 consists of the differential cross section measurements in the diffraction region with an incident neutron momenta between 5 and 30 GeV/c from an optical-spark-chamber-counter experiment conducted at the BNL-AGS \cite{RefDataset3}. And Dataset 4 consists of the experimental measurements of the differential cross section from an experiment over a ${\mid t\mid}$-range from $0.15$ to $\sim 3.6~(\textrm{GeV/c})^{2}$ for incident neutron momenta from 70 to 400 GeV/c at FNAL energies \cite{RefDataset4}. The available experimental data are taken in the ${\mid t\mid}$ ranges that are mentioned in Table \ref{tab:expdata} at their corresponding center-of-mass (c.m.) energies, where the data most likely contain the forward peak, dip and most part of the large-${\mid t\mid}$ region.
	\renewcommand{\arraystretch}{1.0}
	\setlength{\tabcolsep}{2.30pt}
	\begin{table}[h]
		\centering
		\footnotesize
		\caption{Summary of the experimental data for the neutron-proton ($np$) elastic scattering with $\mid t \mid$-ranges at each $\sqrt{s}$ value}
		\label{tab:expdata}
			\begin{tabular}{cccccccccccc}
				\toprule
				\multirow{1}{*}{}  & \multicolumn{2}{c}{\textbf{Dataset 1 \cite{RefDataset1}}} & \multicolumn{3}{c}{\textbf{Dataset 2 \cite{RefDataset2}}} & \multicolumn{3}{c}{\textbf{Dataset 3 \cite{RefDataset3}}} & \multicolumn{3}{c}{\textbf{Dataset 4 \cite{RefDataset4}}} \\
				\midrule
				$\sqrt{s}$ & $|t|$ & No. of & $\sqrt{s}$ & $|t|$ & No. of & $\sqrt{s}$ & $|t|$ & No. of & $\sqrt{s}$ & $|t|$ & No. of \\
				
				${\textrm{GeV}}$ & ${\textrm{GeV}^{2}}$ & points & ${\textrm{GeV}}$ & ${\textrm{GeV}^{2}}$ & points & ${\textrm{GeV}}$ & ${\textrm{GeV}^{2}}$ & points & ${\textrm{GeV}}$ & ${\textrm{GeV}^{2}}$ & points \\
				
				3.363  & $0.134 \leq {\mid t\mid} \leq 5.281$ & 36 & 4.74  & $0.065 \leq {\mid t\mid} \leq 1.68$ & 42 & 3.466-4.409  & $0.174 \leq {\mid t\mid} \leq 0.934$ & 12 & 13.748 & $0.152 \leq {\mid t\mid} \leq 2.093$ & 29 \\
				3.628  & $0.134 \leq {\mid t\mid} \leq 5.341$ & 38 & 5.120 & $0.065 \leq {\mid t\mid} \leq 2.48$ & 47 & 4.409-5.187  & $0.175 \leq {\mid t\mid} \leq 1.476$ & 14 & 16.823 & $0.152 \leq {\mid t\mid} \leq 3.404$ & 33 \\
				3.876 & $0.134 \leq {\mid t\mid} \leq 4.355$ & 42 & 5.473 & $0.065 \leq {\mid t\mid} \leq 2.80$ & 49 & 5.187-5.863 & $0.175 \leq {\mid t\mid} \leq 1.515$ & 14 & 19.416  & $0.152 \leq {\mid t\mid} \leq 3.401$ & 33 \\
				4.109 & $0.134 \leq {\mid t\mid} \leq 5.207$ & 43 & 5.805 & $0.065 \leq {\mid t\mid} \leq 2.80$ & 49 & 5.863-6.47 & $0.223 \leq {\mid t\mid} \leq 1.638$ & 13 & 21.471 & $0.151 \leq {\mid t\mid} \leq 3.606$ & 33 \\
				4.329 & $0.13 \leq {\mid t\mid} \leq 5.113$ & 44 & 6.119  & $0.065 \leq {\mid t\mid} \leq 2.80$ & 49 & 6.47-7.024 & $0.227 \leq {\mid t\mid} \leq 1.185$ & 12 & 22.956 & $0.151 \leq {\mid t\mid} \leq 3.328$ & 33 \\
				4.54 & $0.147 \leq {\mid t\mid} \leq 5.214$ & 44 & 6.418  & $0.065 \leq {\mid t\mid} \leq 2.8$ & 49 & 7.024-7.538 & $0.325 \leq {\mid t\mid} \leq 0.881$ & 8 & 24.536 & $0.151 \leq {\mid t\mid} \leq 3.479$ & 33 \\
				4.741 & $0.175 \leq {\mid t\mid} \leq 5.283$ & 41 & 6.704 & $0.37 \leq {\mid t\mid} \leq 2.80$ & 26 &  & & & 26.019 & $0.175 \leq {\mid t\mid} \leq 3.662$ & 32 \\
				4.935 & $0.175 \leq {\mid t\mid} \leq 5.208$ & 39 & & & & & & & & & \\
				\bottomrule
			\end{tabular}
	\end{table}
	\subsection{Models for the Differential Cross Section}
	The proposed models of the elastic differential cross section, comprising various composite exponential forms, reflect the approximate behavior of the elastic scattering amplitude to fit the data. In light of the description of $np$ experimental data, mentioned in Section \ref{sec:I}, it is understood that the $pp$ system is closely related to the $np$ system. The differential cross section model ref. \cite{Donnachie:2002en} describes the experimental data of elastic $pp$ and $p\bar{p}$ scattering in terms of a single Pomeron ($\mathbb{P}$) and double Pomeron exchanges ($\mathbb{P}\mathbb{P}$) by their trajectories at low-$\mid t\mid$ ($\mid t\mid < 1~\textrm{GeV}^{2}$). This model describes $pp$ elastic scattering in terms of Pomeron coupling with the separate valence quarks of the proton. We mention its differential cross section \cite{Donnachie:2002en} as follows:
	\begin{equation}
		\frac{d\sigma}{dt} = \frac{(3 \beta_\mathbb{P} F_1(t))^{4}}{4\pi}\bigg(\frac{s}{s_0}\bigg)^{{2\alpha_\mathbb{P} }(t)-2},
	\end{equation}
	where the proton form factor is parameterized as $(F_1 (t))^{2} = A e^{a t} + B e^{b t} + C e^{c t} $, the trajectory for the Pomeron exchange is $\alpha_\mathbb{P} (t) = 1 + \epsilon_\mathbb{P} + \alpha_\mathbb{P} ' t$, and $\beta_\mathbb{P}$ is a constant that originated from the Pomeron-exchange contribution to quark-quark scattering. Another notable model is a simplified Regge pole model of ref. \cite{RefDataset3}, which has been used to study the shrinkage of the forward peak in high-energy $pp$ data and low-energy $np$ data in the low-$ | t | $ region. We mention the differential cross section of this Regge pole model \cite{RefDataset3} as follows:
	\begin{equation}
		\frac{d\sigma}{dt} = \frac{1}{16\pi s (s-4m^{2})}\beta_\mathbb{P}^{2}\bigg(\frac{s}{s_0}\bigg)^{2\alpha_\mathbb{P} (t)},
	\end{equation}
	where $\alpha_\mathbb{P} (t)$ is linear Pomeron trajectory, $m$ is mass of the nucleon, and $\beta_\mathbb{P}(t)$ is the residue of the Pomeron pole $\mathbb{P}$. The model under the assumption that $s >> 4m^{2}$, and with fixed value $s_0$ = 1 $\textrm{(GeV/c)}^{2}$, reduces to the following form:
	\begin{equation}
		\frac{d\sigma}{dt} \approx \frac{\beta_\mathbb{P} ^{2}}{16\pi} e^{[2\alpha_\mathbb{P} (t)-2](ln s)}.
	\end{equation}
	The aforementioned models assume that at high energies only the Pomeron contribution is dominant, and other Regge pole contributions ($f_0$, $\omega$, $\rho$, $a_2$) which depend on charge conjugation and isospin in $pp$, $p\bar{p}$, and $n\bar{p}$ elastic processes are suppressed \cite{RefDataset3}. These models are useful in the study of the isolated energy growth of a Pomeron trajectory and capture the universal logarithmic shrinkage of the forward diffractive peak but are insufficient to fit detailed experimental differential cross-sections at large kinematic domains. These models inherently lack the dynamical flexibility required to reproduce fine-structure phenomena across extended $\mid t\mid$ ranges. Keeping these facts in view, we introduce our models for the effective description of the features of experimental and extrapolated data (theoretical results of other models) in different kinematic ranges. The proposed phenomenological models consist of a leading exponential term with subleading terms to describe the experimental data by efficient fitting from low to high-$\mid t\mid$ regions. The scaling-like factors ($s/s_0$) are introduced with dimensionless exponents ($\alpha$, $\delta_O$, and $\delta_I$) that give power-law energy dependence in the models. Such energy-dependent factors have been used to approximate the differential cross sections of various scattering phenomena in ref. \cite{Brodsky:1973kr}. For leading slopes, energy dependence of a logarithmic form $(B_0 + 2\beta ln(s/s_0))$ is introduced in the leading exponential terms. In this study, the analysis relies on non-linear fitting of the $np$ data of Table \ref{tab:expdata} with the proposed models; the outcome profoundly depends on the choice of initial values of the free parameters, which are not known. The fitting procedure relies on finding the optimal parameter values within an assumed range by minimizing $\chi ^{2}$, yielding calculations in close agreement with the reference values. The models are parameterized with a small number of parameters, which makes the models different from other models \cite{Pancheri1,Jenkovszky2011} that often have many parameters and are computationally intensive. We now describe our proposed models in the following:
	\subsubsection{Model 1}
	This exponential model is written as follows:
	\begin{equation}
		\frac{d\sigma(s,t)}{dt} = A_0 \left(\frac{s}{s_{0}}\right)^{\alpha}e^{(B_0 + 2\beta ln(s/s_0))t}\Big[1 + \varepsilon_0 (1 - e^{-(t - t_d)^{2}/\Delta^{2}})\Big] + O_O \left(\frac{s}{s_{0}}\right)^{\delta_O} \xi_C e^{D_O (s/s_0)^{\mu}t} + I_0 \left(\frac{s}{s_{0}}\right)^{\delta_I} \xi_I e^{E_0 (s/s_0)^{\mu}t},
	\end{equation}
	where $s$ is the squared center-of-mass energy, $t$ is the four-momentum transfer, and $s_{0}$ is the Regge parameter that is taken as a fixed scale ($1~\mathrm{GeV^{2}}$ refs. \cite{RefDataset3,Donnachie:2002en,Collins:1977jy}). The first term on the right-hand side governs the forward diffractive peak at low-$\mid t\mid$, while the second and third terms capture the transition into the intermediate and large-$\mid t\mid$ regions. The factor $A_{0}(s/s_{0})^{\alpha}$ represents the energy evolution of the differential cross-section with $\alpha$ as a free parameter. This choice of the power law is treated as an effective description for the energy regimes under consideration, rather than an asymptotic statement for $s\rightarrow \infty$. In the parameterization for the effective slope $(B_0 + 2\beta ln(s/s_0))$ \cite{Donnachie:2002en,Okorokov2015,Okorokov2008}, $\beta$ is a free parameter that quantifies the \emph{shrinkage} of the diffraction cone with energy. The term $e^{(B_0 + 2\beta ln(s/s_0))t}$ describes the exponential fall of the differential cross section in the forward (small-$|t|$) region and is differentiable and increasing in $\ln s$ for $\beta>0$, ensuring a progressive narrowing of the $t$–distribution at higher energies. In the Gaussian modulation factor $(1 + \varepsilon_0 (1 - e^{-(t - t_d)^{2}/\Delta^{2}}))$, the parameters $\varepsilon_0$, $t_{d}$, and $\Delta$ control the depth and width of dip. The sub-leading exponential terms represent the contributions from charge conjugation and isospin channels, respectively, across elastic $np$, $n\bar{p}$, $pp$ and $p\bar{p}$ scattering processes. In these terms, the parameters $\xi_C$ and $\xi_I$ (mentioned in Table \ref{tab:parameters}) are discrete sign factors associated with charge conjugation and isospin exchange channels \cite{Barone:2002cv,Collins:1977jy,Donnachie:2002en,Irving1977,Carlo2003}. The slopes of the subleading terms, $D_O (s/s_0)^{\mu}$ and $E_0 (s/s_0)^{\mu}$ have a different behavior from the slope of the leading term and follow power laws in $s$. Because the sub-leading terms enter additively, they contribute linearly independent exponential curvatures in $t$ and produce a multi–slope profile of the model. The free parameters $\delta_O$ and $\delta_I$ control the power law scales in the subleading terms. The parameter values of this model, determined by fitting the $np$ data of Table \ref{tab:expdata} are written in Table \ref{tab:pmodel1}.
	\renewcommand{\arraystretch}{1.2}
	\setlength{\tabcolsep}{5.50pt}
	\begin{table}[H]
		\centering
		\footnotesize
		\caption{$\xi_C$ and $\xi_I$ parameters for all the models of this study}
		\label{tab:parameters}
		\begin{tabular}{ccc}
			\toprule
			\textbf{Process} & \textbf{$\xi_C$} & \textbf{$\xi_I$} \\
			\midrule
			$np \rightarrow np$ & +1 & -1 \\
			$n\bar{p} \rightarrow n\bar{p}$ & -1 & -1 \\
			$pp \rightarrow pp$ & +1 & +1 \\
			$p\bar{p} \rightarrow p\bar{p}$ & -1 & +1 \\
			\bottomrule
		\end{tabular}
	\end{table}
	\subsubsection{Model 2}
	This exponential model, containing a double-exponential structure instead of the Gaussian modification, is written as follows,
	\begin{equation}
		\frac{d\sigma (s,t)}{dt} = A_0 \left(\frac{s}{s_{0}}\right)^{\alpha} \Big[a_1 e^{(B_{1}^{(0)} + 2\beta ln(s/s_0))t} + a_2 e^{(B_{2}^{(0)} + 2\beta ln(s/s_0))t}\Big] + O_O \left(\frac{s}{s_{0}}\right)^{\delta_O} \xi_C e^{D_O (s/s_0)^{\mu}t} + I_0 \left(\frac{s}{s_{0}}\right)^{\delta_I} \xi_I e^{E_0 (s/s_0)^{\mu}t}.
	\end{equation}
	In this model, the first term is a linear combination of two exponentials, each analytic and convex in $t$. Because the slopes $(B_{1,2}(s)= B_{1,2}^{(0)} + 2\beta\ln(s/s_{0}))$ increase monotonically with $\ln s$ for $\beta>0$, the exponentials become progressively narrower in $t$ as $s$ increases, due to which the full leading term shrinks in width while rising in forward height. For $t<0$ both exponentials satisfy $0<e^{B_{1,2}(s)t}\le 1$ and vanish as $t\to -\infty$ and ensure absolute integrability. The coefficients $a_{1}$ and $a_{2}$ are real, satisfy $a_{1} > a_{2} > 0$, and control the relative strength of the two slopes. The difference between the slopes $B_{1}(s)$ and $B_{2}(s)$ introduces curvature into $\ln(d\sigma/dt)$ as a function of $t$. Every term in the model is an entire function of $t$ and $s$ and ensures analyticity and differentiability in the physical domain. Convexity in $t$ is ensured for each exponential, and all curvature variations originate from the interplay of different slopes in the diffractive sector, giving the model the flexibility to describe the characteristic curvature of $np$ elastic scattering without introducing singularities. The parameter values of this model, determined by fitting the $np$ data, are written in Table \ref{tab:pmodel2}.
	\subsubsection{Model 3}
	This exponential model, with a triple-component exponential structure in its leading term, is written as follows,
	\begin{align}
		\frac{d\sigma (s,t)}{dt} = A_0 \left(\frac{s}{s_{0}}\right)^{\alpha} \Big[f e^{(B_{1}^{(0)} + 2\beta ln(s/s_0))t} + (1 - f) e^{(B_{2}^{(0)} + 2\beta ln(s/s_0))t}\Big] \Big[1 - \varepsilon_0 e^{\frac{-(t-t_d)^2}{2\Delta^{2}}}\Big]+ O_O \left(\frac{s}{s_{0}}\right)^{\delta_O} \xi_C e^{D_O (s/s_0)^{\mu}t} \nonumber \\ + I_0 \left(\frac{s}{s_{0}}\right)^{\delta_I} \xi_I e^{E_0 (s/s_0)^{\mu}t}.
	\end{align}
	The leading part is a convex mixture of two energy-dependent exponentials. The parameter $f\in[0,1.5]$, which controls the relative weight of the two scales, makes the local logarithmic slope smooth and continuous in both $t$ and $\ln s$, and interpolates between $B_{1}(s)$ and $B_{2}(s)$ as $|t|$ grows. This interpolation produces a controlled two-scale curvature and ensures a positive curvature unless $B_{1}=B_{2}$. The multiplicative Gaussian factor $(1-\varepsilon_0\,e^{-(t-t_{d})^{2}/(2\Delta^{2})})$ is entire, bounded and tends to unity for $|t|\to\infty$. So, it may implement a localized modulation (a smooth suppression when $\varepsilon_0 >0$) of the diffractive envelope near $t=t_{d}$ without altering the exponential asymptotics at large $|t|$. The fitted parameters of this model are written in Table \ref{tab:pmodel3}.
	\subsubsection{Model 4}
	This exponential model with a log-enhanced slope is given as follows:
	\begin{align}
		\frac{d\sigma(s,t)}{dt}
		= A_{0}\left(\frac{s}{s_{0}}\right)^{\alpha}
		\exp\Big[\big(B_{0}+2\beta\ln(s/s_{0})+b_{1}\ln(1+t/t_{0})\big)t\Big]
		\Big[1-\varepsilon_{0}e^{-(t-t_{d})^{2}/(2\Delta^{2})}\Big]
		+O_{0}\Big(\frac{s}{s_{0}}\Big)^{\delta_{O}}\xi_{C}e^{D_{0}(s/s_{0})^{\mu}t} \nonumber \\ +I_{0}\Big(\frac{s}{s_{0}}\Big)^{\delta_{I}}\xi_{I}e^{E_{0}(s/s_{0})^{\mu}t}.
	\end{align}
	In this model, the term ($b_{1}\ln(1+t/t_{0})$) in the effective slope introduces a slow, saturating variation of the slope with $t$. This modification can be interpreted as a logarithmic saturation effect to account for the curvature of the Regge trajectory or the onset of partonic saturation at moderate momentum transfers. It allows the exponential slope to vary smoothly with $|t|$ while preserving analyticity and stability of the fit. The fitted parameter values of this model are determined by fitting the $np$ data of Table \ref{tab:expdata} and are written in Table \ref{tab:pmodel4}.
	\subsubsection{Model 5}
	This exponential model with a threshold suppression is written as follows:
	\begin{equation}
		\frac{d\sigma(s,t)}{dt}
		= A_{0}\left(\frac{s}{s_{0}}\right)^{\alpha} e^{(B_{0}+2\beta\ln(s/s_{0}))t}\Big[1-\varepsilon_{0}e^{-\frac{(t-t_{d})^{2}}{2\Delta^{2}}}\Big]
		+ O_{0}\left(\frac{s}{s_{0}}\right)^{\delta_{O}}\xi_{C} e^{D_{O}(s/s_{0})^{\mu}t}\bigl(1-e^{-\gamma|t|}\bigr)
		+ I_{0}\left(\frac{s}{s_{0}}\right)^{\delta_{I}}\xi_{I} e^{E_{0}(s/s_{0})^{\mu}t}.
	\end{equation}
	The third subleading term is the same as for the previous models. When $t \to 0$, the suppression factor $(1 - e^{-\mu|t|}) \to 0$, which eliminates the contribution to the forward cross section. At the optical point $(t=0)$, the behavior is controlled by the forward limits of the components. The suppression factor provides two major benefits: \textbf{(i)} It can enforce a threshold behavior where Odderon contributions vanish smoothly at $t=0$ and produce a forward limit for applications related to the optical theorem. \textbf{(ii)} It can separate central (Pomeron-dominated) and peripheral ($C$-odd and isospin-dependent) effects. The fitted parameter values of this model, determined by fitting the $np$ data of Table \ref{tab:expdata}, are written in Table \ref{tab:pmodel5}.
	
	In the next section, we employ these proposed models with their fitted parameters to calculate the total, elastic and inelastic cross sections, logarithmic slope parameter and the interaction radii, and the cross-section ratios in $np$ elastic scattering. Least $\chi^{2}$ estimation, parameter correlations, statistical information criteria, and a multi-stage error analysis (for the estimates of the derived quantities) have been used to determine the accuracy of the models, and the best model with the most accurate results has been identified.
	\section{Results and Discussion}
	\label{sec:III}
\begin{figure}[t!]
		\centering
		\begin{subfigure}[b]{0.48\textwidth}
			\centering
			\includegraphics[width=\textwidth]{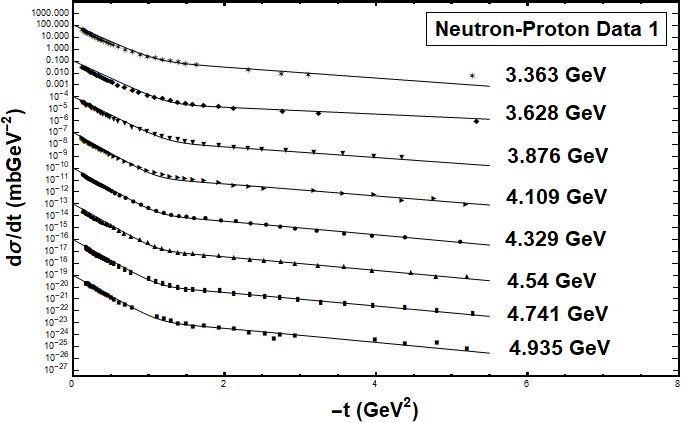}
			\label{fig:a}
		\end{subfigure}
		\hspace{0.05cm}
		\begin{subfigure}[b]{0.48\textwidth}
			\centering
			\includegraphics[width=\textwidth]{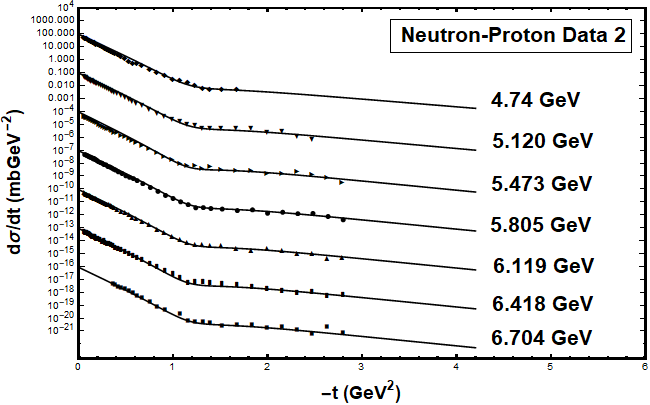}
			\label{fig:b}
		\end{subfigure}
		\vspace{-0.1cm}
		\begin{subfigure}[b]{0.48\textwidth}
			\centering
			\includegraphics[width=\textwidth]{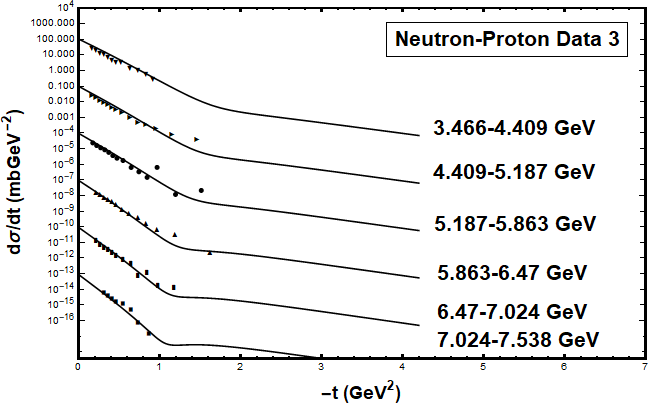}
			\label{fig:c}
		\end{subfigure}
		\hspace{0.05cm}
		\begin{subfigure}[b]{0.48\textwidth}
			\centering
			\includegraphics[width=\textwidth]{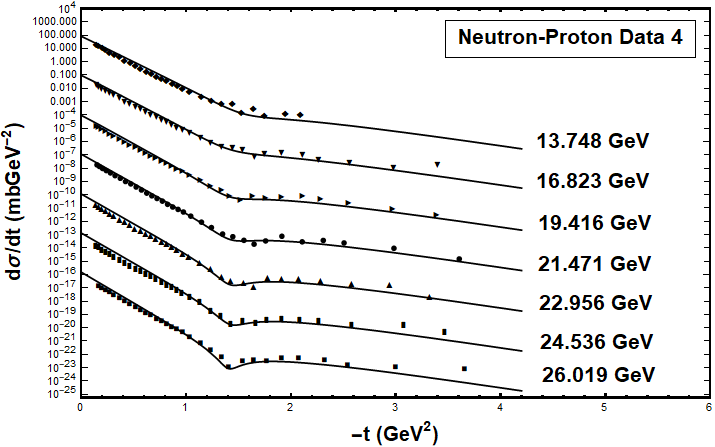}
			\label{fig:d}
		\end{subfigure}
		\caption{The fits on the experimental data of Table \ref{tab:expdata} by our model. Data and model values are multiplied by $10^{-3(n-1)}$, where $n$ is the number of curve and corresponding data set starting from the top. The solid line represents the fit of our model to the data. (a) Model 1 fits on the data of Dataset 1. (b) Model 1 fits on the data of Dataset 2. (c) Model 1 fits on the data of Dataset 3. (d) Model 1 fits on the data of Dataset 4}
		\label{fig:model1fits}
	\end{figure}

\begin{figure}[t!]
		\centering
		\begin{subfigure}[b]{0.48\textwidth}
			\centering
			\includegraphics[width=\textwidth]{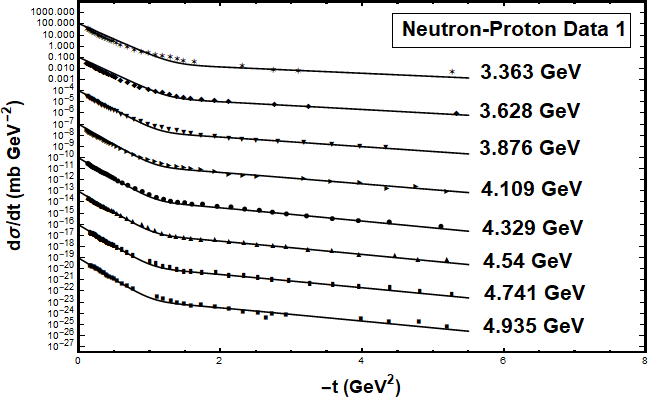}
			\label{fig:a}
		\end{subfigure}
		\hspace{0.05cm}
		\begin{subfigure}[b]{0.48\textwidth}
			\centering
			\includegraphics[width=\textwidth]{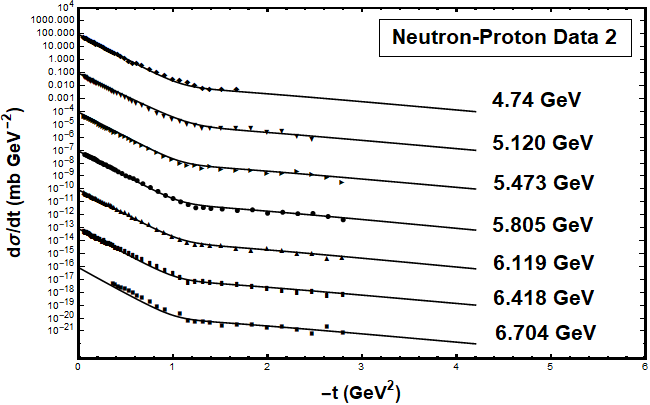}
			\label{fig:b}
		\end{subfigure}
		\vspace{-0.1cm}
		\begin{subfigure}[b]{0.48\textwidth}
			\centering
			\includegraphics[width=\textwidth]{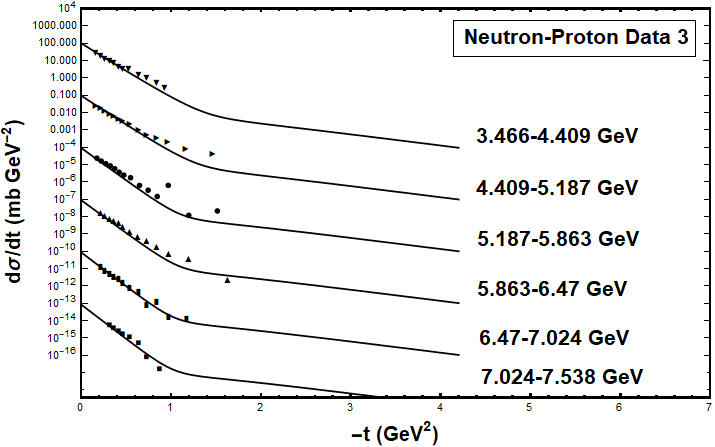}
			\label{fig:c}
		\end{subfigure}
		\hspace{0.05cm}
		\begin{subfigure}[b]{0.48\textwidth}
			\centering
			\includegraphics[width=\textwidth]{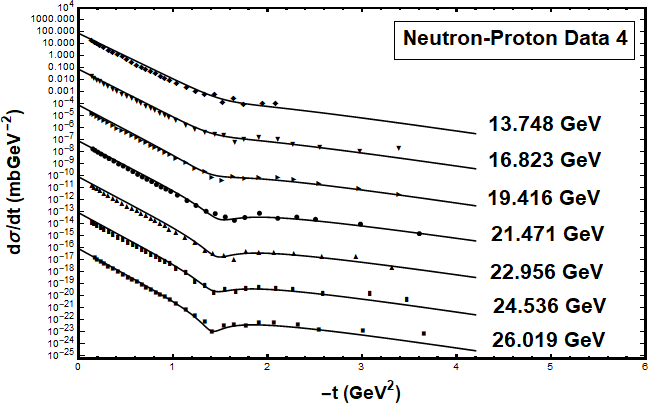}
			\label{fig:d}
		\end{subfigure}
		\caption{Same legend for parts (a), (b), (c), and (d) as in the Figure \ref{fig:model1fits} but with Model 2}
		\label{fig:model2fits}
	\end{figure}

\begin{figure}[t!]
		\centering
		\begin{subfigure}[b]{0.48\textwidth}
			\centering
			\includegraphics[width=\textwidth]{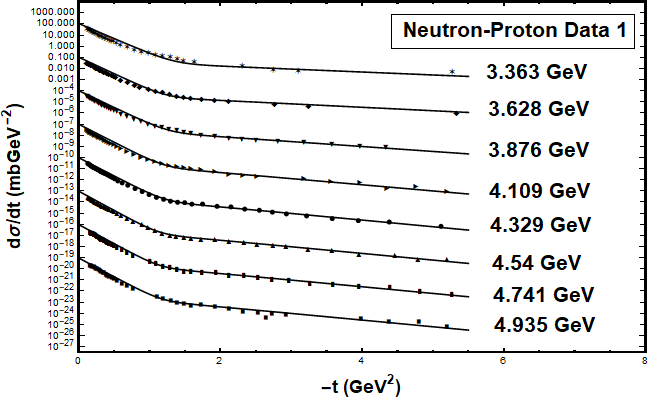}
			\label{fig:a}
		\end{subfigure}
		\hspace{0.05cm}
		\begin{subfigure}[b]{0.48\textwidth}
			\centering
			\includegraphics[width=\textwidth]{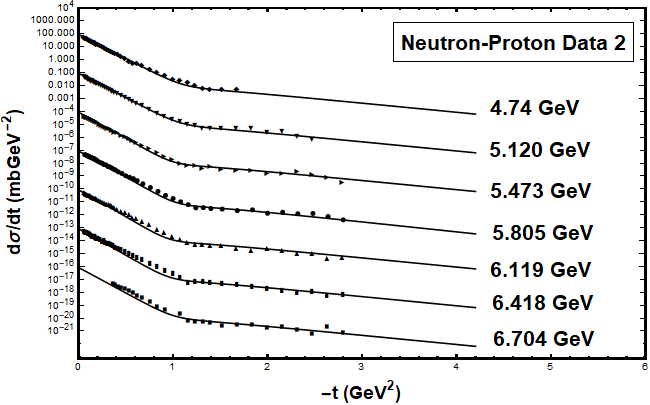}
			\label{fig:b}
		\end{subfigure}
		\vspace{-0.1cm}
		\begin{subfigure}[b]{0.48\textwidth}
			\centering
			\includegraphics[width=\textwidth]{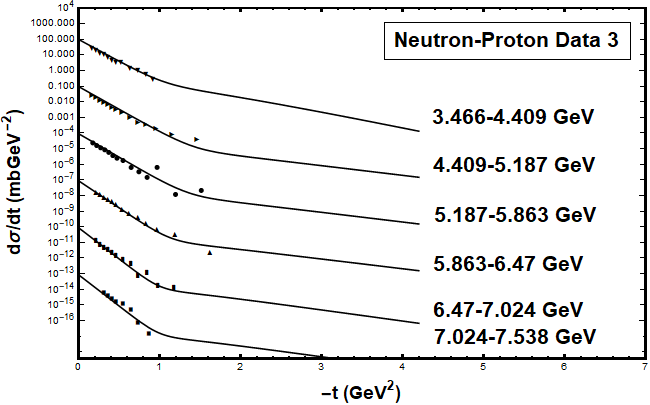}
			\label{fig:c}
		\end{subfigure}
		\hspace{0.05cm}
		\begin{subfigure}[b]{0.48\textwidth}
			\centering
			\includegraphics[width=\textwidth]{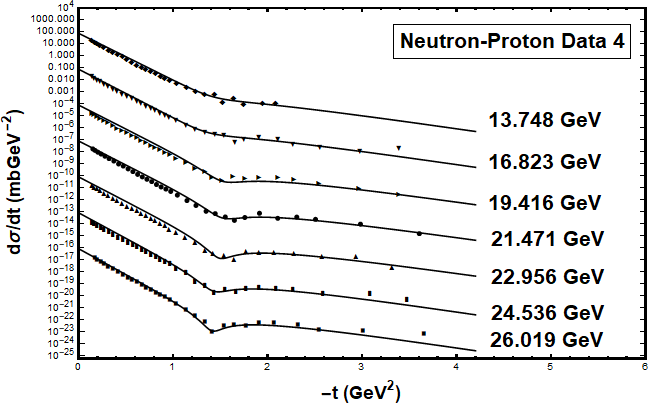}
			\label{fig:d}
		\end{subfigure}
		\caption{Same legend for parts (a), (b), (c), and (d) as in the Figure \ref{fig:model1fits} with Model 3}
		\label{fig:model3fits}
	\end{figure}

\begin{figure}[t!]
		\centering
		\begin{subfigure}[b]{0.48\textwidth}
			\centering
			\includegraphics[width=\textwidth]{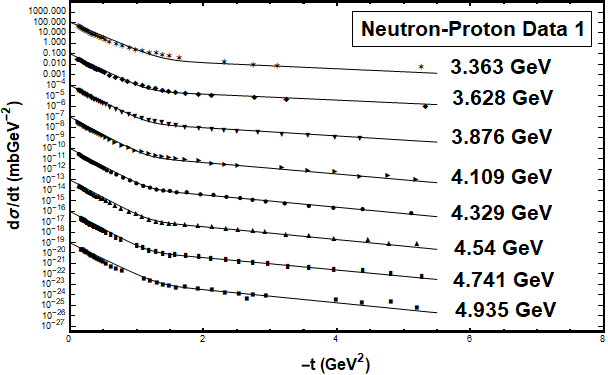}
			\label{fig:a}
		\end{subfigure}
		\hspace{0.05cm}
		\begin{subfigure}[b]{0.48\textwidth}
			\centering
			\includegraphics[width=\textwidth]{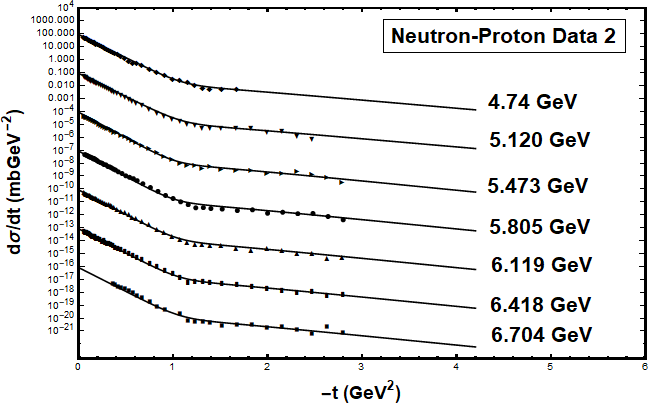}
			\label{fig:b}
		\end{subfigure}
		\vspace{-0.1cm}
		\begin{subfigure}[b]{0.48\textwidth}
			\centering
			\includegraphics[width=\textwidth]{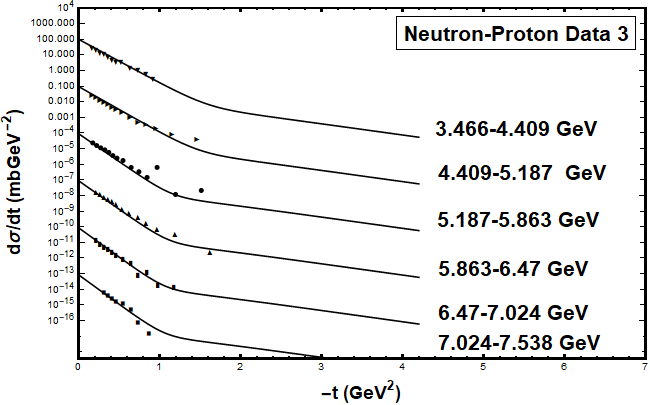}
			\label{fig:c}
		\end{subfigure}
		\hspace{0.05cm}
		\begin{subfigure}[b]{0.48\textwidth}
			\centering
			\includegraphics[width=\textwidth]{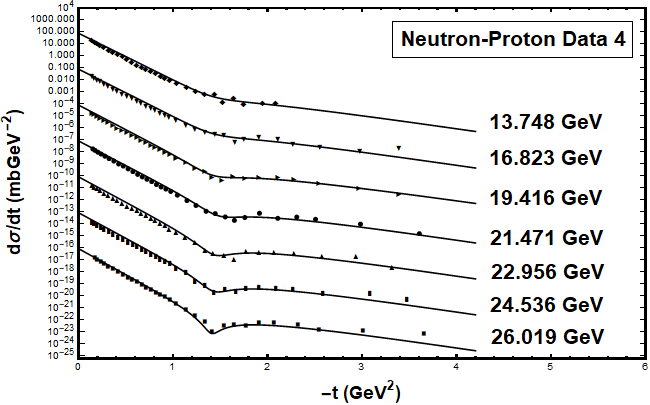}
			\label{fig:d}
		\end{subfigure}
		\caption{Same legend for parts (a), (b), (c), and (d) as in the Figure \ref{fig:model1fits} with Model 4}
		\label{fig:model4fits}
	\end{figure}

\begin{figure}[t!]
		\centering
		\begin{subfigure}[b]{0.48\textwidth}
			\centering
			\includegraphics[width=\textwidth]{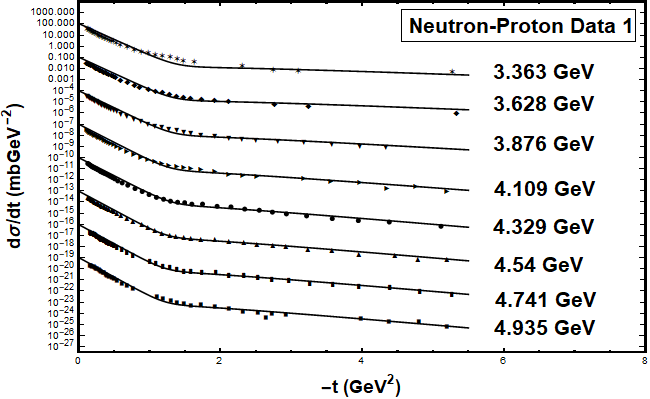}
			\label{fig:a}
		\end{subfigure}
		\hspace{0.05cm}
		\begin{subfigure}[b]{0.48\textwidth}
			\centering
			\includegraphics[width=\textwidth]{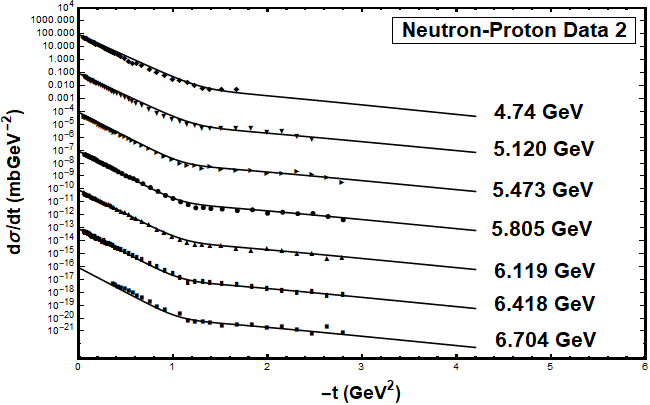}
			\label{fig:b}
		\end{subfigure}
		\vspace{-0.1cm}
		\begin{subfigure}[b]{0.48\textwidth}
			\centering
			\includegraphics[width=\textwidth]{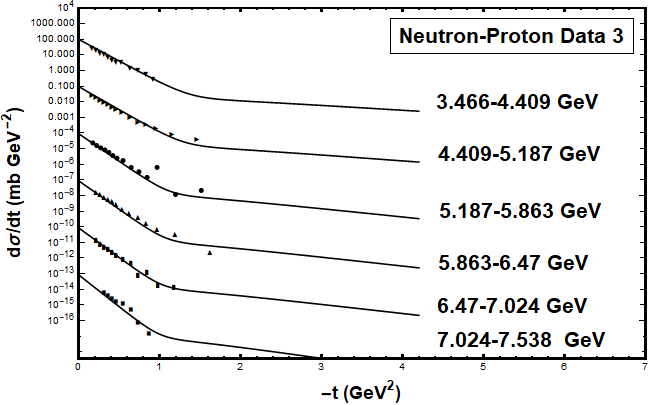}
			\label{fig:c}
		\end{subfigure}
		\hspace{0.05cm}
		\begin{subfigure}[b]{0.48\textwidth}
			\centering
			\includegraphics[width=\textwidth]{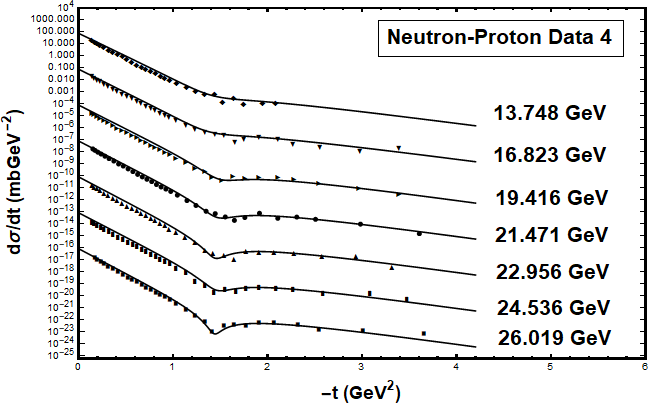}
			\label{fig:d}
		\end{subfigure}
		\caption{Same legend for parts (a), (b), (c), and (d) as in the Figure \ref{fig:model1fits} with Model 5}
		\label{fig:model5fits}
	\end{figure}

	In this study, we have performed a systematic fitting of elastic differential cross-section data mentioned in Table \ref{tab:expdata} by using five distinct composite exponential models. Each model is designed to capture different kinematic features of the data, reflecting the complex internal structure of the proton and neutron, and the energy-dependent nature of elastic scattering. Here, we discuss the results with emphasis on the possible physical insights extracted from the fitting behavior of our models with the data. The results of the obtained fits are shown in Figures \ref{fig:model1fits}, \ref{fig:model2fits}, \ref{fig:model3fits}, \ref{fig:model4fits}, and \ref{fig:model5fits}. The differential cross section for $np$ elastic scattering measured in the four energy domains probes distinct dynamical regimes of the nucleon interaction (mentioned in Section \ref{sec:I}) and can be used to study complementary aspects of the structure of proton and neutron \cite{RefDataset1,RefDataset2,RefDataset3,RefDataset4}. 
	
	The proposed models for the differential cross section obtained through this work show good fitting agreement and exhibit the same features as present in the data. In Dataset 1 \cite{RefDataset1}, while the dip is not sharply visible, the forward peak is less pronounced and exhibits shrinkage as the slope increases from 3.363 GeV to 4.935 GeV. At this lowest energy range, the elastic scattering may possess a partial-wave character that exhibits strong angular structure at moderate $\mid t\mid$ and sensitivity to spin-isospin components, indicating that meson-exchange and the resonance contributions still shape the profile of the differential cross section \cite{Bryan1970}. The forward cone is relatively broad (smaller $B$), reflecting a compact interaction region of order $\sim 1$ fm, which is also expected from nucleon form factors and potential models~\cite{Collins:1977jy,Donnachie:2002en,Barone:2002cv}. 
	
	In Dataset 2 \cite{RefDataset2}, the diffractive cone is more pronounced at small $\mid t\mid$, while a nascent dip appears at moderate $\mid t\mid$. In Dataset 3 \cite{RefDataset3}, the models predict the differential cross section in the regions beyond the highest $\mid t\mid$-values of the data, also showing the shrinkage of the diffraction peak with increasing slope $B$ as energy increases from 3.466-4.409 GeV to 7.024-7.538 GeV. The fits by our models show that the diffraction peak narrows, and the dip becomes more pronounced and shifts towards smaller $\mid t\mid$-values with energy. This collective behavior is consistent with Regge-trajectory expectations and geometric interpretations of the impact-parameter profile in eikonal frameworks ~\cite{Donnachie:2002en,Collins:1977jy,BSW2014}. 
	
	In Dataset 4 \cite{RefDataset4}, a dip is observed at 1.4 $(\text{GeV/c})^{2}$, resembling the dip of elastic $pp$ scattering at ISR energies \cite{ABreakstone}, but not as pronounced as a dip-bump structure which is observed at LHC energies in elastic $pp$ scattering \cite{Antchev_2019}. The dip is less prominent at $13.748$ GeV and $16.823$ GeV and appears at $19.416$ GeV. From $19.416$ GeV onward, the dip becomes sharper and narrower up to $26.019$ GeV. Between $13.748$ GeV and $26.019$ GeV, an increase in the slope and the shrinkage of the forward peak can be observed. The position of the dip slightly shifts towards lower $\mid t\mid$-values with increasing energy, while the forward cone is sharp and Pomeron dynamics largely controls the small-$\mid t\mid$ behavior. The depth and exact $\mid t\mid$ location of the dip reflect the interference of C-even and possible C-odd contributions as well as changing absorptive corrections in the eikonal unitarization that have been significantly investigated in the $pp$ and $p\bar{p}$ data~\cite{Selyugin2021,PhysRevLett.127.062003,BSW2014}. These features are also important in spin-related investigations of the $np$ scattering \cite{Selugin2024}. At large $\mid t\mid$, significant information is encoded in short-range components and nucleon substructure, related to hard scattering or the falloff of nucleon form factors \cite{Donnachie:2002en,Arrington2007,Perdrisat2007}.
	\subsection{Comparisons of the Models}
	The proposed phenomenological models possess a compact form and encode key dynamical features of high-energy elastic scattering while maintaining flexibility to capture detailed structures of the data. In Models 1, 3, 4, and 5, an efficient description of the dip-bump structure (observed in the TeV energy domain) can be obtained by the Gaussian modulation terms, consistent with interference in the scattering amplitude and possible transverse inhomogeneities such as hot-spot structures inside the nucleons that are relevant across $pp$, $p\bar{p}$, $np$, and $n\bar{p}$ channels with sensitivity to the interplay of crossing-even and crossing-odd exchanges \cite{Saleem1979,TOTEM:2013vij}. In Models 2 and 3, the multi-exponential structures emulate non-factorizable behavior reminiscent of squared form factors that enable multiple diffractive scales associated with distinct gluonic configurations or hotspot structures inside hadrons \cite{Baldenegro:2022,Bialas:2007}. In Model 4, the logarithmic t-dependent term, $\ln(1+t/t_0)$, is effective as it may capture deviations from pure exponential behavior by accounting for nonlinear Regge trajectories, including multi-Pomeron (unitarity) effects and threshold contributions (e.g., two-pion loops) that introduce branch-point singularities, consistent with expectations from Reggeon field theory \cite{Kaidalov:2003}. Model 5 incorporates a damping factor $(1-e^{-\gamma|t|})$ that vanishes at $t\to 0$ and suppresses forward contributions. Because this enhances the sensitivity to $C=-1$ exchanges, one can isolate the differences between $pp$ and $p\bar p$ scattering through interference of $C=+1$ and $C=-1$ amplitudes. This approach is fully consistent with analyticity and forward-scattering constraints~\cite{Lukaszuk1973,Royon2022}. 
	
	The proposed models may yield useful phenomenological insights when subjected to model comparisons with other models of elastic scattering. A simple Regge pole model for $np$ scattering \cite{Saleem1979} writes the independent helicity amplitudes as sums of exchanges with linear trajectories, exponential residues, and signature factors. Our models can describe the same behavior through energy dependences $(s/s_0)^{\alpha,\delta_O,\delta_I}$ acting as effective intercepts, exponential $t$-dependence encoding linear trajectories and shrinkage of the diffraction cone. While the discrete sign factors $(\xi_C,\xi_I)$ allowing to accommodate isoscalar and isovector exchanges across $pp$, $p\bar{p}$, $np$, and $n\bar{p}$ channels. A Regge theory-based framework \cite{Donnachie2013} includes soft Pomeron, secondary Reggeons, double-Pomeron screening, and perturbative three-gluon contributions to explain the elastic $pp$ and $p\bar{p}$ scattering. The multi-component exponential structures in our models can provide an observable-level realization of the same hierarchy of the Regge exchanges, where Pomeron dominance and the Reggeon contributions can be described by leading and subleading terms, respectively. 
	
	Another prominent approach in the Regge limit ($s\gg t$) is Holographic QCD \cite{Watanabe2023} that models the Pomeron and Reggeon exchanges through Reggeized spin-2 glueball and vector meson propagators, respectively. Our models do not explicitly implement the propagator structure or the underlying holographic dual dynamics. However, the energy exponents can be treated as effective trajectory intercepts, while the slope parameters $(B_0,\beta)$ accommodate transverse diffusion and trajectory slopes. The coefficients $(A_0,O_0,I_0)$ can describe overlap functions or coupling strengths, capturing the contribution of different exchange mechanisms. The multi-exponential and Gaussian terms can be used to represent multiple trajectories, describing the data in regions where the Holographic approach only provides qualitative predictions. 
	
	The Phillips-Barger (PB) model \cite{PHILLIPS1973412} is an empirical parametrization which describes elastic scattering through a sum of spin coherent exponential amplitudes whose interference reproduces the diffraction cone and dip-bump structure. In our models, this structure is retained, where each component can be attributed to specific quantum numbers (e.g., Pomerons, Odderons, and Reggeons), providing a physically motivated extension of the PB model by giving these interference effects a dynamical interpretation based on crossing-even and crossing-odd contributions. 
	
	In the approach of Bourrely-Soffer-Wu (BSW) \cite{BSW2014}, a refined matter distribution is tied to the charge distribution of the proton by incorporating Regge background contributions and a non-trivial hadronic matter current, which account for the spatial overlap and opacity function in the transverse impact parameter space. In our models, multiple exponential terms allow for an approximate inverse Fourier representation of complex impact parameter distributions. The good fit agreement of the models with differential cross-section data, including the forward peak, dip-bump structure, and high-$|t|$ region, shows their utility as effective phenomenological tools for probing the underlying complex QCD dynamics of elastic hadronic scattering.
	\subsection{Model Parameters, Correlations, and Information Criteria}
	The parameter values of the models which have been found by fitting all the data at the c.m. energies of Table \ref{tab:expdata} are mentioned in the Tables \ref{tab:pmodel1}, \ref{tab:pmodel2}, \ref{tab:pmodel3}, \ref{tab:pmodel4}, and \ref{tab:pmodel5}. The estimated bounds or fixed values of the parameters, along with the assumed bounds, are mentioned in Table \ref{tab:pranges}. These estimated bounds of the parameters, obtained by the $np$ data fitting can be used to use the models to fit the data of other elastic scattering processes. These bounds can be useful to investigate the collective behavior of the parameters across all the scattering processes. These bounds can be used to identify intervals for the forward slopes, energy exponents, deformation strengths, and the corrections related to short-range dynamics, so that stable and mutually consistent trends can be extracted from the data to find universal phenomenological constraints. Fractional error in the fitting has been determined by chi-square ($\chi^{2}$) that is a normalized error measure, particularly useful to determine goodness-of-fit. The dimensionless $\chi^{2}$ is given as follows,
	\begin{equation}
		\chi^2 = \frac{\sum_{i=1}^{N}[(d\sigma/dt)^{\text{model}}_i - (d\sigma/dt)^{\text{data}}_i]^{2}}{\sum_{i=1}^{N} [(d\sigma/dt)^{\text{data}}_i]^{2}}.
	\end{equation}
	with i = 1, 2, 3,...,N, number of data points. The $\chi^{2}$ values for each model across all the energies are written in Tables \ref{tab:pmodel1}, \ref{tab:pmodel2}, \ref{tab:pmodel3}, \ref{tab:pmodel4}, and \ref{tab:pmodel5}. 
	
	An analysis based on the product-moment correlation matrix has been carried out to statistically evaluate and compare correlations and stability of the parameters of all the models, across 28 distinct c.m. energies. The correlation framework \cite{Joseph1988} evaluates the localized linear dependency and phase-space coupling between any pair of extracted model parameters ($X, Y$) across the complete energy domain. The covariance of the parameters against the product of their respective standard deviations is scaled by the correlation coefficient $\rho_{X, Y}$ (or $r$), formulated as:
	\begin{equation}
		\rho_{X,Y} = \frac{\text{cov}(X,Y)}{\sigma_{X}\sigma_{Y}} = \frac{\sum_{i=1}^{n}(X_{i}-\bar{X})(Y_{i}-\bar{Y})}{\sqrt{\sum_{i=1}^{n}(X_{i}-\bar{X})^2}\sqrt{\sum_{i=1}^{n}(Y_{i}-\bar{Y})^2}}
	\end{equation}
	where $n = 28$ denotes the total number of c.m. energy bins; $X_{i}$ and $Y_{i}$ represent the localized and optimized parameter values determined by data fitting at the $i$-th energy point; $\bar{X}$ and $\bar{Y}$ indicate the respective means across the entire energy range; and $\sigma_{X}, \sigma_{Y}$ represent the sample standard deviations. It first establishes baseline averages, next, it tracks structural covariation using direction-dependent deviations, and finally, it normalizes the result to produce a dimensionless value bounded within $[-1.0, +1.0]$. The analysis of the topography of the correlation matrices across the five potential models provides a framework for optimal model selection based on: \textbf{(i)} Parametric Orthogonality and Minimization of Redundancy: An off-diagonal correlation landscape approaching zero ($|\rho_{X,Y}| \ll 0.5$) should be exhibited by an ideal model, which maximizes parameter independence. Over-parameterization is revealed by high correlations ($|\rho_{X,Y}| \to 1.0$). \textbf{(ii)} Sensitivity vs. Absolute Global Constraints: Parameters that maintain constant values across all the energies result in zero-variance columns and produce undefined correlations. The best model isolates global scaling parameters from localized, energy-dependent active parameters. \textbf{(iii)} Statistical Parsimony: The optimal model is identified by comparing the matrices of 5 models, as the one that achieves excellent data description ($\chi^2/\text{d.o.f.}$) while minimizing global parameter couplings. Models for which the multi-parameter correlation clusters (e.g., $|\rho| > 0.8$) are widespread are structurally unstable models. The correlation matrices for all the models are shown in the Tables \ref{tab:matrix1}, \ref{tab:matrix2}, \ref{tab:matrix3}, \ref{tab:matrix4}, and \ref{tab:matrix5}, and their statistical metrics are shown in Table \ref{tab:aic_bic}. In these matrices the parameters that are fixed during fitting and show no correlations are excluded. According to the discussed viewpoints, Model 4 is the optimal model as per the statistical evaluation across the 28 c.m. energy points, achieving the lowest penalty score of 1.1387. While a marginally lower mean absolute correlation ($\langle \vert{}r\vert{} \rangle = 0.3324$) is presented by Model 3, Model 4 offers superior parametric parsimony as it tracks the data fitting with only 4 active parameters while maintaining 12 stable constants. Models 4 and 3 exhibit zero perfect redundancies ($\ge 0.98$) and only a single high parameter dependency ($>0.80$). According to this approach, Models 1, 2, and 5 demonstrate structural instability, characterized by widespread multi-parameter correlation clusters, culminating in  Model 5, which has 6 perfect redundancies, 20 high dependencies, and a high penalty score of $70.1147$.
	\renewcommand{\arraystretch}{0.720}
	\setlength{\tabcolsep}{0.650pt}
	\begin{table}[H]
		\centering
		\footnotesize
		\caption{Parameter values of Model 1 for the neutron-proton ($np$) elastic scattering data of Table \ref{tab:expdata}}
		\label{tab:pmodel1}
		\begin{tabular}{cccccccccccccccc}
			\toprule
			\textbf{$\sqrt{s}$} & \textbf{$A_0$} & \textbf{$\alpha$} & \textbf{$B_0$} & \textbf{$\beta$} & \textbf{$\epsilon_0$} & \textbf{$t_d$} & \textbf{$\Delta$} & \textbf{$O_O$} & \textbf{$\delta_O$} & \textbf{$D_O$} & \textbf{$I_0$} & \textbf{$\delta_I$} & \textbf{$E_0$} & \textbf{$\mu$} & \textbf{$\chi^{2}$} \\
			\midrule
			& ($\text{mb}/\textrm{GeV}^{2}$) &  & ($\textrm{GeV}^{-2}$) & ($\textrm{GeV}^{-2}$) &  & ($\textrm{GeV}^{2}$) & ($\textrm{GeV}^{2}$) & ($\text{mb}/\textrm{GeV}^{2}$) &  & ($\textrm{GeV}^{-2}$) & ($\text{mb}/\textrm{GeV}^{2}$) &  & ($\textrm{GeV}^{-2}$) &  & \\
			Dataset 1 &&&&&&&&&&&&&&& \\
			3.363 ${\textrm{GeV}}$ & 89.25 & 0.085 & 4.6 & 0.4 & 0.08 & -0.8 & 0.2 & 0.2 & 0.12 & 0.85 & 0.06 & 0.12 & 2.9 & 0.09 & 0.189 \\
			3.628 ${\textrm{GeV}}$ & 86.0 & 0.075 & 4.7 & 0.4 & 0.08 & -0.2 & 0.2 & 0.04 & 0.1 & 0.65 & 0.06 & 0.1 & 2.9 & 0.009 & 0.3151 \\
			3.876 ${\textrm{GeV}}$ & 82.5 & 0.085 & 5.0 & 0.4 & 0.08 & -0.2 & 0.2 & 0.04 & 0.1 & 1.02 & 0.06 & 0.1 & 2.9 & 0.009 & 0.1623 \\
			4.109 ${\textrm{GeV}}$ & 71.5 & 0.12 & 5.9 & 0.4 & 0.08 & -0.2 & 0.2 & 0.04 & 0.1 & 1.15 & 0.06 & 0.1 & 2.9 & 0.009 & 0.0404 \\
			4.329 ${\textrm{GeV}}$ & 69.25 & 0.12 & 5.8 & 0.4 & 0.08 & -0.2 & 0.2 & 0.04 & 0.1 & 1.3 & 0.06 & 0.1 & 2.9 & 0.009 & 0.0736 \\
			4.54 ${\textrm{GeV}}$ & 66.8 & 0.12 & 5.8 & 0.4 & 0.08 & -0.2 & 0.2 & 0.04 & 0.1 & 1.3 & 0.06 & 0.1 & 2.9 & 0.009 & 0.0752 \\
			4.741 ${\textrm{GeV}}$ & 65.0 & 0.12 & 5.6 & 0.4 & 0.08 & -0.2 & 0.2 & 0.04 & 0.1 & 1.3 & 0.06 & 0.1 & 2.9 & 0.009 & 0.1084 \\
			4.935 ${\textrm{GeV}}$ & 64.0 & 0.12 & 5.5 & 0.4 & 0.08 & -0.2 & 0.2 & 0.04 & 0.12 & 1.35 & 0.06 & 0.12 & 2.9 & 0.009 & 0.2733 \\
			\midrule
			Dataset 2 &&&&&&&&&&&&&&& \\
			4.74 ${\textrm{GeV}}$ & 61.5 & 0.13 & 5.5 & 0.4 & 0.32 & -0.2 & 0.5 & 0.04 & 0.125 & 1.15 & 0.42 & 0.125 & 3.0 & 0.06 & 0.0709 \\
			5.12 ${\textrm{GeV}}$ & 58.5 & 0.135 & 5.8 & 0.4 & 0.32 & -0.2 & 0.5 & 0.04 & 0.125 & 1.25 & 0.42 & 0.125 & 3.0 & 0.06 & 0.0348 \\
			5.473 ${\textrm{GeV}}$ & 56.473 & 0.135 & 6.02 & 0.4 & 0.32 & -0.2 & 0.5 & 0.04 & 0.125 & 1.35 & 0.4 & 0.125 & 3.0 & 0.06 & 0.0735 \\
			5.805 ${\textrm{GeV}}$ & 55.9 & 0.13 & 5.65 & 0.4 & 0.32 & -0.2 & 0.5 & 0.04 & 0.125 & 1.35 & 0.42 & 0.125 & 3.0 & 0.06 & 0.0285\\
			6.119 ${\textrm{GeV}}$ & 54.2 & 0.13 & 5.65 & 0.4 & 0.32 & -0.2 & 0.5 & 0.04 & 0.125 & 1.35 & 0.42 & 0.125 & 3.0 & 0.06 & 0.019 \\
			6.418 ${\textrm{GeV}}$ & 52.8 & 0.13 & 5.65 & 0.4 & 0.32 & -0.2 & 0.5 & 0.04 & 0.125 & 1.35 & 0.42 & 0.125 & 3.0 & 0.06 & 0.0268\\
			6.704 ${\textrm{GeV}}$ & 51.20 & 0.13 & 5.45 & 0.4 & 0.32 & -0.2 & 0.5 & 0.04 & 0.125 & 1.35 & 0.42 & 0.125 & 3.0 & 0.06 & 0.0054 \\
			\midrule
			Dataset 3 &&&&&&&&&&&&&&& \\			
			3.466-4.409 ${\textrm{GeV}}$ & 71.45 & 0.13 & 4.2 & 0.4 & 0.32 & -0.2 & 0.5 & 0.04 & 0.125 & 1.35 & 0.42 & 0.125 & 3.0 & 0.06 & 0.1987 \\
			4.409-5.187 ${\textrm{GeV}}$ & 64.0 & 0.13 & 4.45 & 0.4 & 0.32 & -0.2 & 0.5 & 0.04 & 0.125 & 1.35 & 0.42 & 0.125 & 3.0 & 0.06 & 0.0861 \\
			5.187-5.863 ${\textrm{GeV}}$ & 60.15 & 0.13 & 4.65 & 0.4 & 0.32 & -0.2 & 0.5 & 0.04 & 0.125 & 1.35 & 0.42 & 0.125 & 3.0 & 0.06 & 0.0629 \\
			5.863-6.4 ${\textrm{GeV}}$ & 56.7 & 0.13 & 5.45 & 0.4 & 0.32 & -0.2 & 0.5 & 0.04 & 0.125 & 1.35 & 0.42 & 0.125 & 3.0 & 0.06 & 0.0003 \\
			6.47-7.024 ${\textrm{GeV}}$ & 50.650 & 0.13 & 5.85 & 0.4 & 0.32 & -0.2 & 0.5 & 0.04 & 0.125 & 1.35 & 0.42 & 0.125 & 3.0 & 0.06 & 0.0085 \\
			7.024-7.538 ${\textrm{GeV}}$ & 49.0 & 0.13 & 6.25 & 0.4 & 0.32 & -0.2 & 0.5 & 0.04 & 0.125 & 1.35 & 0.42 & 0.125 & 3.0 & 0.06 & 0.0141 \\
			\midrule
			Dataset 4 &&&&&&&&&&&&&&& \\
			13.748 ${\textrm{GeV}}$ & 41.8 & 0.123 & 4.7 & 0.4 & 0.08 & -0.2 & 0.2 & 0.04 & 0.15 & 2.73 & 0.06 & 0.15 & 3.0 & 0.009 & 0.0447 \\
			16.823 ${\textrm{GeV}}$ & 40.2 & 0.123 & 4.9 & 0.4 & 0.3 & -0.2 & 0.2 & 0.04 & 0.1 & 2.68 & 0.06 & 0.1 & 3.0 & 0.009 & 0.1919 \\
			19.416 ${\textrm{GeV}}$ & 38.2 & 0.123 & 5.25 & 0.4 & 0.3 & -0.2 & 0.2 & 0.04 & 0.1 & 2.73 & 0.06 & 0.1 & 3.0 & 0.009 & 0.1963 \\
			21.471 ${\textrm{GeV}}$ & 38.5 & 0.123 & 5.25 & 0.4 & 0.4 & -0.5 & 0.2 & 0.04 & 0.1 & 2.74 & 0.06 & 0.1 & 3.0 & 0.009 & 0.1651 \\
			22.956 ${\textrm{GeV}}$ & 38.3 & 0.123 & 5.3 & 0.4 & 0.4 & -0.5 & 0.2 & 0.04 & 0.1 & 2.74 & 0.06 & 0.1 & 2.99 & 0.01 & 0.3596 \\
			24.536 ${\textrm{GeV}}$ & 38.0 & 0.125 & 5.3 & 0.4 & 0.6 & -0.5 & 0.2 & 0.04 & 0.1 & 2.74 & 0.06 & 0.1 & 2.99 & 0.01 & 0.9788 \\
			26.019 ${\textrm{GeV}}$ & 32.0 & 0.15 & 5.45 & 0.4 & 0.8 & -0.5 & 0.2 & 0.04 & 0.1 & 2.74 & 0.06 & 0.1 & 2.99 & 0.01 & 1.1303 \\
			\bottomrule
		\end{tabular}
	\end{table}
	\renewcommand{\arraystretch}{0.720}
	\setlength{\tabcolsep}{0.650pt}
	\begin{table}[H]
		\centering
		\footnotesize
		\caption{Parameter values of Model 2 for the neutron-proton ($np$) elastic scattering data of Table \ref{tab:expdata}}
		\label{tab:pmodel2}
		\begin{tabular}{cccccccccccccccc}
			\toprule
			\textbf{$\sqrt{s}$} & \textbf{$A_0$} & \textbf{$\alpha$} & \textbf{$a_1$} & \textbf{$B_1^{(0)}$} & \textbf{$a_2$} & \textbf{$B_2^{(0)}$} & \textbf{$\beta$} & \textbf{$O_O$} & \textbf{$\delta_O$} & \textbf{$D_O$} & \textbf{$I_0$} & \textbf{$\delta_I$} & \textbf{$E_0$} & \textbf{$\mu$} & \textbf{$\chi^{2}$} \\
			\midrule
			& ($\text{mb}/\textrm{GeV}^{2}$) &  &  & ($\textrm{GeV}^{-2}$) &  & ($\textrm{GeV}^{-2}$) & ($\textrm{GeV}^{-2}$) & ($\text{mb}/\textrm{GeV}^{2}$) &  & ($\textrm{GeV}^{-2}$) & ($\text{mb}/\textrm{GeV}^{2}$) &  & ($\textrm{GeV}^{-2}$) &  & \\
			Dataset 1 &&&&&&&&&&&&&&& \\
			3.363 ${\textrm{GeV}}$ & 33.6 & 0.12 & 1.5 & 4.6 & 1.0 & 4.4 & 0.4 & 0.04 & 0.1 & 0.65 & 0.04 & 0.12 & 2.75 & 0.001 & 0.138 \\
			3.628 ${\textrm{GeV}}$ & 31.25 & 0.12 & 1.5 & 4.75 & 1.0 & 4.8 & 0.4 & 0.04 & 0.1 & 0.8 & 0.04 & 0.12 & 2.75 & 0.001 & 0.2182 \\
			3.876 ${\textrm{GeV}}$ & 29.7 & 0.12 & 1.5 & 5.7 & 1.0 & 5.4 & 0.4 & 0.04 & 0.1 & 1.0 & 0.04 & 0.12 & 2.75 & 0.001 & 0.0368 \\
			4.109 ${\textrm{GeV}}$ & 28.65 & 0.12 & 1.5 & 6.75 & 1.0 & 5.43 & 0.4 & 0.04 & 0.1 & 1.2 & 0.04 & 0.12 & 2.75 & 0.001 & 0.0065 \\
			4.329 ${\textrm{GeV}}$ & 26.35 & 0.12 & 1.5 & 6.85 & 1.0 & 5.48 & 0.4 & 0.04 & 0.1 & 1.4 & 0.04 & 0.12 & 2.75 & 0.001 & 0.003 \\
			4.54 ${\textrm{GeV}}$ & 25.6 & 0.12 & 1.5 & 6.9 & 1.0 & 5.485 & 0.4 & 0.04 & 0.1 & 1.4 & 0.04 & 0.12 & 2.75 & 0.001 & 0.0026 \\
			4.741 ${\textrm{GeV}}$ & 24.9 & 0.12 & 1.5 & 6.95 & 1.0 & 5.495 & 0.4 & 0.04 & 0.1 & 1.4 & 0.04 & 0.12 & 2.75 & 0.001 & 0.0079 \\
			4.935 ${\textrm{GeV}}$ & 25.6 & 0.12 & 1.5 & 7.0 & 1.0 & 5.5 & 0.4 & 0.04 & 0.1 & 1.4 & 0.04 & 0.12 & 2.75 & 0.001 & 0.0447 \\
			\midrule
			Dataset 2 &&&&&&&&&&&&&&& \\
			4.74 ${\textrm{GeV}}$ & 24.5 & 0.13 & 1.5 & 5.9 & 1.0 & 6.55 & 0.4 & 0.04 & 0.1 & 1.5 & 0.06 & 0.12 & 2.75 & 0.001 & 0.016 \\
			5.12 ${\textrm{GeV}}$ & 23.6 & 0.13 & 1.5 & 5.0 & 1.0 & 6.55 & 0.4 & 0.04 & 0.1 & 1.5 & 0.06 & 0.12 & 2.75 & 0.001 & 0.0156 \\
			5.473 ${\textrm{GeV}}$ & 22.7 & 0.13 & 1.5 & 5.7 & 1.0 & 6.55 & 0.4 & 0.04 & 0.1 & 1.5 & 0.06 & 0.12 & 2.75 & 0.001 & 0.0295 \\
			5.805 ${\textrm{GeV}}$ & 22.0 & 0.13 & 1.5 & 5.8 & 1.0 & 6.5 & 0.4 & 0.04 & 0.1 & 1.6 & 0.06 & 0.12 & 2.75 & 0.001 & 0.0023 \\
			6.119 ${\textrm{GeV}}$ & 21.5 & 0.13 & 1.5 & 5.4 & 1.0 & 6.75 & 0.4 & 0.04 & 0.1 & 1.6 & 0.06 & 0.12 & 2.75 & 0.001 & 0.0018 \\
			6.418 ${\textrm{GeV}}$ & 20.8 & 0.13 & 1.5 & 5.0 & 1.0 & 6.95 & 0.4 & 0.04 & 0.1 & 1.5 & 0.06 & 0.12 & 2.75 & 0.001 & 0.0035 \\
			6.704 ${\textrm{GeV}}$ & 20.35 & 0.13 & 1.5 & 5.5 & 1.0 & 7.55 & 0.4 & 0.04 & 0.1 & 1.5 & 0.06 & 0.12 & 2.75 & 0.001 & 0.1155 \\
			\midrule
			Dataset 3 &&&&&&&&&&&&&&& \\
			3.466-4.409 ${\textrm{GeV}}$ & 28.43 & 0.13 & 1.5 & 4.6 & 1.0 & 6.55 & 0.4 & 0.04 & 0.1 & 1.5 & 0.06 & 0.12 & 2.75 & 0.001 & 0.0006 \\
			4.409-5.187 ${\textrm{GeV}}$ & 25.54 & 0.13 & 1.5 & 4.6 & 1.0 & 6.55 & 0.4 & 0.04 & 0.1 & 1.5 & 0.06 & 0.12 & 2.75 & 0.001 & 0.0022 \\
			5.187-5.863 ${\textrm{GeV}}$ & 23.64 & 0.13 & 1.5 & 5.5 & 1.0 & 6.55 & 0.4 & 0.04 & 0.1 & 1.5 & 0.06 & 0.12 & 2.75 & 0.001 & 0.0216 \\
			5.863-6.47 ${\textrm{GeV}}$ & 22.27 & 0.13 & 1.5 & 5.4 & 1.0 & 6.55 & 0.4 & 0.04 & 0.1 & 1.5 & 0.06 & 0.12 & 2.75 & 0.001 & 0.0551 \\
			6.47-7.024 ${\textrm{GeV}}$ & 20.85 & 0.13 & 1.5 & 5.5 & 1.0 & 6.55 & 0.4 & 0.04 & 0.1 & 1.5 & 0.06 & 0.12 & 2.75 & 0.001 & 0.0102 \\
			7.024-7.538 ${\textrm{GeV}}$ & 20.35 & 0.13 & 1.5 & 5.6 & 1.0 & 6.55 & 0.4 & 0.04 & 0.1 & 1.5 & 0.06 & 0.12 & 2.75 & 0.001 & 0.0185 \\
			\midrule
			Dataset 4 &&&&&&&&&&&&&&& \\
			13.748 ${\textrm{GeV}}$ & 21.22 & 0.12 & 1.001 & 4.2 & 1.0 & 5.08 & 0.4 & 0.04 & 0.06 & 2.65 & 0.06 & 0.06 & 3.0 & 0.008 & 0.0202 \\
			16.823 ${\textrm{GeV}}$ & 19.47 & 0.12 & 1.001 & 4.3 & 1.0 & 5.08 & 0.4 & 0.04 & 0.08 & 2.64 & 0.06 & 0.08 & 3.0 & 0.008 & 0.027 \\
			19.416 ${\textrm{GeV}}$ & 20.1 & 0.12 & 1.001 & 4.6 & 1.0 & 5.08 & 0.4 & 0.04 & 0.14 & 2.72 & 0.06 & 0.14 & 3.0 & 0.008 & 0.0572 \\
			21.471 ${\textrm{GeV}}$ & 19.8 & 0.12 & 1.001 & 4.2 & 1.0 & 5.08 & 0.4 & 0.04 & 0.14 & 2.75 & 0.06 & 0.14 & 3.0 & 0.002 & 0.0153 \\
			22.956 ${\textrm{GeV}}$ & 19.7 & 0.12 & 1.001 & 4.3 & 1.0 & 5.08 & 0.4 & 0.04 & 0.14 & 2.75 & 0.06 & 0.14 & 3.0 & 0.004 & 0.0758 \\
			24.536 ${\textrm{GeV}}$ & 19.6 & 0.12 & 1.001 & 4.4 & 1.0 & 5.08 & 0.4 & 0.04 & 0.14 & 2.75 & 0.06 & 0.14 & 3.0 & 0.008 & 0.163 \\
			26.019 ${\textrm{GeV}}$ & 19.5685 & 0.12 & 1.001 & 4.5 & 1.0 & 5.08 & 0.4 & 0.04 & 0.14 & 2.75 & 0.06 & 0.14 & 3.0 & 0.008 & 0.0924 \\
			\bottomrule
		\end{tabular}
	\end{table}
	\renewcommand{\arraystretch}{0.720}
	\setlength{\tabcolsep}{0.650pt}
	\begin{table}[H]
		\centering
		\footnotesize
		\caption{Parameter values of Model 3 for the neutron-proton ($np$) elastic scattering data of Table \ref{tab:expdata}}
		\label{tab:pmodel3}
		\begin{tabular}{cccccccccccccccccc}
			\toprule
			\textbf{$\sqrt{s}$} & \textbf{$A_0$} & \textbf{$\alpha$} & \textbf{$f$} & \textbf{$B_1^{(0)}$} & \textbf{$B_2^{(0)}$} & \textbf{$\beta$} & \textbf{$\epsilon_0$}
			& \textbf{$t_d$} & \textbf{$\Delta$} & \textbf{$O_O$} & \textbf{$\delta_O$} & \textbf{$D_O$} & \textbf{$I_0$} & \textbf{$\delta_I$} & \textbf{$E_0$} & \textbf{$\mu$} & \textbf{$\chi^{2}$} \\
			\midrule
			& ($\text{mb}/\textrm{GeV}^{2}$) &  &  & ($\textrm{GeV}^{-2}$) & ($\textrm{GeV}^{-2}$) & ($\textrm{GeV}^{-2}$) &
			& ($\textrm{GeV}^{2}$) & ($\textrm{GeV}^{2}$) & ($\text{mb}/\textrm{GeV}^{2}$) &  & ($\textrm{GeV}^{-2}$) & ($\text{mb}/\textrm{GeV}^{2}$) &  & ($\textrm{GeV}^{-2}$) & & \\
			Dataset 1 &&&&&&&&&&&&&&&&& \\
			3.363 ${\textrm{GeV}}$ & 88.2 & 0.12 & 1.1 & 4.3 & 4.3 & 0.4 & 0.045 & -0.1 & 0.2 & 0.04 & 0.14 & 0.6 & 0.06 & 0.14 & 3.0 & 0.008 & 0.1745 \\
			3.628 ${\textrm{GeV}}$ & 81.9 & 0.12 & 1.1 & 4.8 & 5.08 & 0.4 & 0.045 & -0.1 & 0.2 & 0.04 & 0.14 & 0.7 & 0.06 & 0.14 & 3.0 & 0.008 & 0.2191 \\
			3.876 ${\textrm{GeV}}$ & 78.6 & 0.12 & 1.1 & 4.8 & 5.08 & 0.4 & 0.045 & -0.1 & 0.2 & 0.04 & 0.14 & 1.0 & 0.06 & 0.14 & 3.0 & 0.008 & 0.1205 \\
			4.109 ${\textrm{GeV}}$ & 75.0 & 0.12 & 1.1 & 5.2 & 5.08 & 0.4 & 0.045 & -0.1 & 0.2 & 0.04 & 0.14 & 1.25 & 0.06 & 0.14 & 3.0 & 0.008 & 0.0535 \\
			4.329 ${\textrm{GeV}}$ & 72.25 & 0.12 & 1.1 & 5.2 & 5.08 & 0.4 & 0.045 & -0.1 & 0.2 & 0.04 & 0.14 & 1.35 & 0.06 & 0.14 & 3.0 & 0.008 & 0.0766 \\
			4.54 ${\textrm{GeV}}$ & 69.7 & 0.12 & 1.1 & 5.5 & 5.08 & 0.4 & 0.045 & -0.1 & 0.2 & 0.04 & 0.14 & 1.35 & 0.06 & 0.14 & 3.0 & 0.008 & 0.0514 \\
			4.741 ${\textrm{GeV}}$ & 68.8 & 0.12 & 1.1 & 5.4 & 5.08 & 0.4 & 0.045 & -0.1 & 0.2 & 0.04 & 0.14 & 1.35 & 0.06 & 0.14 & 3.0 & 0.008 & 0.1206 \\
			4.935 ${\textrm{GeV}}$ & 67.15 & 0.12 & 1.1 & 5.2 & 5.08 & 0.4 & 0.045 & -0.1 & 0.2 & 0.04 & 0.14 & 1.35 & 0.06 & 0.14 & 3.0 & 0.008 & 0.2420 \\
			\midrule
			Dataset 2 &&&&&&&&&&&&&&&&& \\
			4.74 ${\textrm{GeV}}$ & 66.0 & 0.12 & 1.1 & 5.8 & 4.3 & 0.4 & 0.045 & -0.1 & 0.2 & 0.04 & 0.14 & 1.6 & 0.06 & 0.14 & 3.0 & 0.008 & 0.0193 \\
			5.12 ${\textrm{GeV}}$ & 63.4 & 0.12 & 1.1 & 5.8 & 4.3 & 0.4 & 0.045 & -0.1 & 0.2 & 0.04 & 0.14 & 1.6 & 0.06 & 0.14 & 3.0 & 0.008 & 0.0099 \\
			5.473 ${\textrm{GeV}}$ & 61.0 & 0.12 & 1.1 & 6.2 & 4.3 & 0.4 & 0.045 & -0.1 & 0.2 & 0.04 & 0.14 & 1.6 & 0.06 & 0.14 & 3.0 & 0.008 & 0.0183 \\
			5.805 ${\textrm{GeV}}$ & 59.5 & 0.12 & 1.1 & 6.1 & 4.2 & 0.4 & 0.045 & -0.1 & 0.2 & 0.04 & 0.14 & 1.8 & 0.06 & 0.14 & 3.0 & 0.0001 & 0.0012 \\
			6.119 ${\textrm{GeV}}$ & 57.8 & 0.12 & 1.1 & 6.3 & 4.3 & 0.4 & 0.045 & -0.1 & 0.2 & 0.04 & 0.14 & 1.6 & 0.06 & 0.14 & 3.0 & 0.008 & 0.0013 \\
			6.418 ${\textrm{GeV}}$ & 56.4 & 0.12 & 1.1 & 6.15 & 4.3 & 0.4 & 0.045 & -0.1 & 0.2 & 0.04 & 0.14 & 1.6 & 0.06 & 0.14 & 3.0 & 0.008 & 0.0025 \\
			6.704 ${\textrm{GeV}}$ & 55.4 & 0.12 & 1.1 & 5.8 & 4.3 & 0.4 & 0.045 & -0.1 & 0.2 & 0.04 & 0.14 & 1.6 & 0.06 & 0.14 & 3.0 & 0.008 & 0.0804 \\
			\midrule
			Dataset 3 &&&&&&&&&&&&&&&&& \\
			3.466-4.409 ${\textrm{GeV}}$ & 76.45 & 0.12 & 1.1 & 4.2 & 4.3 & 0.4 & 0.045 & -0.1 & 0.2 & 0.04 & 0.14 & 2.4 & 0.06 & 0.14 & 3.0 & 0.008 & 0.0510 \\
			4.409-5.187 ${\textrm{GeV}}$ & 69.0 & 0.12 & 1.1 & 4.2 & 4.3 & 0.4 & 0.045 & -0.1 & 0.2 & 0.04 & 0.14 & 1.4 & 0.06 & 0.14 & 3.0 & 0.008 & 0.0377 \\
			5.187-5.863 ${\textrm{GeV}}$ & 64.0 & 0.12 & 1.1 & 4.2 & 4.3 & 0.4 & 0.045 & -0.1 & 0.2 & 0.04 & 0.14 & 1.4 & 0.06 & 0.14 & 3.0 & 0.008 & 0.0334 \\
			5.863-6.47 ${\textrm{GeV}}$ & 60.4 & 0.12 & 1.1 & 5.0 & 4.3 & 0.4 & 0.045 & -0.1 & 0.2 & 0.04 & 0.14 & 1.4 & 0.06 & 0.14 & 3.0 & 0.008 & 0.0073 \\
			6.47-7.024 ${\textrm{GeV}}$ & 57.0 & 0.12 & 1.1 & 5.8 & 4.3 & 0.4 & 0.045 & -0.1 & 0.2 & 0.04 & 0.14 & 1.6 & 0.06 & 0.14 & 3.0 & 0.008 & 0.0125 \\
			7.024-7.538 ${\textrm{GeV}}$ & 55.4 & 0.12 & 1.1 & 5.8 & 4.3 & 0.4 & 0.045 & -0.1 & 0.2 & 0.04 & 0.14 & 1.6 & 0.06 & 0.14 & 3.0 & 0.008 & 0.0230 \\
			\midrule
			Dataset 4 &&&&&&&&&&&&&&&&& \\
			13.748 ${\textrm{GeV}}$ & 44.45 & 0.12 & 1.1 & 4.8 & 5.08 & 0.4 & 0.045 & -0.1 & 0.2 & 0.04 & 0.14 & 2.65 & 0.06 & 0.14 & 3.0 & 0.008 & 0.0136 \\
			16.823 ${\textrm{GeV}}$ & 42.77 & 0.12 & 1.1 & 4.8 & 5.08 & 0.4 & 0.045 & -0.1 & 0.2 & 0.04 & 0.14 & 2.65 & 0.06 & 0.14 & 3.0 & 0.008 & 0.0225 \\
			19.416 ${\textrm{GeV}}$ & 42.1 & 0.12 & 1.1 & 4.5 & 5.08 & 0.4 & 0.045 & -0.1 & 0.2 & 0.04 & 0.14 & 2.75 & 0.06 & 0.14 & 3.0 & 0.001 & 0.1021 \\
			21.471 ${\textrm{GeV}}$ & 41.5 & 0.12 & 1.1 & 4.35 & 5.08 & 0.4 & 0.045 & -0.1 & 0.2 & 0.04 & 0.14 & 2.75 & 0.06 & 0.14 & 3.0 & 0.0001 & 0.0352 \\
			22.956 ${\textrm{GeV}}$ & 41.25 & 0.12 & 1.1 & 4.45 & 5.08 & 0.4 & 0.045 & -0.1 & 0.2 & 0.04 & 0.14 & 2.75 & 0.06 & 0.14 & 3.0 & 0.0001 & 0.1124 \\
			24.536 ${\textrm{GeV}}$ & 41.0 & 0.12 & 1.1 & 4.7 & 5.08 & 0.4 & 0.045 & -0.1 & 0.2 & 0.04 & 0.14 & 2.75 & 0.06 & 0.14 & 3.0 & 0.008 & 0.1743 \\
			26.019 ${\textrm{GeV}}$ & 40.85 & 0.12 & 1.1 & 4.75 & 5.08 & 0.4 & 0.045 & -0.1 & 0.2 & 0.04 & 0.14 & 2.75 & 0.06 & 0.14 & 3.0 & 0.008 & 0.0996 \\
			\bottomrule
		\end{tabular}
	\end{table}
	\renewcommand{\arraystretch}{0.720}
	\setlength{\tabcolsep}{0.650pt}
	\begin{table}[H]
		\centering
		\footnotesize
		\caption{Parameter values of Model 4 for the neutron-proton ($np$) elastic scattering data of Table \ref{tab:expdata}}
		\label{tab:pmodel4}
		\begin{tabular}{cccccccccccccccccc}
			\toprule
			\textbf{$\sqrt{s}$} & \textbf{$A_0$} & \textbf{$\alpha$} & \textbf{$B_0$} & \textbf{$\beta$} & \textbf{$B_1$} & \textbf{$t_0$} & \textbf{$\epsilon_0$} & \textbf{$t_d$} & \textbf{$\Delta$} & \textbf{$O_O$} & \textbf{$\delta_O$} & \textbf{$D_O$} & \textbf{$I_0$} & \textbf{$\delta_I$} & \textbf{$E_0$} & \textbf{$\mu$} & \textbf{$\chi^{2}$} \\
			\midrule
			& ($\text{mb}/\textrm{GeV}^{2}$) &  & ($\textrm{GeV}^{-2}$) & ($\textrm{GeV}^{-2}$) & ($\textrm{GeV}^{-2}$) & ($\textrm{GeV}^{2}$) &  & ($\textrm{GeV}^{2}$) & ($\textrm{GeV}^{2}$) & ($\text{mb}/\textrm{GeV}^{2}$) & & ($\textrm{GeV}^{-2}$) & ($\text{mb}/\textrm{GeV}^{2}$) &   & ($\textrm{GeV}^{-2}$) &  & \\
			Dataset 1 &&&&&&&&&&&&&&&&& \\
			3.363 ${\textrm{GeV}}$ & 145.0 & 0.18 & 4.2 & 0.4 & 0.1 & 10 & 0.5 & -0.1 & 0.2 & 0.04 & 0.14 & 0.65 & 0.06 & 0.14 & 3.0 & 0.008 & 0.1887 \\
			3.628 ${\textrm{GeV}}$ & 133.9 & 0.18 & 5.0 & 0.4 & 0.1 & 10 & 0.5 & -0.1 & 0.2 & 0.04 & 0.14 & 0.65 & 0.06 & 0.14 & 3.0 & 0.008 & 0.1800 \\
			3.876 ${\textrm{GeV}}$ & 127.8 & 0.18 & 5.1 & 0.4 & 0.1 & 10 & 0.5 & -0.1 & 0.2 & 0.04 & 0.14 & 0.89 & 0.06 & 0.14 & 3.0 & 0.008 & 0.0853 \\
			4.109 ${\textrm{GeV}}$ & 120.5 & 0.18 & 5.5 & 0.4 & 0.1 & 10 & 0.5 & -0.1 & 0.2 & 0.04 & 0.14 & 1.25 & 0.06 & 0.14 & 3.0 & 0.008 & 0.0333 \\
			4.329 ${\textrm{GeV}}$ & 116.5 & 0.18 & 5.4 & 0.4 & 0.1 & 10 & 0.5 & -0.1 & 0.2 & 0.04 & 0.14 & 1.35 & 0.06 & 0.14 & 3.0 & 0.008 & 0.0651 \\
			4.54 ${\textrm{GeV}}$ & 112.95 & 0.18 & 5.3 & 0.4 & 0.1 & 10 & 0.5 & -0.1 &  0.2 & 0.04 & 0.14 & 1.4 & 0.06 & 0.14 & 3.0 & 0.008 & 0.0892 \\
			4.741 ${\textrm{GeV}}$ & 108.95 & 0.18 & 5.2 & 0.4 & 0.1 & 10 & 0.5 & -0.1 & 0.2 & 0.04 & 0.14 & 1.35 & 0.06 & 0.14 & 3.0 & 0.008 & 0.1669 \\
			4.935 ${\textrm{GeV}}$ & 105.95 & 0.18 & 4.45 & 0.4 & 0.1 & 10 & 0.5 & -0.1 & 0.2 & 0.04 & 0.14 & 1.425 & 0.06 & 0.14 & 3.0 & 0.008 & 0.5176 \\
			\midrule
			Dataset 2 &&&&&&&&&&&&&&&&& \\
			4.74 ${\textrm{GeV}}$ & 104.7 & 0.18 & 5.8 & 0.4 & 0.1 & 10 & 0.5 & -0.1 & 0.2 & 0.04 & 0.14 & 1.425 & 0.06 & 0.14 & 3.0 & 0.008 & 0.0235 \\
			5.12 ${\textrm{GeV}}$ & 99.8 & 0.18 & 5.7 & 0.4 & 0.1 & 10 & 0.5 & -0.1 & 0.2 & 0.04 & 0.14 & 1.425 & 0.06 & 0.14 & 3.0 & 0.008 & 0.0136 \\
			5.473 ${\textrm{GeV}}$ & 96.2 & 0.18 & 6.4 & 0.4 & 0.1 & 10 & 0.5 & -0.1 & 0.2 & 0.04 & 0.14 & 1.625 & 0.06 & 0.14 & 3.0 & 0.008 & 0.0226 \\
			5.805 ${\textrm{GeV}}$ & 92.7 & 0.18 & 6.3 & 0.4 & 0.1 & 10 & 0.5 & -0.1 & 0.2 & 0.04 & 0.14 & 1.625 & 0.06 & 0.14 & 3.0 & 0.008 & 0.0017 \\
			6.119 ${\textrm{GeV}}$ & 89.5 & 0.18 & 5.9 & 0.4 & 0.1 & 10 & 0.5 & -0.1 & 0.2 & 0.04 & 0.14 & 1.625 & 0.06 & 0.14 & 3.0 & 0.008 & 0.0019 \\
			6.418 ${\textrm{GeV}}$ & 86.2 & 0.18 & 5.7 & 0.4 & 0.1 & 10 & 0.5 & -0.1 & 0.2 & 0.04 & 0.14 & 1.625 & 0.06 & 0.14 & 3.0 & 0.008 & 0.0059 \\
			6.704 ${\textrm{GeV}}$ & 84.0 & 0.18 & 5.5 & 0.4 & 0.1 & 10 & 0.5 & -0.1 & 0.2 & 0.04 & 0.14 & 1.625 & 0.06 & 0.14 & 3.0 & 0.008 & 0.0171 \\
			\midrule
			Dataset 3 &&&&&&&&&&&&&&&&& \\
			3.466-4.409 ${\textrm{GeV}}$ & 124.0 & 0.18 & 4.25 & 0.4 & 0.1 & 10 & 0.5 & -0.1 & 0.2 & 0.04 & 0.14  & 1.625 & 0.06 & 0.14 & 3.0 & 0.008 & 0.0655 \\
			4.409-5.187 ${\textrm{GeV}}$ & 109.0 & 0.18 & 4.25 & 0.4 & 0.1 & 10 & 0.5 & -0.1 & 0.2 & 0.04 & 0.14 & 1.625 & 0.06 & 0.14 & 3.0 & 0.008 & 0.0332 \\
			5.187-5.863 ${\textrm{GeV}}$ & 99.5 & 0.18 & 5.5 & 0.4 & 0.1 & 10 & 0.5 & -0.1 & 0.2 & 0.04 & 0.14 & 1.625 & 0.06 & 0.14 & 3.0 & 0.008 & 0.0061 \\
			5.863-6.47 ${\textrm{GeV}}$ & 93.0 & 0.18 & 5.5 & 0.4 & 0.1 & 10 & 0.5 & -0.1 & 0.2 & 0.04 & 0.14 & 1.625 & 0.06 & 0.14 & 3.0 & 0.008 & 0.0283 \\
			6.47-7.024 ${\textrm{GeV}}$ & 86.0 & 0.18 & 5.5 & 0.4 & 0.1 & 10 & 0.5 & -0.1 & 0.2 & 0.04 & 0.14 & 1.625 & 0.06 & 0.14 & 3.0 & 0.008 & 0.0072 \\
			7.024-7.538 ${\textrm{GeV}}$ & 83.25 & 0.18 & 5.5 & 0.4 & 0.1 & 10 & 0.5 & -0.1 & 0.2 & 0.04 & 0.14 & 1.625 & 0.06 & 0.14 & 3.0 & 0.008 & 0.0003 \\
			\midrule
			Dataset 4 &&&&&&&&&&&&&&&&& \\
			13.748 ${\textrm{GeV}}$ & 61.95 & 0.18 & 4.75 & 0.4 & 0.1 & 10 & 0.5 & -0.1 & 0.2 & 0.04 & 0.14 & 2.65 & 0.06 & 0.14 & 3.0 & 0.008 & 0.0149 \\
			16.823 ${\textrm{GeV}}$ & 58.2 & 0.18 & 4.55 & 0.4 & 0.1 & 10 & 0.5 & -0.1 & 0.2 & 0.04 & 0.14 & 2.67 & 0.06 & 0.14 & 3.0 & 0.008 & 0.0367 \\
			19.416 ${\textrm{GeV}}$ & 56.0 & 0.18 & 4.7 & 0.4 & 0.1 & 10 & 0.5 & -0.1 & 0.2 & 0.04 & 0.14 & 2.72 & 0.06 & 0.14 & 3.0 & 0.001 & 0.0690 \\
			21.471 ${\textrm{GeV}}$ & 55.0 & 0.18 & 4.55 & 0.4 & 0.1 & 10 & 0.5 & -0.1 & 0.2 & 0.04 & 0.14 & 2.75 & 0.06 & 0.14 & 3.0 & 0.008 & 0.0207 \\
			22.956 ${\textrm{GeV}}$ & 54.25 & 0.18 & 4.75 & 0.4 & 0.1 & 10 & 0.5 & -0.1 & 0.2 & 0.04 & 0.14 & 2.75 & 0.06 & 0.14 & 3.0 & 0.008 & 0.0716 \\
			24.536 ${\textrm{GeV}}$ & 53.4 & 0.18 & 4.68 & 0.4 & 0.1 & 10 & 0.5 & -0.1 & 0.2 & 0.04 & 0.14 & 2.75 & 0.06 & 0.14 & 3.0 & 0.008 & 0.1756 \\
			26.019 ${\textrm{GeV}}$ & 52.8 & 0.18 & 4.78 & 0.4 & 0.1 & 10 & 0.5 & -0.1 & 0.2 & 0.04 & 0.14 & 2.75 & 0.06 & 0.14 & 3.0 & 0.008 & 0.0931 \\
			\bottomrule
		\end{tabular}
	\end{table}
	\renewcommand{\arraystretch}{0.720}
	\setlength{\tabcolsep}{0.650pt}
	\begin{table}[H]
		\centering
		\footnotesize
		\caption{Parameter values of Model 5 for the neutron-proton ($np$) elastic scattering data of Table \ref{tab:expdata}}
		\label{tab:pmodel5}
		\begin{tabular}{ccccccccccccccccc}
			\toprule
			\textbf{$\sqrt{s}$} & \textbf{$A_0$} & \textbf{$\alpha$} & \textbf{$B_0$} & \textbf{$\beta$} & \textbf{$\epsilon_0$} & \textbf{$t_d$} & \textbf{$\Delta$} & \textbf{$O_O$}  & \textbf{$\delta_O$} & \textbf{$D_O$} & \textbf{$\gamma$} & \textbf{$I_0$} & \textbf{$\delta_I$} & \textbf{$E_0$} & \textbf{$\mu$} &  \textbf{$\chi^{2}$} \\
			\midrule
			& ($\text{mb}/\textrm{GeV}^{2}$) &  & ($\textrm{GeV}^{-2}$) & ($\textrm{GeV}^{-2}$) &  & ($\textrm{GeV}^{2}$) & ($\textrm{GeV}^{2}$) &  ($\text{mb}/\textrm{GeV}^{2}$) &  & ($\textrm{GeV}^{-2}$) & ($\textrm{GeV}^{-2}$) & ($\text{mb}/\textrm{GeV}^{2}$) &  & ($\textrm{GeV}^{-2}$) &  & \\
			Dataset 1 &&&&&&&&&&&&&&&& \\
			3.363 ${\textrm{GeV}}$ & 561.7 & 0.12 & 4.65 & 0.4 & 0.85 & -0.1 & 0.1 & 0.04 & 0.14 & 0.452 & 0.5 & 0.06 & 0.14 & 3.0 & 0.08 & 0.1254 \\
			3.628 ${\textrm{GeV}}$ & 522.7 & 0.12 & 4.7 & 0.4 & 0.85 & -0.1 & 0.1 & 0.04 & 0.14 & 0.49 & 0.5 & 0.06 & 0.14 & 3.0 & 0.08 & 0.2403 \\
			3.876 ${\textrm{GeV}}$ & 496.0 & 0.12 & 4.9 & 0.4 & 0.85 & -0.1 & 0.1 & 0.04 & 0.14 & 0.69 & 0.5 & 0.06 & 0.14 & 3.0 & 0.08 & 0.0992 \\
			4.109 ${\textrm{GeV}}$ & 476.7 & 0.12 & 5.25 & 0.4 & 0.85 & -0.1 & 0.1 & 0.04 & 0.14 & 0.9 & 0.5 & 0.06 & 0.14 & 3.0 & 0.08 & 0.0510 \\
			4.329 ${\textrm{GeV}}$ & 463.0 & 0.12 & 5.15 & 0.4 & 0.85 & -0.1 & 0.1 & 0.04 & 0.14 & 1.0 & 0.5 & 0.06 & 0.14 & 3.0 & 0.08 & 0.0894 \\
			4.54 ${\textrm{GeV}}$ & 448.0 & 0.12 & 5.425 & 0.4 & 0.85 & -0.1 & 0.1 & 0.04 & 0.14 & 1.0 & 0.5 & 0.06 & 0.14 & 3.0 & 0.08 & 0.0699 \\
			4.741 ${\textrm{GeV}}$ & 437.0 & 0.12 & 5.45 & 0.4 & 0.85 & -0.1 & 0.1 & 0.04 & 0.14 & 1.0 & 0.5 & 0.06 & 0.14 & 3.0 & 0.08 & 0.1183 \\
			4.935 ${\textrm{GeV}}$ & 428.0 & 0.15 & 5.5 & 0.4 & 0.85 & -0.1 & 0.1 & 0.04 & 0.14 & 1.0 & 0.5 & 0.06 & 0.14 & 3.0 & 0.08 & 0.3190 \\
			\midrule
			Dataset 2 &&&&&&&&&&&&&&&& \\
			4.74 ${\textrm{GeV}}$ & 252.0 & 0.12 & 5.0 & 0.4 & 0.75 & -0.1 & 0.1 & 0.04 & 0.14 & 1.25 & 1.0 & 0.06 & 0.14 & 3.0 & 0.08 & 0.0461 \\
			5.12 ${\textrm{GeV}}$ & 242.0 & 0.12 & 5.2 & 0.4 & 0.75 & -0.1 & 0.1 & 0.04 & 0.14 & 1.25 & 1.0 & 0.06 & 0.14 & 3.0 & 0.08 & 0.0207 \\
			5.473 ${\textrm{GeV}}$ & 234.0 & 0.12 & 5.9 & 0.4 & 0.75 & -0.1 & 0.1 & 0.04 & 0.14 & 1.25 & 1.0 & 0.06 & 0.14 & 3.0 & 0.08 & 0.0317 \\
			5.805 ${\textrm{GeV}}$ & 227.0 & 0.12 & 5.95 & 0.4 & 0.75 & -0.1 & 0.1 & 0.04 & 0.14 & 1.25 & 1.0 & 0.06 & 0.14 & 3.0 & 0.08 & 0.0027 \\
			6.119 ${\textrm{GeV}}$ & 221.0 & 0.12 & 5.6 & 0.4 & 0.75 & -0.1 & 0.1 & 0.04 & 0.14 & 1.25 & 1.0 & 0.06 & 0.14 & 3.0 & 0.08 & 0.0030 \\
			6.418 ${\textrm{GeV}}$ & 215.5 & 0.12 & 5.7 & 0.4 & 0.75 & -0.1 & 0.1 & 0.04 & 0.14 & 1.25 & 1.0 & 0.06 & 0.14 & 3.0 & 0.08 & 0.0058 \\
			6.704 ${\textrm{GeV}}$ & 211.0 & 0.12 & 5.8 & 0.4 & 0.75 & -0.1 & 0.1 & 0.04 & 0.14 & 1.25 & 1.0 & 0.06 & 0.14 & 3.0 & 0.08 & 0.0516 \\
			\midrule
			Dataset 3 &&&&&&&&&&&&&&&& \\
			3.466-4.409 ${\textrm{GeV}}$ & 292.3 & 0.12 & 4.2 & 0.4 & 0.75 & -0.1 & 0.1 & 0.04 & 0.14 & 0.6 & 1.0  & 0.06 & 0.14 & 3.0 & 0.08 & 0.0708 \\
			4.409-5.187 ${\textrm{GeV}}$ & 263.5 & 0.12 & 4.3 & 0.4 & 0.75 & -0.1 & 0.1 & 0.04 & 0.14 & 0.7 & 1.0  & 0.06 & 0.14 & 3.0 & 0.08 & 0.0292 \\
			5.187-5.863 ${\textrm{GeV}}$ & 244.5 & 0.12 & 5.4 & 0.4 & 0.75 & -0.1 & 0.1 & 0.04 & 0.14 & 0.95 & 1.0 & 0.06 & 0.14 & 3.0 & 0.08 & 0.0037 \\
			5.863-6.47 ${\textrm{GeV}}$ & 231.5 & 0.12 & 5.5 & 0.4 & 0.75 & -0.1 & 0.1 & 0.04 & 0.14 & 1.0 & 1.0 & 0.06 & 0.14 & 3.0 & 0.08 & 0.0289 \\
			6.47-7.024 ${\textrm{GeV}}$ & 217.35 & 0.12 & 5.6 & 0.4 & 0.75 & -0.1 & 0.1 & 0.04 & 0.14 & 1.0 & 1.0 & 0.06 & 0.14 & 3.0 & 0.08 & 0.0060 \\
			7.024-7.538 ${\textrm{GeV}}$ & 211.5 & 0.12 & 6.2 & 0.4 & 0.75 & -0.1 & 0.1 & 0.04 & 0.14 & 1.25 & 1.0 & 0.06 & 0.14 & 3.0 & 0.08 & 0.0430 \\
			\midrule
			Dataset 4 &&&&&&&&&&&&&&&& \\
			13.748 ${\textrm{GeV}}$ & 430.5 & 0.16 & 5.65 & 0.3 & 0.92 & -0.1 & 0.1 & 0.04 & 0.02 & 2.4 & 1.0 & 0.06 & 0.02 & 3.0 & 0.001 & 0.0241 \\
			16.823 ${\textrm{GeV}}$ & 408.5 & 0.16 & 5.65 & 0.3 & 0.92 & -0.1 & 0.1 & 0.04 & 0.02 & 2.4 & 1.0 & 0.06 & 0.02 & 3.0 & 0.001 & 0.0425 \\
			19.416 ${\textrm{GeV}}$ & 397.0 & 0.16 & 5.65 & 0.3 & 0.92 & -0.1 & 0.1 & 0.04 & 0.02 & 2.6 & 1.0 & 0.06 & 0.02 & 3.0 & 0.001 & 0.1076 \\
			21.471 ${\textrm{GeV}}$ & 387.0 & 0.16 & 5.7 & 0.3 & 0.92 & -0.1 & 0.1 & 0.04 & 0.02 & 2.6 & 1.0 & 0.06 & 0.02 & 3.0 & 0.001 & 0.0254 \\
			22.956 ${\textrm{GeV}}$ & 383.0 & 0.16 & 5.8 & 0.3 & 0.92 & -0.1 & 0.1 & 0.04 & 0.02 & 2.6 & 1.0 & 0.06 & 0.02 & 3.0 & 0.001 & 0.0973 \\
			24.536 ${\textrm{GeV}}$ & 379.0 & 0.16 & 5.6 & 0.3 & 0.92 & -0.1 & 0.1 & 0.04 & 0.02 & 2.6 & 1.0 & 0.06 & 0.02 & 3.0 & 0.001 & 0.2601 \\
			26.019 ${\textrm{GeV}}$ & 376.0 & 0.16 & 5.72 & 0.3 & 0.92 & -0.1 & 0.1 & 0.04 & 0.02 & 2.6 & 1.0 & 0.06 & 0.02 & 3.0 & 0.001 & 0.1646 \\
			\bottomrule
		\end{tabular}
	\end{table}
	\renewcommand{\arraystretch}{1.0}
	\setlength{\tabcolsep}{1.20pt}
	\begin{table}[!htbp]
		\centering
		\footnotesize
		\caption{Parameter bounds of the models estimated by fitting the elastic $np$ scattering data of Table \ref{tab:expdata}}
		\label{tab:pranges}
		\begin{tabular}{cccccccccccc}
			\toprule
			\textbf{Model 1} &  &  &  & \textbf{Model 2} &  &  &  & \textbf{Model 3} &  &  &  \\
			\midrule
			Parameter & Dimension & Estimated & Assumed  & Parameter & Dimension & Estimated & Assumed & Parameter & Dimension & Estimated & Assumed \\
			$A_0$ & $\text{mb}/\textrm{GeV}^{2}$ & [32.0,89.25] & [$10^{-6}$,$10^{3}$] & $A_0$ & $\text{mb}/\textrm{GeV}^{2}$ & [19.347.33.6] & [$10^{-6}$,$10^{3}$] & $A_0$ & $\text{mb}/\textrm{GeV}^{2}$ & [40.85,88.20] & [$10^{-6}$,$10^{3}$] \\
			$\alpha$ & & [0.075,0.15] & [0,0.2] & $\alpha$ & & [0.12,0.13] & [0,0.2] &  $\alpha$ &  & 0.12 & [0,0.2] \\
			$B_0$ & $\textrm{GeV}^{-2}$ & [4.20,6.25] & [2,30] & $a_1$ & & [1.0,1.5] & [0,10] & $f$ & & 1.1 & [0,1.5] \\
			$\beta$ & $\textrm{GeV}^{-2}$ & 0.4 & [0,1.0] &  $B_1^{(0)}$ & $\textrm{GeV}^{-2}$ & [4.2,7.0] & [2,30] & $B_1^{(0)}$ & $\textrm{GeV}^{-2}$ & [4.2,6.3] & [2,30] \\
			$\epsilon_0$ & & [0.08,0.8] & [0,1.0] & $a_2$ & & [1,1.1] & [0,10] &  $B_2^{(0)}$ & $\textrm{GeV}^{-2}$ & [4.3,5.08] & [2,30] \\
			$t_d$ & $\textrm{GeV}^{2}$ & [-0.8,-0.2] & [-1.2,-0.2] & $B_2^{(0)}$ & $\textrm{GeV}^{-2}$ & [4.4,7.55] & [2,30] & $\beta$ & $\textrm{GeV}^{-2}$  & 0.4 & [0,1.0] \\
			$\Delta$ & $\textrm{GeV}^{2}$ & [0.2,0.5] & [0.01,2] & $\beta$ & $\textrm{GeV}^{-2}$ & 0.4 & [0,1.0] & $\epsilon_0$ &  & 0.045 & [0,1.0] \\
			$O_O$ & $\text{mb}/\textrm{GeV}^{2}$ & [0.04,0.2] & [$10^{-8}$,10] & $O_O$ & $\text{mb}/\textrm{GeV}^{2}$ & 0.04 & [$10^{-8}$,10] & $t_d$ & $\textrm{GeV}^{2}$ & -0.1 & [-0.1,-0.2] \\
			$\delta_O$ & & [0.1,0.125] & [-0.1,1] & $\delta_O$ & & [0.06,0.14] & [-0.1,1] & $\Delta$ & $\textrm{GeV}^{2}$ & 0.2 & [0.01,2] \\
			$D_O$ & $\textrm{GeV}^{-2}$ & [0.65,2.74] & [0.1,25] & $D_O$ & $\textrm{GeV}^{-2}$ & [0.65,2.75] & [0.1,25] & $O_O$ & $\text{mb}/\textrm{GeV}^{2}$ & 0.04 & [$10^{-8}$,10] \\
			$I_0$ & $\text{mb}/\textrm{GeV}^{2}$ & [0.06,0.42] & [$10^{-8}$,10] & $I_0$ & $\text{mb}/\textrm{GeV}^{2}$ & [0.04,0.06] & [$10^{-8}$,10] & $\delta_O$ & & 0.14 & [-0.1,1] \\
			$\delta_I$ & & [0.1,0.125] & [-0.1,1] & $\delta_I$ &  & [0.06,0.14] & [-0.1,1] & $D_O$ & $\textrm{GeV}^{-2}$ & [0.6,2.75] & [0.1,25] \\
			$E_0$ & $\textrm{GeV}^{-2}$ &  [2.9,3.0] & [0.1,25] & $E_0$ & $\textrm{GeV}^{-2}$ & [2.75,3.0] & [0.1,25] & $I_0$ & $\text{mb}/\textrm{GeV}^{2}$ & 0.06 & [$10^{-8}$,10] \\
			$\mu$ & & [0.009,0.09] & [0,1] & $\mu$ & & [0.001,0.008] & [0,1] & $\delta_I$ & & 0.14 & [-0.1,1] \\
			& & & & & & & & $E_0$ & $\textrm{GeV}^{-2}$ & 3.0 & [0.1,25] \\
			& & & & & & & & $\mu$ & & [0.0001,0.008] & [0,1] \\
			\midrule
			\textbf{Model 4}  &  &  &  & \textbf{Model 5} &  &  &  &  &  &  & \\
			\toprule
			Parameter & Dimension & Estimated & Assumed & Parameter & Dimension & Estimated & Assumed &  &  &  &  \\
			$A_0$ & $\text{mb}/\textrm{GeV}^{2}$ & [52.8,145] & [$10^{-6}$,$10^{3}$] & $A_0$ & $\text{mb}/\textrm{GeV}^{2}$ & [211,561.7] & [$10^{-6}$,$10^{3}$] &  &  &  &  \\
			$\alpha$ &  & 0.18 & [0,0.2] & $\alpha$ &  & [0.12,0.16] & [0,0.2] &  &  & & \\
			$B_0$ & $\textrm{GeV}^{-2}$ & [4.2,6.4] & [2,30] & $B_0$ & $\textrm{GeV}^{-2}$ & [4.2,6.4] & [2,30] &  &  &  & \\
			$\beta$ & $\textrm{GeV}^{-2}$ & 0.4 & [0,1.0] & $\beta$ & $\textrm{GeV}^{-2}$ & [0.3,0.4] & [0,1.0] &  &  &  & \\
			$B_1$ & $\textrm{GeV}^{-2}$ & 0.1 & [0,5] & $\epsilon_0$ &  & [0.75,0.92] & [0,1.0] &  &  &  &    \\
			$t_0$ & $\textrm{GeV}^{2}$ & 10 & [0.1,10] & $t_d$ & $\textrm{GeV}^{2}$ & -0.1 & [-1.2,-0.2] &  &  &  &  \\
			$\epsilon_0$ &  & 0.5 & [0,1.0] & $\Delta$ & $\textrm{GeV}^{2}$ & 0.1 & [0.01,2] &  &  &  & \\
			$t_d$ & $\textrm{GeV}^{2}$ & -0.1 & [-1.2,-0.2] & $O_O$ & $\text{mb}/\textrm{GeV}^{2}$ & 0.04 & [$10^{-8}$,10] &  &  &  & \\
			$\Delta$ & $\textrm{GeV}^{2}$ & 0.2 & [0.01,2] & $\delta_O$ & & [0.02,0.14] & [-0.1,1] &  &  &  &  \\
			$O_O$ & $\text{mb}/\textrm{GeV}^{2}$ & 0.04 & [$10^{-8}$,10] & $D_O$ & $\textrm{GeV}^{-2}$ & [0.425,2.60] & [0.1,25] &  &  &  &  \\
			$\delta_O$ &  & 0.14 & [-0.1,1] & $\gamma$ & $\textrm{GeV}^{-2}$ & [0.5,1.0] & [0.01,5.0] &  &  &  & \\
			$D_O$ & $\textrm{GeV}^{-2}$ &  [0.65,2.75] & [0.1,25] & $I_0$ & $\text{mb}/\textrm{GeV}^{2}$ & 0.06 & [$10^{-8}$,10] &  &  &  & \\
			$I_0$ & $\text{mb}/\textrm{GeV}^{2}$ & 0.06 & [$10^{-8}$,10] & $\delta_I$ &  & [0.02,0.14] & [-0.1,1] &  &  &  &  \\
			$\delta_I$ &  & 0.14 & [-0.1,1] & $E_0$ & $\textrm{GeV}^{-2}$ & 3.0 & [0.1,25] &  &  &  &  \\
			$E_0$ & $\textrm{GeV}^{-2}$ & 3.0 & [0.1,25] & $\mu$ & & [0.001,0.008] & [0,1] &  &  &  &    \\
			$\mu$ &  & [0.001,0.008] &  [0,1] &  &  &  &  &  &  &  &  \\
			\bottomrule
		\end{tabular}
	\end{table}
	\renewcommand{\arraystretch}{1.0}
	\setlength{\tabcolsep}{1.20pt}
	\begin{table}[!htbp]
		\centering
		\small
		\begin{minipage}[t]{0.45\textwidth}
			\centering
			\footnotesize
			\caption{Correlation matrix of Model 1}
			\label{tab:matrix1}
			\vspace{0.1mm} 
			\begin{tabular}{lrrrrrrrrrrrrr}
				\toprule
				& \textbf{$A_0$} & \textbf{$\alpha$} & \textbf{$B_0$} & \textbf{$\epsilon_0$} & \textbf{$t_d$} & \textbf{$\Delta$} & \textbf{$O_O$} & \textbf{$\delta_O$} & \textbf{$D_O$} & \textbf{$I_0$} & \textbf{$\delta_I$} & \textbf{$E_0$} & \textbf{$\mu$} \\
				\midrule
				$A_0$ & 1.00 & -0.69 & -0.23 & -0.71 & 0.09 & -0.01 & 0.42 & -0.01 & -0.86 & -0.01 & -0.01 & -0.70 & 0.22 \\
				$\alpha$ & -0.69 & 1.00 & 0.37 & 0.64 & 0.21 & 0.48 & -0.47 & 0.31 & 0.37 & 0.48 & 0.31 & 0.69 & 0.20 \\
				$B_0$ & -0.23 & 0.37 & 1.00 & 0.07 & 0.22 & 0.16 & -0.28 & -0.08 & -0.11 & 0.15 & -0.08 & -0.02 & -0.01 \\
				$\epsilon_0$ & -0.71 & 0.64 & 0.07 & 1.00 & -0.31 & 0.25 & -0.22 & -0.04 & 0.54 & 0.25 & -0.04 & 0.70 & 0.12 \\
				$t_d$ & 0.09 & 0.21 & 0.22 & -0.31 & 1.00 & 0.41 & -0.70 & 0.26 & -0.33 & 0.41 & 0.26 & 0.09 & -0.01 \\
				$\Delta$ & -0.01 & 0.48 & 0.16 & 0.25 & 0.41 & 1.00 & -0.18 & 0.69 & -0.41 & 1.00 & 0.69 & 0.62 & 0.84 \\
				$O_O$ & 0.42 & -0.47 & -0.28 & -0.22 & -0.70 & -0.18 & 1.00 & 0.07 & -0.22 & -0.18 & 0.07 & -0.30 & 0.39 \\
				$\delta_O$ & -0.01 & 0.31 & -0.08 & -0.04 & 0.26 & 0.69 & 0.07 & 1.00 & -0.23 & 0.69 & 1.00 & 0.48 & 0.69 \\
				$D_O$ & -0.86 & 0.37 & -0.11 & 0.54 & -0.33 & -0.41 & -0.22 & -0.23 & 1.00 & -0.41 & -0.23 & 0.45 & -0.50 \\
				$I_0$ & -0.01 & 0.48 & 0.15 & 0.25 & 0.41 & 1.00 & -0.18 & 0.69 & -0.41 & 1.00 & 0.69 & 0.62 & 0.84 \\
				$\delta_I$ & -0.01 & 0.31 & -0.08 & -0.04 & 0.26 & 0.69 & 0.07 & 1.00 & -0.23 & 0.69 & 1.00 & 0.48 & 0.69 \\
				$E_0$ & -0.70 & 0.69 & -0.02 & 0.70 & 0.09 & 0.62 & -0.30 & 0.48 & 0.45 & 0.62 & 0.48 & 1.00 & 0.42 \\
				$\mu$ & 0.22 & 0.20 & -0.01 & 0.12 & -0.01 & 0.84 & 0.39 & 0.69 & -0.50 & 0.84 & 0.69 & 0.42 & 1.00 \\
				\bottomrule
			\end{tabular}
		\end{minipage}
		\hfill
		\begin{minipage}[t]{0.45\textwidth}
			\centering
			\footnotesize
			\caption{Correlation matrix of Model 2}
			\label{tab:matrix2}
			\vspace{0.1mm} 
			\begin{tabular}{lrrrrrrrrrrr}
				\toprule
				& \textbf{$A_0$} & \textbf{$\alpha$} & \textbf{$a_1$} & \textbf{$B_1^{(0)}$} & \textbf{$B_2^{(0)}$} & \textbf{$\delta_O$} & \textbf{$D_O$} & \textbf{$I_0$} & \textbf{$\delta_I$} & \textbf{$E_0$} & \textbf{$\mu$} \\
				\midrule
				\textbf{$A_0$} & 1.00 & -0.20 & 0.56 & 0.30 & -0.26 & -0.31 & -0.77 & -0.76 & -0.04 & -0.56 & -0.50 \\
				\textbf{$\alpha$} & -0.20 & 1.00 & 0.54 & 0.01 & 0.93 & -0.26 & -0.30 & 0.59 & -0.00 & -0.54 & -0.48 \\
				\textbf{$a_1$} & 0.56 & 0.54 & 1.00 & 0.64 & 0.57 & -0.48 & -0.94 & -0.37 & -0.00 & -1.00 & -0.90 \\
				\textbf{$B_1^{(0)}$} & 0.30 & 0.01 & 0.64 & 1.00 & 0.20 & -0.27 & -0.55 & -0.60 & 0.04 & -0.64 & -0.56 \\
				\textbf{$B_2^{(0)}$} & -0.26 & 0.93 & 0.57 & 0.20 & 1.00 & -0.27 & -0.29 & 0.49 & -0.00 & -0.57 & -0.51 \\
				\textbf{$\delta_O$} & -0.31 & -0.26 & -0.48 & -0.27 & -0.27 & 1.00 & 0.48 & 0.17 & 0.88 & 0.48 & 0.28 \\
				\textbf{$D_O$} & -0.77 & -0.30 & -0.94 & -0.55 & -0.29 & 0.48 & 1.00 & 0.57 & 0.04 & 0.94 & 0.84 \\
				\textbf{$I_0$} & -0.76 & 0.59 & -0.37 & -0.60 & 0.49 & 0.17 & 0.57 & 1.00 & 0.00 & 0.37 & 0.33 \\
				\textbf{$\delta_I$} & -0.04 & -0.00 & -0.00 & 0.04 & -0.00 & 0.88 & 0.04 & 0.00 & 1.00 & 0.00 & -0.17 \\
				\textbf{$E_0$} & -0.56 & -0.54 & -1.00 & -0.64 & -0.57 & 0.48 & 0.94 & 0.37 & 0.00 & 1.00 & 0.90 \\
				\textbf{$\mu$} & -0.50 & -0.48 & -0.90 & -0.56 & -0.51 & 0.28 & 0.84 & 0.33 & -0.17 & 0.90 & 1.00 \\
				\bottomrule
			\end{tabular}
		\end{minipage}
	\end{table}
	\renewcommand{\arraystretch}{1.0}
	\setlength{\tabcolsep}{1.20pt}
	\begin{table}[!htbp]
		\centering
		\small
		\begin{minipage}[t]{0.45\textwidth}
			\centering
			\footnotesize
			\caption{Correlation matrix of Model 3}
			\label{tab:matrix3}
			\vspace{0.1mm} 
			\begin{tabular}{lrrrrr}
				\toprule
				& \textbf{$A_0$} & \textbf{$B_1^{(0)}$} & \textbf{$B_2^{(0)}$} & \textbf{$D_O$} & \textbf{$\mu$} \\
				\midrule
				\textbf{$A_0$} & 1.00 & 0.00 & -0.22 & -0.87 & 0.44 \\
				\textbf{$B_1^{(0)}$} & 0.00 & 1.00 & -0.39 & -0.26 & 0.17 \\
				\textbf{$B_2^{(0)}$} & -0.22 & -0.39 & 1.00 & 0.31 & -0.17 \\
				\textbf{$D_O$} & -0.87 & -0.26 & 0.31 & 1.00 & -0.48 \\
				\textbf{$\mu$} & 0.44 & 0.17 & -0.17 & -0.48 & 1.00 \\
				\bottomrule
			\end{tabular}
			\hfill
			\newline
			\centering
			\footnotesize
			\caption{Correlation matrix of Model 4}
			\label{tab:matrix4}
			\vspace{0.1mm} 
			\begin{tabular}{lrrrr}
				\toprule
				& \textbf{$A_0$} & \textbf{$B_0$} & \textbf{$D_O$} & \textbf{$\mu$} \\
				\midrule
				\textbf{$A_0$} & 1.00 & 0.08 & -0.95 & 0.27 \\
				\textbf{$B_0$} & 0.08 & 1.00 & -0.28 & 0.15 \\
				\textbf{$D_O$} & -0.95 & -0.28 & 1.00 & -0.30 \\
				\textbf{$\mu$} & 0.27 & 0.15 & -0.30 & 1.00 \\
				\bottomrule
			\end{tabular}
		\end{minipage}
		\begin{minipage}[t]{0.45\textwidth}
			\centering
			\footnotesize
			\caption{Correlation matrix of Model 5}
			\label{tab:matrix5}
			\vspace{0.1mm} 
			\begin{tabular}{lrrrrrrrrrr}
				\toprule
				& \textbf{$A_0$} & \textbf{$\alpha$} & \textbf{$B_0$} & \textbf{$\beta$} & \textbf{$\epsilon_0$} & \textbf{$\delta_O$} & \textbf{$D_O$} & \textbf{$\gamma$} & \textbf{$\delta_I$} & \textbf{$\mu$} \\
				\midrule
				\textbf{$A_0$} & 1.00 & 0.30 & -0.32 & -0.26 & 0.76 & -0.26 & 0.03 & -0.77 & -0.26 & -0.26 \\
				\textbf{$\alpha$} & 0.30 & 1.00 & 0.35 & -0.95 & 0.82 & -0.95 & 0.90 & 0.26 & -0.95 & -0.95 \\
				\textbf{$B_0$} & -0.32 & 0.35 & 1.00 & -0.34 & 0.13 & -0.34 & 0.56 & 0.36 & -0.34 & -0.34 \\
				\textbf{$\beta$} & -0.26 & -0.95 & -0.34 & 1.00 & -0.81 & 1.00 & -0.95 & -0.37 & 1.00 & 1.00 \\
				\textbf{$\epsilon_0$} & 0.76 & 0.82 & 0.13 & -0.81 & 1.00 & -0.81 & 0.66 & -0.26 & -0.81 & -0.81 \\
				\textbf{$\delta_O$} & -0.26 & -0.95 & -0.34 & 1.00 & -0.81 & 1.00 & -0.95 & -0.37 & 1.00 & 1.00 \\
				\textbf{$D_O$} & 0.03 & 0.90 & 0.56 & -0.95 & 0.66 & -0.95 & 1.00 & 0.50 & -0.95 & -0.95 \\
				\textbf{$\gamma$} & -0.77 & 0.26 & 0.36 & -0.37 & -0.26 & -0.37 & 0.50 & 1.00 & -0.37 & -0.37 \\
				\textbf{$\delta_I$} & -0.26 & -0.95 & -0.34 & 1.00 & -0.81 & 1.00 & -0.95 & -0.37 & 1.00 & 1.00 \\
				\textbf{$\mu$} & -0.26 & -0.95 & -0.34 & 1.00 & -0.81 & 1.00 & -0.95 & -0.37 & 1.00 & 1.00 \\
				\bottomrule
			\end{tabular}
		\end{minipage}
	\end{table}
	\renewcommand{\arraystretch}{1.0}
	\setlength{\tabcolsep}{1.50pt}
	\begin{table}[!htbp]
		\centering
		\footnotesize
		\caption{Correlation and Statistical Metrics}
		\label{tab:aic_bic}
		\begin{tabular}{lccccccccc}
			\toprule
			\textbf{Model} & \textbf{Total Baseline} & \textbf{Active Count} & \textbf{Fixed Count} & \textbf{Redundancies ($\ge 0.98$)} & \textbf{Dependencies ($> 0.80$)} & \textbf{Mean $|r|$} & \textbf{Penalty Score} & \textbf{AIC} & \textbf{BIC} \\
			\midrule
			Model 4 & 16 & 4 & 12 & 0 & 1 & 0.3387 & \textbf{1.1387} & 673.09 & 692.46 \\
			Model 3 & 16 & 5 & 11 & 0 & 1 & 0.3324 & 1.2324 & 619.13 & 643.34 \\
			Model 1 & 14 & 13 & 1 & 2 & 5 & 0.3678 & 20.2678 & 1521.50 & 1584.45 \\
			Model 2 & 14 & 11 & 3 & 1 & 8 & 0.4399 & 21.1399 & \textbf{193.62} & \textbf{246.89}\\
			Model 5 & 15 & 10 & 5 & 6 & 20 & 0.6147 & 70.1147 & 749.07 & 797.49 \\
			\bottomrule
		\end{tabular}
	\end{table}
	
	The independent normalized $\chi^2$ values, calculated at each of the c.m. energies, are incorporated into linear information criteria. For this purpose, least-squares regression framework \cite{burnham2002model} has been employed. The information criteria have been evaluated by Akaike's Information Criterion (AIC) and Bayesian Information Criterion (BIC):
	\begin{equation}
		\text{AIC} = N_{\text{total}} \ln\left(\frac{\text{RSS}}{N_{\text{total}}}\right) + 2k,
	\end{equation}
	\begin{equation}
		\text{BIC} = N_{\text{total}} \ln\left(\frac{\text{RSS}}{N_{\text{total}}}\right) + k \cdot \ln(N_{\text{total}}),
	\end{equation}
	where RSS is the Residual Sum of Squares ($\text{RSS} = \sum_{i=1}^{N}[(d\sigma/dt)^{\text{model}}_i - (d\sigma/dt)^{\text{data}}_i]^{2}$), $k$ represents the active parameter count, and $N_{\text{total}} = 937$ denotes the total combined experimental data points across all 28 energy distributions. The results of the AIC and BIC criteria are shown in Table \ref{tab:aic_bic}. According to these criteria, Model 2 represents the superior framework that achieves the minimum scores of $\text{AIC} = 193.62$ and $\text{BIC} = 246.89$. Although Model 2 maintains a moderately rich parameter set ($k = 11$), it has exceptional capacity for $\chi^{2}$ error minimization. Under the Bayesian Information Criterion (BIC), this reduction in the $\chi^{2}$ strongly offsets the complexity penalty, outperforming the restricted parsimony of Model 3 ($\Delta\text{BIC} = +396.45$) and Model 4 ($\Delta\text{BIC} = +445.57$). Conversely, models such as Model 1 ($k = 13$, $\text{BIC} = 1584.45$) may struggle to justify their degrees of freedom due to low error convergence. In contrast, Model 2 avoids overfitting while maximizing predictive capability and can potentially provide a statistically supported description of the scattering dynamics.
	\subsection{Derived Quantities from the Models}
	We derive important observables of the scattering phenomena, including the total cross section $\sigma_{\text{tot}}$, logarithmic slope parameter $B(s)$, interaction radius $R(s)$, elastic cross section $\sigma_{\mathrm{el}}(s)$, inelastic cross section $\sigma_{\mathrm{inel}}(s)$, and their scaling ratios within the c.m. energy interval $\sqrt{s} = 3.36$ to $26.02$ GeV using the proposed models. Modern analyses, measurements \cite{TOTEM:2013vij,Block2005,Block2005.1,Refsigmael1,Sigmaelref2}, and recent holographic QCD approaches \cite{Watanabe2023} emphasize that precise simultaneous predictions of these physical observables are important to analyze the scattering processes.
	\subsubsection{Total Cross Section}
\begin{figure}[H]
		\centering
		\includegraphics[width=0.45\textwidth]{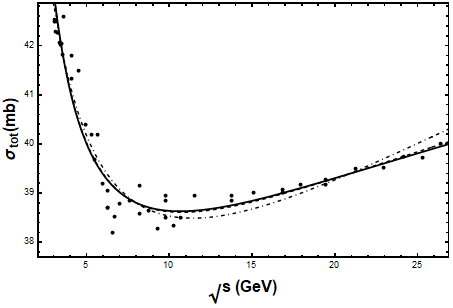}
		\caption{Fits on the $np$ total cross section data of ref. \cite{WorkmanPDG}. Fit 1, Fit 2, and Fit 3 are shown by solid line, dashed line, and dotted dashed line, respectively. The data points are represented by filled black circles}
		\label{fig:totalplot}
	\end{figure}

	The total cross section ${\sigma_{\text{tot}}}$ is calculated by the fitted models by their extrapolation at $t=0$, using the optical theorem \cite{TOTEM:2013lle}:
	\begin{equation}
		\sigma_{\text{tot}}^{2} = \frac{16 \pi (\hbar c)^{2}}{1 + \rho^{2}} \frac{d\sigma}{dt}\bigg|_{t=0}.
	\end{equation}
	The conversion factor $(\hbar c)^{2}\simeq 0.38$ $ \textrm{GeV}^{2} mbarn$ \cite{PDG2023} is used in the ${\sigma_{\text{tot}}}$ calculations. The $\rho$ parameter values used for each c.m. energy along with the calculated total cross sections are written in Table \ref{tab:totalcrosssection2}. For Datasets 1 and 3, the $\rho$ values are from refs. \cite{RefDataset1} and \cite{RefDataset3}, respectively. For Dataset 2, the values of the $\rho$ parameter are obtained by linear interpolation of the values reported in ref. \cite{RefDataset2} at the corresponding c.m. energies. According to ref. \cite{RefDataset4}, the $\rho$ parameter for elastic $np$ scattering has the same values as for elastic $pp$ scattering; therefore, the $\rho$ parameter values are obtained across Dataset 4 by linear interpolation of the $\rho$ for elastic $pp$ scattering in ref. \cite{Refrhodataset4}. In these calculations, ${\sigma_{\text{tot}}}$ relies mostly on the outcome of the models after their extrapolation at $t=0$ and obtains a small contribution from $\rho$ values. The $t=0$ extrapolation results in a major contribution to $\sigma_{\mathrm{tot}}$ from the leading components of the models, which have a power-law energy dependence $(s/s_0)^{\alpha}$. In this study, our fitting optimization yields the exponent $\alpha$ within an assumed interval $[0, 0.2]$ for each model (as mentioned in Table \ref{tab:pranges}), leading to good agreement with the reference values of $np$ total cross sections, satisfying the effective Froissart growth rate locally within the currently available $np$ experimental data. In the absence of exact reference values, especially for Datasets 2 and 3 \cite{RefDataset2,RefDataset3}, for which only fixed total normalization values $\sigma _{\text{tot}}$ of $39\text{ mb}$ and $38\text{ mb}$ are available, we perform individual fits using a few parametrizations with data of np total cross sections \cite{WorkmanPDG} in the c.m. energy range $3.0\text{ GeV} \leqslant \sqrt{s} \leqslant 26.4\text{ GeV}$, and extract the $\sigma _{\text{tot}}$ reference values by interpolation. We employ the following parametrizations, 
	\textbf{(i)} Parametrization 1: The extended form of the parametrization of ref. \cite{DONNACHIE1992227} is used:
	\begin{equation}
		\sigma_{\text{tot}}(s) = A + B s^{-\eta} + C s^{\epsilon},
	\end{equation}where $s^{\epsilon}$ and $s^{-\eta}$ give Pomeron rising contribution and Reggeon decreasing contribution, respectively. 
	\textbf{(ii)} Parametrization 2: It contains Froissart-like $\ln^{2}(s)$ growth term, a Reggeon term that ensures correct low-$s$ behavior, and a constant A:
	\begin{equation}
		\sigma_{\text{tot}}(s) = A + B Log^{2}(s) + C s^{-\eta}.
	\end{equation}
	\textbf{(iii)} Parametrization 3: It contains linear and quadratic terms in $ln(s)$ \cite{Block2005.1}:
	\begin{equation}
		\sigma_{\text{tot}}(s) = A + B \ln(s) + C \, \ln^{2}(s).
	\end{equation}
	The results from these fits are very close to each other, as shown in Fig. \ref{fig:totalplot}. The best fit parameter values of these parametrizations along with their $\chi^{2}$ are mentioned in Table \ref{tab:totalcrosssection1}. The dimensionless $\chi^{2}$ used for these fits is given as follows:
	\begin{equation}
		\chi^2 = \frac{\sum_{i=1}^{N}[(\sigma_{\text{tot}})^{\text{model}}_i - (\sigma_{\text{tot}})^{\text{data}}_i]^{2}}{\sum_{i=1}^{N} [(\sigma_{\text{tot}})^{\text{data}}_i]^{2}},
	\end{equation}with i = 1, 2, 3,...,N, number of data points. The $\sigma_{\rm tot}$ results, shown in Table \ref{tab:totalcrosssection2}, obtained by the proposed models of the differential cross section show good agreement with the reference values and with the results obtained by separate fits.
	\renewcommand{\arraystretch}{0.85}
	\setlength{\tabcolsep}{2.5pt}
	\begin{table}[!htbp]
		\centering
		\footnotesize
		\caption{Parameter values of the parametrizations of the total cross section $\sigma_{\text{tot}}(s)$ through fitting the $np$ $\sigma_{\text{tot}}(s)$ data}
		\label{tab:totalcrosssection1}
			\begin{tabular}{cccccccc}
				\toprule
				\textbf{} \textbf{No.} & \textbf{Parametrization} & \textbf{A} & \textbf{B} & \textbf{C} & \textbf{$\epsilon$} & \textbf{$\eta$} & \textbf{$\chi^{2}$} \\
				\midrule
				&  & (mb) & (mb) & (mb) &  & & Approx. \\
				1 & $\sigma_{\text{tot}}(s) = A + B s^{\epsilon} + C s^{-\eta}$ & 17.5118 & 12.3638 & 26.2255 & 0.0808 & 0.4525 & 0.00025 \\
				2 & $\sigma_{\text{tot}}(s) = A + B \ln^{2}(s) + C s^{-\eta}$ & 33.2611 & 0.1283 & 21.8285 &  & 0.4525 & 0.00026 \\
				3 & $\sigma_{\text{tot}}(s) = A + B \ln(s) + C \ln^{2}(s)$ & 49.5012 & -4.5454 & 0.4743 &  &  & 0.00036 \\
				\bottomrule
			\end{tabular}
	\end{table}
	\renewcommand{\arraystretch}{0.69}
	\setlength{\tabcolsep}{1.50pt}
	\begin{table}[!htbp]
		\centering
		\footnotesize
		\caption{Results of the $np$ total cross section ${\sigma_{\text{tot}}}$ by extrapolation of our models at $t=0$ and with the separate fits}
		\label{tab:totalcrosssection2}
			\begin{tabular}{ccccccccccc}
				\toprule
				\textbf{$\sqrt{s}$} & \textbf{$\rho$} & \textbf{Model 1} & \textbf{Model 2} & \textbf{Model 3} & \textbf{Model 4} & \textbf{Model 5} & \textbf{Ref. values} & \textbf{Fit 1} & \textbf{Fit 2} & \textbf{Fit 3} \\
				\midrule
				 & & $\sigma_{tot}$ & $\sigma_{tot}$ & $\sigma_{tot}$ & $\sigma_{tot}$ & $\sigma_{tot}$ &  $\sigma_{tot}$ &  $\sigma_{tot}$ &  $\sigma_{tot}$ &  $\sigma_{tot}$ \\
				& & (mb) & (mb) & (mb) & (mb) & (mb) & (mb) & (mb) & (mb) & (mb)  \\
				Dataset 1 &&&&&&&&&& \\
				3.363 ${\textrm{GeV}}$ & -0.4875 \cite{RefDataset1} & 41.1787 & 41.6455 & 41.698 & 41.6056 & 41.6936 & 41.69 \cite{RefDataset1}  & 41.3032 & 41.2993 & 41.2663 \\
				3.628 ${\textrm{GeV}}$ & -0.475 \cite{RefDataset1} & 40.3198 & 40.7284 & 40.7468 & 40.7291 & 40.7854 & 40.77 \cite{RefDataset1} & 40.9082 & 40.9135 & 40.9367 \\
				3.876 ${\textrm{GeV}}$ & -0.4625 \cite{RefDataset1} & 40.1488 & 40.2139 & 40.4284 & 40.4598 & 40.2522 & 40.25 \cite{RefDataset1} & 40.5972 & 40.6082 & 40.6672 \\
				4.109 ${\textrm{GeV}}$ & -0.45 \cite{RefDataset1} & 39.9234 & 39.963 & 39.9577 & 39.8906 & 39.9272 & 39.92 \cite{RefDataset1} & 40.347 & 40.3615 & 40.4431 \\
				4.329 ${\textrm{GeV}}$ & -0.4375 \cite{RefDataset1} & 39.7204 & 39.745 & 39.6476 & 39.7766 & 39.7801 & 39.70 \cite{RefDataset1} & 40.142 & 40.1584 & 40.2538 \\
				4.54 ${\textrm{GeV}}$ & -0.425 \cite{RefDataset1} & 39.4135 & 39.5833 & 39.3429 & 39.6827 & 39.5181 & 39.53 \cite{RefDataset1}  & 39.9694 & 39.9868 & 40.0901 \\
				4.741 ${\textrm{GeV}}$ & -0.4125 \cite{RefDataset1} & 39.2557 & 39.4211 & 39.467 & 39.4539 & 39.4079 & 39.40 \cite{RefDataset1} & 39.8239 & 39.8416 & 39.9485 \\
				4.935 ${\textrm{GeV}}$ & -0.40 \cite{RefDataset1} & 39.311 & 39.3164 & 39.35 & 39.3598 & 39.3589 & 39.30 \cite{RefDataset1} & 39.6987  & 39.7162 & 39.8238 \\
				\midrule
				Dataset 2 &&&&&&&&&& \\
				4.74 ${\textrm{GeV}}$ & -0.31 & 39.9545 & 39.9879 & 39.9388 & 39.9601 & 39.9158 & 39.0 \cite{RefDataset2} & 39.8246 & 39.8423 & 39.9492 \\
				5.12 ${\textrm{GeV}}$ & -0.293 & 39.8659 & 39.8284 & 39.6939 & 39.7452 & 39.6645 & 39.0 \cite{RefDataset2}  & 39.5916 & 39.6086 & 39.7147 \\
				5.473 ${\textrm{GeV}}$ & -0.277 & 39.7024 & 39.5678 & 39.4137 & 39.6595 & 39.4821 & 39.0 \cite{RefDataset2} & 39.4157  & 39.4311 & 39.5302 \\
				5.805 ${\textrm{GeV}}$ & -0.26 & 39.6157 & 39.4195 & 39.3692 & 39.5139 & 39.3292 & 39.0 \cite{RefDataset2} & 39.279  & 39.2923 & 39.3813 \\
				6.119 ${\textrm{GeV}}$ & -0.243 & 39.4309 & 39.3947 & 39.2059 & 39.3535 & 39.2087 & 39.0 \cite{RefDataset2} & 39.1709 & 39.1819 & 39.2592 \\
				6.418 ${\textrm{GeV}}$ & -0.227 & 39.3336 & 39.1648 & 39.1233 & 39.144 & 39.1123 & 39.0 \cite{RefDataset2} & 39.084 & 39.0927 & 39.1577 \\
				6.704 ${\textrm{GeV}}$ & -0.21 & 39.0505 & 39.0604 & 39.0826 & 39.033 & 39.0086 & 39.0 \cite{RefDataset2} & 39.0134 & 39.0199 & 39.0726 \\
				\midrule
				Dataset 3 &&&&&&&&&& \\
				3.466-4.409 ${\textrm{GeV}}$ & -0.42 \cite{RefDataset3} & 40.5963 & 40.5901 & 40.5792 & 40.5996 & 40.5856 & 38.0 \cite{RefDataset3} & 40.5275 & 40.5396 & 40.6055 \\
				4.409-5.187 ${\textrm{GeV}}$ & -0.39 \cite{RefDataset3} & 39.8232 & 39.8865 & 39.8901 & 39.8566 & 39.8711 & 38.0 \cite{RefDataset3} & 39.7857 & 39.8034 & 39.9107 \\
				5.187-5.863 ${\textrm{GeV}}$ & -0.36 \cite{RefDataset3} & 39.7038 & 39.4713 & 39.4602 & 39.4461 & 39.448 & 38.0 \cite{RefDataset3} & 39.3926 & 39.4077 & 39.5054 \\
				5.863-6.47 ${\textrm{GeV}}$ & -0.33 \cite{RefDataset3} & 39.4587 & 39.222 & 39.2035 & 39.2584 & 39.2544 & 38.0 \cite{RefDataset3} & 39.1561 & 39.1668 & 39.2422 \\
				6.47-7.024 ${\textrm{GeV}}$ & -0.30 \cite{RefDataset3} & 39.0703 & 39.0154 & 39.0949 & 39.0969 & 39.0443 & 38.0 \cite{RefDataset3} & 39.0038 & 39.0099 & 39.0607 \\
				7.024-7.538 ${\textrm{GeV}}$ & -0.27 \cite{RefDataset3} & 38.9195 & 38.9484 & 38.9384 & 38.9072 & 38.9106 & 38.0 \cite{RefDataset3} & 38.9018 & 38.9037 & 38.931 \\
				\midrule
				Dataset 4 &&&&&&&&&& \\
				13.748 ${\textrm{GeV}}$ & -0.094 & 38.8228 & 38.8165 & 38.8175 & 38.8052 & 38.8225 & 38.8 \cite{RefDataset4} & 38.8426 & 38.8225 & 38.7074 \\
				16.823 ${\textrm{GeV}}$ & -0.060 & 39.1336 & 39.1332 & 39.111 & 39.1053 & 39.1607 & 39.1 \cite{RefDataset4} & 39.0602 & 39.0462 & 38.9572 \\
				19.416 ${\textrm{GeV}}$ & -0.027 & 39.3658 & 39.5311 & 39.5338 & 39.4179 & 39.5581 & 39.5 \cite{RefDataset4} & 39.2694 & 39.2656 & 39.2284 \\
				21.471 ${\textrm{GeV}}$ & -0.012 & 39.5317 & 39.7229 & 39.7392 & 39.7898 & 39.7021 & 39.7 \cite{RefDataset4} & 39.4409 & 39.4474 & 39.4653 \\
				22.956 ${\textrm{GeV}}$ & 0.0221 & 39.7477 & 39.9347 & 39.9316 & 39.9891 & 39.9145 & 39.9 \cite{RefDataset4} & 39.5654 & 39.5803 & 39.6441 \\
				24.536 ${\textrm{GeV}}$ & 0.024 & 40.1718 & 40.1508 & 40.1278 & 40.1509 & 40.1291 & 40.1 \cite{RefDataset4} & 39.6974 & 39.7217 & 39.8388 \\
				26.019 ${\textrm{GeV}}$ & 0.028 & 40.2835 & 40.3978 & 40.3331 & 40.3444 & 40.3429 & 40.3 \cite{RefDataset4} & 39.82 & 39.8538 & 40.0245 \\
				\bottomrule
			\end{tabular}
	\end{table}
	\subsubsection{Slope Parameter and Interaction Radius}
	The logarithmic slope parameter $B(s)$ is a key dynamical quantity in elastic scattering of hadrons \cite{RefSlope,Okorokov2015,Okorokov2008}. The parameter $B(s)$ \cite{Jenkovszky2011,TOTEM_2018psk} characterizes the fall-off of the differential cross section in the forward region. It is calculated by the following relation:
	\begin{equation}
		B(s,t) = \frac{d}{dt} \left[ \ln \left( \frac{d\sigma (s,t)}{dt}\bigg|_{t=0} \right) \right].
	\end{equation}
	The results of the slope parameter are shown in Table \ref{tab:slopeparameter}. The slope parameter characterizes the width of the diffractive cone and encodes information about the spatial distribution of the strong-interaction matter inside the nucleon. Across all five models, the slope parameter exhibits a continuous, monotonic growth across the c.m. energy interval $\sqrt{s} = 3.36$ to $26.02$ GeV. It agrees with the well-established "shrinkage" of the diffractive peak \cite{Collins:1977jy,Donnachie:2002en}. Within the composite exponential frameworks of our models, this growth is governed by the energy-dependent slope factors $B_i(s) \propto \ln(s/s_0)$. The broadening of the effective interaction region \cite{RefDataset1} is indicated by this growth as the c.m. energy increases, reflecting the increasing relevance of gluon-mediated processes and the broadening of the impact-parameter distribution \cite{BSW2014}. The interaction radius $R$ \cite{RefDataset1} is calculated from the extracted values of the slope parameter $B$:
	\begin{equation}
		R = 2 \sqrt{B}.
	\end{equation}
	The results of the interaction radius are shown in Table \ref{tab:slopeparameter}. For $np$ scattering, the interaction radius increases from approximately $0.9~\mathrm{fm}$ at $\sqrt{s}\approx2~\mathrm{GeV}$ to about $1.2~\mathrm{fm}$ at $\sqrt{s}\approx6~\mathrm{GeV}$, consistent with the expansion of the diffractive cone and the geometric growth of the hadronic overlap region.
	\renewcommand{\arraystretch}{0.65}
	\setlength{\tabcolsep}{1.0pt}
	\begin{table}[!htbp]
		\centering
		\footnotesize
		\caption{Results of the slope parameter $B(s)$ and the interaction radius $R$ calculated by our models with the $np$ data of Table \ref{tab:expdata}}
		\label{tab:slopeparameter}
			\begin{tabular}{ccccccccccccc}
				\toprule
				\textbf{$\sqrt{s}$} & \textbf{Model 1} & \textbf{Model 2} & \textbf{Model 3} & \textbf{Model 4} & \textbf{Model 5} & \textbf{Ref. values} & \textbf{Model 1} & \textbf{Model 2} & \textbf{Model 3} & \textbf{Model 4} & \textbf{Model 5} & \textbf{Ref. values} \\
				\midrule
				& $B$ & $B$ & $B$ & $B$ & $B$ & $B$ & $R$ & $R$ & $R$ & $R$ & $R$ & $R$\\
				& $\text{(GeV/c)}^{-2}$ & $\text{(GeV/c)}^{-2}$ & $\text{(GeV/c)}^{-2}$ & $\text{(GeV/c)}^{-2}$ & $\text{(GeV/c)}^{-2}$ & $\text{(GeV/c)}^{-2}$ & (fm) & (fm) & (fm) & (fm) & (fm) & (fm) \\
				Dataset 1 &&&&&&&&&&&& \\
				3.363 ${\textrm{GeV}}$  & 6.5306 & 6.45966 & 6.2392 & 6.1201 & 6.5641 & 7.10 $\pm$ 0.12 \cite{RefDataset1}  & 1.0068 & 1.0013 & 0.9840 & 0.9746 & 1.0093 & 1.05 \cite{RefDataset1} \\
				3.628 ${\textrm{GeV}}$  & 6.4815 & 6.8311 & 6.8327 & 7.0417 & 6.7357 & 7.54 $\pm$ 0.10 \cite{RefDataset1} & 1.0030 & 1.0297 & 1.0297 & 1.0454 & 1.0224 & 1.08  \cite{RefDataset1} \\
				3.876 ${\textrm{GeV}}$  & 6.8876 & 7.7469 & 6.9387 & 7.2476 & 7.0419 & 7.87 $\pm$ 0.08 \cite{RefDataset1} & 1.0339 & 1.0965 & 1.0377 & 1.0606 & 1.0454 & 1.105 \cite{RefDataset1} \\
				4.109 ${\textrm{GeV}}$  & 7.8813 & 8.4824 & 7.4724 & 7.7414 & 7.4858 & 7.90 $\pm$ 0.08 \cite{RefDataset1} & 1.1060 & 1.1474 & 1.0769 & 1.0961 & 1.0779 & 1.107 \cite{RefDataset1} \\
				4.329 ${\textrm{GeV}}$  & 7.8648 & 8.6460 & 7.5559 & 7.7249 & 7.4693 & 7.80 $\pm$ 0.07 \cite{RefDataset1} & 1.1048 & 1.1584 & 1.0829 & 1.0950 & 1.0767 & 1.10 \cite{RefDataset1} \\
				4.54 ${\textrm{GeV}}$  & 7.9410 & 8.7541 & 7.9622 & 7.7011 & 7.8208 & 7.99 $\pm$ 0.07 \cite{RefDataset1} & 1.1102 & 1.1656 & 1.1116 & 1.0933 & 1.1017 & 1.11 \cite{RefDataset1} \\
				4.741 ${\textrm{GeV}}$ & 7.8103 & 8.8574 & 7.9215 & 7.6704 & 7.9144 & 8.26 $\pm$ 0.07 \cite{RefDataset1} & 1.1010 & 1.1725 & 1.1088 & 1.0911 & 1.1083 & 1.13 \cite{RefDataset1} \\
				4.935 ${\textrm{GeV}}$ & 7.7745 & 8.9536 & 7.7657 & 6.9844 & 8.0293 & 8.66 $\pm$ 0.10 \cite{RefDataset1} & 1.0985 & 1.1788 & 1.0978 & 1.0411 & 1.1163 & 1.16 \cite{RefDataset1} \\
				\midrule
				Dataset 2 &&&&&&&&&&&& \\
				4.74 ${\textrm{GeV}}$ & 7.5948 & 8.651 & 8.4396 & 8.2703 & 7.4777 & 8.24 $\pm$ 0.26 \cite{RefDataset2} & 1.0857 & 1.1587 & 1.1445 & 1.1329 & 1.0773 & 1.1309 \cite{RefDataset2} \\
				5.12 ${\textrm{GeV}}$ & 8.0214 & 8.2343 & 8.5630 & 8.2937 & 7.8015 & 8.54 $\pm$ 0.23 \cite{RefDataset2} & 1.1158 & 1.13047 & 1.1528 & 1.1345 & 1.1004 & 1.1513 \cite{RefDataset2} \\
				5.473 ${\textrm{GeV}}$ & 8.3493 & 8.7612 & 9.1099 & 9.1009 & 8.6091 & 8.93 $\pm$ 0.20 \cite{RefDataset2} & 1.1383 & 1.16608 & 1.1891 & 1.1885 & 1.1559 & 1.1773 \cite{RefDataset2} \\
				5.805 ${\textrm{GeV}}$ & 8.0739 & 8.8956 & 9.1044 & 9.0951 & 8.7536 & 8.54 $\pm$ 0.17 \cite{RefDataset2} & 1.1194 & 1.1750 & 1.1887 & 1.1881 & 1.1656 & 1.1513 \cite{RefDataset2} \\
				6.119 ${\textrm{GeV}}$ & 8.1594 & 8.8400 & 9.3986 & 8.7793 & 8.4877 & 8.05 $\pm$ 0.15 \cite{RefDataset2} & 1.1253 & 1.1713 & 1.2078 & 1.1673 & 1.1477 & 1.1178 \cite{RefDataset2} \\
				6.418 ${\textrm{GeV}}$ & 8.2368 & 9.0564 & 9.3100 & 8.6557 & 8.6643 & 8.39 $\pm$ 0.14 \cite{RefDataset2} & 1.1306 & 1.18556 & 1.2020 & 1.1590 & 1.1596 & 1.1412 \cite{RefDataset2} \\
				6.704 ${\textrm{GeV}}$ & 8.1064 & 9.4263 & 8.9946 & 8.5254 & 8.8344 & - & 1.1217 & 1.2095 & 1.1815 & 1.1503 & 1.1709 & - \\
				\midrule
				Dataset 3 &&&&&&&&&&&& \\
				3.466-4.409 ${\textrm{GeV}}$ & 5.9875 & 7.5736 & 6.3829 & 6.4230 & 6.3796 & 7.01 $\pm$ 0.43 \cite{RefDataset3} & 0.9640 & 1.0841 & 0.9953 & 0.99843 & 0.9950 & 1.0431 \cite{RefDataset3} \\
				4.409-5.187 ${\textrm{GeV}}$ & 6.3566 & 7.8901 & 6.6983 & 6.7394 & 6.7963 & 7.12 $\pm$ 0.33 \cite{RefDataset3} & 0.9933 & 1.1065 & 1.0196 & 1.0227 & 1.0270 & 1.0513 \cite{RefDataset3} \\
				5.187-5.863 ${\textrm{GeV}}$ & 6.9866 & 8.6562 & 6.9241 & 8.2157 & 8.1235 & 7.29 $\pm$ 0.35 \cite{RefDataset3} & 1.0413 & 1.1590 & 1.0366 & 1.1292 & 1.1228 & 1.0638 \cite{RefDataset3} \\
				5.863-6.47 ${\textrm{GeV}}$ & 7.9693 & 8.7721 & 7.9803 & 8.3915 & 8.3998 & 7.31 $\pm$ 0.35 \cite{RefDataset3} & 1.1121 & 1.1668 & 1.1129 & 1.1412 & 1.1417 & 1.0652 \cite{RefDataset3} \\
				6.47-7.024 ${\textrm{GeV}}$ & 8.5182 & 9.0673 & 9.0958 & 8.6265 & 8.7352 & 8.57 $\pm$ 0.53 \cite{RefDataset3} & 1.1498 & 1.1863 & 1.1881 & 1.1571 & 1.1644 & 1.1534 \cite{RefDataset3} \\
				7.024-7.538 ${\textrm{GeV}}$ & 9.0428 & 9.1583 & 9.1268 & 8.6575 & 9.3671 & 8.58 $\pm$ 0.62 \cite{RefDataset3} & 1.1847 & 1.1922 & 1.1902 & 1.1592 & 1.2057 & 1.1541 \cite{RefDataset3} \\
				\midrule
				Dataset 4 &&&&&&&&&&&& \\
				13.748 ${\textrm{GeV}}$ & 8.6158 & 8.8351 & 8.9671 & 8.9260 & 8.7418 & 9.8 \cite{RefDataset4} & 1.1564 & 1.1710 & 1.1797 & 1.177 & 1.1648 & 1.2334 \cite{RefDataset4} \\
				16.823 ${\textrm{GeV}}$ & 8.2903 & 9.2270 & 9.2904 & 9.0492 & 8.9842 & 10.1 \cite{RefDataset4} & 1.1343 & 1.1967 & 1.2008 & 1.1851 & 1.1808 & 1.2521 \cite{RefDataset4} \\
				19.416 ${\textrm{GeV}}$ & 9.0701 & 9.5890 & 9.1900 & 9.4290 & 9.1563 & 10.3 \cite{RefDataset4} & 1.1865 & 1.2199 & 1.1943 & 1.2097 & 1.1921 & 1.2644 \cite{RefDataset4} \\
				21.471 ${\textrm{GeV}}$ & 10.145 & 9.5502 & 9.1860 & 9.4400 & 9.3271 & 10.5 \cite{RefDataset4} & 1.2548 & 1.2175 & 1.1940 & 1.2104 & 1.2032 & 1.2767 \cite{RefDataset4} \\
				22.956 ${\textrm{GeV}}$ & 10.302 & 9.7072 & 9.4581 & 9.7472 & 9.5075 & 10.8 \cite{RefDataset4} & 1.26447 & 1.2274 & 1.2116 & 1.2299 & 1.2147 & 1.2948 \cite{RefDataset4} \\
				24.536 ${\textrm{GeV}}$ & 10.404 & 9.8638 & 9.7848 & 9.7838 & 9.3872 & 10.9 \cite{RefDataset4} & 1.27071 & 1.2373 & 1.2323 & 1.2323 & 1.2070 & 1.3007 \cite{RefDataset4} \\
				26.019 ${\textrm{GeV}}$ & 10.6444 & 10.0078 & 9.9338 & 9.9778 & 9.5778 & 11.1 \cite{RefDataset4} & 1.2853 & 1.2463 & 1.2417 & 1.2444 & 1.2192 & 1.3126 \cite{RefDataset4} \\
				\bottomrule
			\end{tabular}
	\end{table}
	\subsubsection{Elastic and Inelastic Cross Sections}
	The elastic cross section \cite{Refsigmael1,Sigmaelref2} is evaluated through the diffractive-cone relation:
	\begin{equation}
		\sigma_{\text{el}} = \frac{\sigma_{\text{tot}} ^{2}(1 + \rho^{2})}{16 \pi B}.
	\end{equation}
	The values of $\rho$, $\sigma_{\text{tot}}$, and $B$, mentioned in Tables \ref{tab:totalcrosssection2} and \ref{tab:slopeparameter} are used to calculate $\sigma_{\text{el}}$. The calculated $\sigma_{\mathrm{el}}$ values are shown in Table \ref{tab:elcrosssection}. These $\sigma_{\mathrm{el}}$ results are in good agreement with the reference values across each dataset. This relation (Eq. 20) is derived under the assumption that the hadronic amplitude at small momentum transfer is dominated by a single exponential profile that emerges from Regge pole dominance and diffraction models. Its dependence on $\sigma_{\mathrm{tot}}$ stems from the optical theorem and makes $\sigma_{\mathrm{el}}$ extremely responsive to even small variations in the strength of the total interaction. The inverse dependence on the slope parameter $B$ reveals a geometric interpretation \cite{Refsigmael1,Sigmaelref2}. The elastic cross section $\sigma_{\mathrm{el}}$ encodes an interplay between \textbf{(i)} the opacity of the hadronic core (or blackness of the core) that determines the amount of absorption and, the maximum possible elastic scattering, \textbf{(ii)} the diffractive coherence of peripheral mesonic exchanges, and \textbf{(iii)} the spatial localization of the scattering region that refers to the effective size and density profile of the interacting particles \cite{FAGUNDES2016194,Refsigmael1,Sigmaelref2,Adamczyk:2015gfy}. The integrated elastic cross section is calculated in the specified $\mid t\mid$-ranges (mentioned in Table \ref{tab:expdata}) by numerical integration of the models with their parameter values for all the energies by the following relation:
	\begin{equation}
		\sigma_{\text{el}} = \int_{t_{\text{min}}}^{t_{\text{max}}} \frac{d\sigma}{dt} dt.
	\end{equation}
	The results of the integrated elastic cross section are shown in Table \ref{tab:elcrosssection}. The inelastic cross section $\sigma_{\text{inel}}$ \cite{TOTEM:2013lle,TOTEM:2013vij} is calculated by following relation:
	\begin{equation}
		\space \sigma_{\text{inel}} = \sigma_{\text{tot}} - \sigma_{\text{el}}.
	\end{equation}
	The $\sigma_{\text{inel}}$ results are shown in Table \ref{tab:sigmainel}.
	\renewcommand{\arraystretch}{0.65}
	\setlength{\tabcolsep}{2.0pt}
	\begin{table}[!htbp]
		\centering
		\footnotesize
		\caption{Results of the elastic cross section $\sigma_{\mathrm{el}}$ for the $np$ data of Table \ref{tab:expdata}}
		\label{tab:elcrosssection}
		\begin{tabular}{cccccccccccc}
			\toprule
			\multirow{2}{*}{} \textbf{${\sqrt{s}}$} & \multicolumn{5}{c}{\textbf{${\sigma_{\text{el}}}$ by Eq. 20}} & \multicolumn{5}{c}{\textbf{${ \sigma_{\text{el}}}$ by integration}} & \multicolumn{1}{c}{\textbf{Ref. values of ${\sigma_{\text{el}}}$}} \\
			\midrule
			& Model 1 & Model 2 & Model 3 & Model 4 & Model 5 & Model 1 & Model 2 & Model 3 & Model 4 & Model 5 & Ref. \\
			& (mb) & (mb) & (mb) & (mb) & (mb) & (mb) & (mb) & (mb) & (mb) & (mb) & (mb) \\
			Dataset 1 &&&&&&&&&&& \\
			3.363 ${\textrm{GeV}}$ & 6.3933 & 6.6108 & 6.8617 & 6.9643 & 6.5207  & 7.7073 & 7.3785 & 7.8899 & 8.1177 & 7.1562 & 6.0275 $\pm$ 0.10 \cite{RefDataset1} \\
			3.628 ${\textrm{GeV}}$ & 6.1157 & 5.921 & 5.9250 & 5.7441 & 6.0216 & 6.7682 & 6.2807 & 6.2923 & 5.9358 & 6.4580 & 5.3752 $\pm$ 0.07 \cite{RefDataset1} \\
			3.876 ${\textrm{GeV}}$ & 5.6519 & 5.0413 & 5.6886 & 5.4546 & 5.5566  & 6.0072 & 4.7329 & 5.9373 & 5.4672 & 5.7035 & 4.9173 $\pm$ 0.05 \cite{RefDataset1} \\
			4.109 ${\textrm{GeV}}$ & 4.8381 & 4.5041 & 5.1116 & 4.9174 & 5.0947 & 4.4639 & 3.8904 & 4.9633 & 4.6027 & 4.9215 & 4.8258 $\pm$ 0.05 \cite{RefDataset1} \\
			4.329 ${\textrm{GeV}}$ & 4.7508 & 4.3305 & 4.9310 & 4.8546 & 5.0216 & 4.5363 & 3.5975 & 4.8773 & 4.6934 & 5.006 & 4.7894 $\pm$ 0.04 \cite{RefDataset1} \\
			4.54 ${\textrm{GeV}}$ & 4.5947 & 4.2039 & 4.5661 & 4.8028 & 4.6901  & 3.7800 & 2.9788 & 3.7428 & 4.0869 & 3.9150 & 4.5936 $\pm$ 0.04 \cite{RefDataset1} \\
			4.741 ${\textrm{GeV}}$ & 4.59317 & 4.0844 & 4.5776 & 4.7243 & 4.5679 & 3.0742 & 2.2392 & 3.0285 & 3.2624 & 3.0146 & 4.3751 $\pm$ 0.03 \cite{RefDataset1} \\
			4.935 ${\textrm{GeV}}$ & 4.5872 & 3.9842 & 4.6015 & 5.1188 & 4.4524 & 3.0886 & 2.2640 & 3.1298 & 3.9790 & 3.1696 & 4.1158 $\pm$ 0.04 \cite{RefDataset1} \\
			\midrule
			Dataset 2 &&&&&&&&&&& \\
			4.74 ${\textrm{GeV}}$ & 4.3784 & 4.0306 & 4.1215 & 4.2103 & 4.6463 & 7.7329 & 6.0661 & 6.2294 & 6.4811 & 7.5248 & 4.052 $\pm$ 0.12 \cite{RefDataset2} \\
			5.12 ${\textrm{GeV}}$ & 4.1666 & 4.1616 & 3.9748 & 4.1145 & 4.3564 & 7.0204 & 6.5110 & 5.9632 & 6.3270 & 6.9101 & 3.8474 $\pm$ 0.10 \cite{RefDataset2} \\
			5.473 ${\textrm{GeV}}$ & 3.8956 & 3.7383 & 3.5672 & 3.6155 & 3.7879 & 6.4905 & 5.7313 & 5.2692 & 5.4002 & 5.8381 & 3.6485 $\pm$ 0.08 \cite{RefDataset2} \\
			5.805 ${\textrm{GeV}}$ & 4.1285 & 3.7101 & 3.6158 & 3.6461 & 3.7530 & 6.7459 & 5.4979 & 5.2141 & 5.3209 & 5.5964 & 3.7828 $\pm$ 0.07 \cite{RefDataset2} \\
			6.119 ${\textrm{GeV}}$ & 4.0703 & 3.6989 & 3.4458 & 3.7167 & 3.8161 & 6.5228 & 5.5393 & 4.8757 & 5.5360 & 5.7899 & 3.9809 $\pm$ 0.07 \cite{RefDataset2} \\
			6.418 ${\textrm{GeV}}$ & 4.0364 & 3.5431 & 3.4393 & 3.7032 & 3.6936 & 6.3392 & 5.2302 & 4.8954 & 5.5462 & 5.5305 & 3.7924 $\pm$ 0.06 \cite{RefDataset2} \\
			6.704 ${\textrm{GeV}}$ & 4.0905 & 3.3620 & 3.5274 & 3.7121 & 3.5778 & 0.5154 & 0.3083 & 0.3304 & 0.4270 & 0.3657 & - \\
			\midrule
			Dataset 3 &&&&&&&&&&& \\
			3.466-4.409 ${\textrm{GeV}}$ & 6.4418 & 5.0912 & 6.0377 & 6.0061 & 6.0428 & 6.2019 & 3.7487 & 5.6534 & 5.1281 & 5.2059 & 4.8210 $\pm$ 0.29 \cite{RefDataset3} \\
			4.409-5.187 ${\textrm{GeV}}$ & 5.7183 & 4.62154 & 5.4448 & 5.4026 & 5.3612 & 5.1774 & 3.2080 & 4.4489 & 4.3732 & 4.3077 & 4.6484 $\pm$ 0.21 \cite{RefDataset3} \\
			5.187-5.863 ${\textrm{GeV}}$ & 4.8183 & 4.0447 & 5.0537 & 4.2562 & 4.3049 & 4.1138 & 2.3765 & 3.0748 & 2.6656 & 2.7423 & 4.4514 $\pm$ 0.21 \cite{RefDataset3} \\
			5.863-6.47 ${\textrm{GeV}}$ & 4.3101 & 3.8688 & 4.2487 & 4.0518 & 4.0470 & 1.9946 & 1.4753 & 1.8924 & 1.6486 & 1.6453 & 4.3579 $\pm$ 0.20 \cite{RefDataset3} \\
			6.47-7.024 ${\textrm{GeV}}$ & 3.8860 & 3.6404 & 3.6438 & 3.8424 & 3.7844 & 1.5301 & 1.2481 & 1.1994 & 1.4322 & 1.3763 & 3.6538 $\pm$ 0.22 \cite{RefDataset3} \\
			7.024-7.538 ${\textrm{GeV}}$ & 3.5754 & 3.5355 & 3.5459 & 3.7321 & 3.4500 & 0.5259 & 0.4842 & 0.4668 & 0.5880 & 0.4316 & 3.5923 $\pm$ 0.25 \cite{RefDataset3} \\
			\midrule
			Dataset 4 &&&&&&&&&&& \\
			13.748 ${\textrm{GeV}}$ & 3.51097 & 3.4227 & 3.3725 & 3.3859 & 3.4604 & 2.4942 & 2.3691 & 2.2679 & 2.2896 & 2.4053 & 3.0831 \cite{RefDataset4} \\
			16.823 ${\textrm{GeV}}$ & 3.6882 & 3.3138 & 3.2874 & 3.3740 & 3.4081 & 2.7836 & 2.1565 & 2.1044 & 2.2389 & 2.2829 & 3.0222 \cite{RefDataset4} \\
			19.416 ${\textrm{GeV}}$ & 3.4015 & 3.2445 & 3.3859 & 3.2807 & 3.4025 & 2.2424 & 1.9886 & 2.1970 & 2.0535 & 2.2193 & 3.0158 \cite{RefDataset4} \\
			21.471 ${\textrm{GeV}}$ & 3.0650 & 3.2875 & 3.4206 & 3.3371 & 3.3626 & 2.4293 & 2.0564 & 2.2397 & 2.1052 & 2.1568 & 2.9867 \cite{RefDataset4} \\
			22.956 ${\textrm{GeV}}$ & 3.0524 & 3.27 & 3.3556 & 3.26547 & 3.3353 & 2.3624 & 1.9940 & 2.1535 & 1.9662 & 2.0815 & 2.9340 \cite{RefDataset4} \\
			24.536 ${\textrm{GeV}}$ & 3.0876 & 3.2533 & 3.2758 & 3.2799 & 3.4185 & 2.6870 & 1.9347 & 1.9620 & 1.9640 & 2.1704 & 2.9366 \cite{RefDataset4} \\
			26.019 ${\textrm{GeV}}$ & 3.0353 & 3.2467 & 3.2605 & 3.2479 & 3.4003 & 2.2244 & 1.4839 & 1.5035 & 1.4853 & 1.6591 & 2.9131 \cite{RefDataset4} \\
			\bottomrule
		\end{tabular}
	\end{table}
	\renewcommand{\arraystretch}{0.650}
	\setlength{\tabcolsep}{2.0pt}
	\begin{table}[!htbp]
		\centering
		\footnotesize
		\caption{Results of $np$ inelastic cross section ${\sigma_{\text{inel}}}$}
		\label{tab:sigmainel}
		\begin{tabular}{ccccccc}
			\toprule
			\textbf{$\sqrt{s}$} & \textbf{Model 1} & \textbf{Model 2} & \textbf{Model 3} & \textbf{Model 4} & \textbf{Model 5} & \textbf{Ref. values}  \\
			\midrule
			& $\sigma_{inel}$ & $\sigma_{inel}$ & $\sigma_{inel}$ & $\sigma_{inel}$ & $\sigma_{inel}$ & $\sigma_{inel}$  \\
			& (mb) & (mb) & (mb) & (mb) & (mb) & (mb)  \\
			Dataset 1 &&&&&& \\
			3.363 ${\textrm{GeV}}$ & 34.7854 & 35.0347 & 34.8363 & 34.6413 & 35.1729 & 35.6625 $\pm$ 0.10 \cite{RefDataset1} \\
			3.628 ${\textrm{GeV}}$ & 34.2041 & 34.8074 & 34.8218 & 34.985 & 34.7638 & 35.3948 $\pm$ 0.07 \cite{RefDataset1} \\
			3.876 ${\textrm{GeV}}$ & 34.4969 & 35.1726 & 34.7398 & 35.0052 & 34.6956 & 35.3327 $\pm$ 0.05 \cite{RefDataset1} \\
			4.109 ${\textrm{GeV}}$ & 35.0853 & 35.4589 & 34.8461 & 34.9732 & 34.8325 & 35.0942 $\pm$ 0.05 \cite{RefDataset1} \\
			4.329 ${\textrm{GeV}}$ & 34.9696 & 34.4145 & 34.7166 & 34.922 & 34.7585 & 34.9106 $\pm$ 0.04 \cite{RefDataset1} \\
			4.54 ${\textrm{GeV}}$ & 34.8188 & 35.3794 & 34.7768 & 34.8799 & 34.828 & 34.9364 $\pm$ 0.04 \cite{RefDataset1} \\
			4.741 ${\textrm{GeV}}$ & 34.6625 & 34.3367 & 34.8894 & 34.7296 & 34.84 & 35.0249 $\pm$ 0.03 \cite{RefDataset1} \\
			4.935 ${\textrm{GeV}}$ & 34.7238 & 35.3322 & 34.7485 & 34.241 & 34.9065 & 35.1842 $\pm$ 0.04 \cite{RefDataset1}\\
			\midrule
			Dataset 2 &&&&&& \\
			4.74 ${\textrm{GeV}}$ & 35.5761 & 35.9573 & 35.8173 & 35.7498 & 35.2695 & 34.948 $\pm$ 0.12 \cite{RefDataset2} \\
			5.12 ${\textrm{GeV}}$ & 35.6993 & 35.6668 & 35.7191 & 35.6307 & 35.3081 & 35.1526 $\pm$ 0.10 \cite{RefDataset2} \\
			5.473 ${\textrm{GeV}}$ & 35.8068 & 35.8295 & 35.8465 & 36.044 & 35.6942 & 35.3515 $\pm$ 0.08 \cite{RefDataset2} \\
			5.805 ${\textrm{GeV}}$ & 35.4872 & 35.7094 & 35.7534 & 35.8678 & 35.5762 & 35.2172 $\pm$ 0.07 \cite{RefDataset2} \\
			6.119 ${\textrm{GeV}}$ & 35.3606 & 35.6958 & 35.7601 & 35.6368 & 35.3926 & 35.0191 $\pm$ 0.07 \cite{RefDataset2} \\
			6.418 ${\textrm{GeV}}$ & 35.2972 & 35.6217 & 35.684 & 35.4408 & 35.4187 & 35.2076 $\pm$ 0.06 \cite{RefDataset2} \\
			6.704 ${\textrm{GeV}}$ & 34.96 & 35.6984 & 35.5552 & 35.3209 & 35.4308 & - \\
			\midrule
			Dataset 3 &&&&&& \\
			3.466-4.409 ${\textrm{GeV}}$ & 34.1545 & 35.4989 & 34.5415 & 34.5935 & 34.5428 & 33.179 $\pm$ 0.29 \cite{RefDataset3} \\
			4.409-5.187 ${\textrm{GeV}}$ & 34.1049 & 35.265 & 34.4453 & 34.454 & 34.5099 & 33.3516 $\pm$ 0.21 \cite{RefDataset3} \\
			5.187-5.863 ${\textrm{GeV}}$ & 34.8855 & 35.4266 & 34.4065 & 35.1899 & 35.1431 & 33.5486 $\pm$ 0.21 \cite{RefDataset3} \\
			5.863-6.47 ${\textrm{GeV}}$ & 35.1486 & 35.3532 & 34.9548 & 35.2067 & 35.2074 & 33.6421 $\pm$ 0.20 \cite{RefDataset3} \\
			6.47-7.024 ${\textrm{GeV}}$ & 35.1843 & 35.375 & 35.4511 & 35.2545 & 35.2599 & 34.3462 $\pm$ 0.22 \cite{RefDataset3} \\
			7.024-7.538 ${\textrm{GeV}}$ & 35.3441 & 35.4129 & 35.3925 & 35.1751 & 35.4606 & 34.4077 $\pm$ 0.25 \cite{RefDataset3} \\
			\midrule
			Dataset 4 &&&&&& \\
			13.748 ${\textrm{GeV}}$ & 35.7875 & 35.3939 & 35.445 & 35.4193 & 35.3621 & 35.7169 \cite{RefDataset4} \\
			16.823 ${\textrm{GeV}}$ & 36.046 & 35.8194 & 35.8236 & 35.7313 & 35.7526 & 36.0778 \cite{RefDataset4} \\
			19.416 ${\textrm{GeV}}$ & 36.3134 & 36.2866 & 36.1479 & 36.1372 & 36.1556 & 36.4842 \cite{RefDataset4} \\
			21.471 ${\textrm{GeV}}$ & 36.4667 & 36.4354 & 36.3186 & 36.4527 & 36.3395 & 36.7133 \cite{RefDataset4} \\
			22.956 ${\textrm{GeV}}$ & 36.6953 & 36.6648 & 36.576 & 36.7236 & 36.5792 & 36.966 \cite{RefDataset4} \\
			24.536 ${\textrm{GeV}}$ & 37.0842 & 36.8976 & 36.852 & 36.871 & 36.7106 & 37.1634 \cite{RefDataset4} \\
			26.019 ${\textrm{GeV}}$ & 37.3214 & 37.1512 & 37.0726 & 37.0965 & 36.9426 & 37.3869 \cite{RefDataset4} \\
			\bottomrule
		\end{tabular}
	\end{table}
	\subsubsection{Ratios of Elastic and Inelastic Cross Sections to Total Cross Section}
	The ratios $R_{\rm el}\equiv\sigma_{\rm el}/\sigma_{\rm tot}$ and $R_{\rm inel}\equiv\sigma_{\rm inel}/\sigma_{\rm tot}=1-R_{\rm el}$ are compact, model-independent measures of how the total interaction probability is partitioned between elastic scattering and inelastic production. The results of $R_{\rm el}$ and $R_{\rm inel}$ are shown in Table \ref{tab:ratiotable}, which are in good agreement with their reference values across all four datasets. It is evident from Table \ref{tab:ratiotable} that $R_{\rm el}$ values gradually decrease with c.m. energy, while the $R_{\rm inel}$ values increase with c.m. energy. $R_{\rm el}$ quantifies the degree of coherence and opacity of the scattering system \cite{Adamczyk:2015gfy}. In the black-disk limit $R_{\rm el}$ approaches $1/2$, whereas small values of $R_{\rm el}$ signal strong absorption into inelastic processes. $R_{\rm el}$ also links the $\sigma_{\rm tot}$ to the diffraction slope $B$ and the ratio $\rho$ through the forward-cone relation (Eq. 20). Empirically, the energy dependence of $R_{\rm el}$ reveals the changing balance between diffraction and multiparticle production \cite{Barone:2002cv,Adamczyk:2015gfy}. In elastic $pp$ scattering, the ratio $R_{\rm el}$ tends to grow steadily with energy, which reflects increasing diffractive coherence \cite{Oueslati2024,FAGUNDES2016194}. While in $np$ interactions this ratio is found to evolve more slowly, which indicates a stronger absorptive (inelastic) component and persistence of meson-exchange mechanisms even at intermediate energies \cite{Saleem1985}. Correspondingly, the slope parameter $B$ in $np$ scattering increases with $\sqrt{s}$ but at a noticeably reduced rate compared to $pp$ \cite{RefDataset1,RefDataset2,RefDataset3,RefDataset4}, which signals a more gradual transverse expansion of the interaction radius $R$ and a weaker approach to the black-disc limit \cite{FAGUNDES2016194,Adamczyk:2015gfy}. As $\sqrt{s}$ increases, the total cross section exhibits the expected slow logarithmic growth, while the elastic component rises more slowly. The ratio $R_{\rm el}$, remains significantly below the corresponding $pp$ values, confirming that $np$ scattering remains far from saturating the unitarity bound or achieving diffractive dominance over the studied range. Meanwhile, the inelastic fraction $R_{\rm inel}$ stays comparatively large, which reveals that inelastic processes remain the principal contributors to the total scattering phenomena. This indicates that elastic scattering is a less dominant process relative to the total interaction cross-section in $np$ collisions compared to $pp$ collisions within the studied energy range.
	\renewcommand{\arraystretch}{0.70}
	\setlength{\tabcolsep}{1.70pt}
	\begin{table}[!htbp]
		\centering
		\footnotesize
		\caption{Results of the cross section ratios $R_{el}$ and $R_{inel}$}
		\label{tab:ratiotable}
			\begin{tabular}{ccccccccccccc}
				\toprule
				\textbf{$\sqrt{s}$} & \textbf{Model 1} & \textbf{Model 2} & \textbf{Model 3} & \textbf{Model 4} & \textbf{Model 5} & \textbf{Ref. values} & \textbf{Model 1} & \textbf{Model 2} & \textbf{Model 3} & \textbf{Model 4} & \textbf{Model 5} & \textbf{Ref. values} \\
				\midrule
				& $R_{\rm el}$ & $R_{\rm el}$ & $R_{\rm el}$ & $R_{\rm el}$ & $R_{\rm el}$ & $R_{\rm el}$ & $R_{\rm inel}$ & $R_{\rm inel}$ & $R_{\rm inel}$ & $R_{\rm inel}$ & $R_{\rm inel}$ & $R_{\rm inel}$ \\
				Dataset 1 &&&&&&&&&&&& \\
				3.363 ${\textrm{GeV}}$  & 0.1552 & 0.1587 & 0.1645 & 0.1673 & 0.1563 & 0.1445 $\pm$ 0.10 \cite{RefDataset1}  & 0.8448 & 0.8413 & 0.8355 & 0.8327 & 0.8437 & 0.8555 $\pm$ 0.10 \cite{RefDataset1} \\
				3.628 ${\textrm{GeV}}$  & 0.1516 & 0.1453 & 0.1454 & 0.1410 & 0.1476 & 0.1318 $\pm$ 0.07 \cite{RefDataset1} & 0.8484 & 0.8547 & 0.8546 & 0.859 & 0.8524 & 0.8682 $\pm$ 0.07 \cite{RefDataset1} \\
				3.876 ${\textrm{GeV}}$  & 0.1407 & 0.1246 & 0.1465 & 0.1419 & 0.1380 & 0.1222 $\pm$ 0.05 \cite{RefDataset1} & 0.8593 & 0.8754 & 0.8535 & 0.8581 & 0.862 & 0.8778 $\pm$ 0.05 \cite{RefDataset1} \\
				4.109 ${\textrm{GeV}}$  & 0.1211 & 0.1127 & 0.1279 & 0.1232 & 0.1275 & 0.12 $\pm$ 0.05 \cite{RefDataset1} & 0.8789 & 0.8873 & 0.8721 & 0.8768 & 0.8725 & 0.88 $\pm$ 0.05 \cite{RefDataset1} \\
				4.329 ${\textrm{GeV}}$  & 0.1196 & 0.1089 & 0.1243 & 0.1220 & 0.1262 & 0.1206 $\pm$ 0.04 \cite{RefDataset1} & 0.8804 & 0.8911 & 0.8757 & 0.878 & 0.8738 & 0.8794 $\pm$ 0.04 \cite{RefDataset1} \\
				4.54 ${\textrm{GeV}}$  & 0.1165 & 0.1062 & 0.1160 & 0.1210 & 0.1186 & 0.1162 $\pm$ 0.04 \cite{RefDataset1} & 0.8835 & 0.8938 & 0.884 & 0.879 & 0.8814 & 0.8838 $\pm$ 0.04 \cite{RefDataset1} \\
				4.741 ${\textrm{GeV}}$ & 0.1170 & 0.1036 & 0.1159 & 0.1197 & 0.1159 & 0.1110 $\pm$ 0.03 \cite{RefDataset1} & 0.883 & 0.8964 & 0.8841 & 0.8803 & 0.8841 & 0.889 $\pm$ 0.03 \cite{RefDataset1} \\
				4.935 ${\textrm{GeV}}$ & 0.1166 & 0.1013 & 0.1169 & 0.1300 & 0.1131 & 0.1047 $\pm$ 0.10 \cite{RefDataset1} & 0.8834 & 0.8987 & 0.8831 & 0.87 & 0.8869 & 0.8953 $\pm$ 0.10 \cite{RefDataset1} \\
				\midrule
				Dataset 2 &&&&&&&&&&&& \\
				4.74 ${\textrm{GeV}}$ & 0.1095 & 0.1007 & 0.12 & 0.1053 & 0.1164 & 0.1038 $\pm$ 0.26 \cite{RefDataset2} & 0.8905 & 0.8993 & 0.88 & 0.8947 & 0.8836 & 0.8962 $\pm$ 0.26 \cite{RefDataset2} \\
				5.12 ${\textrm{GeV}}$ & 0.1045 & 0.1044 & 0.1001 & 0.1035 & 0.1098 & 0.099 $\pm$ 0.10 \cite{RefDataset2} & 0.8955 & 0.8956 & 0.8999 & 0.8965 & 0.8902 & 0.901 $\pm$ 0.10 \cite{RefDataset2} \\
				5.473 ${\textrm{GeV}}$ & 0.098 & 0.0944 & 0.0905 & 0.0911 & 0.0959 & 0.0935 $\pm$ 0.08 \cite{RefDataset2} & 0.902 & 0.9056 & 0.9095 & 0.9089 & 0.9041 & 0.9065 $\pm$ 0.08 \cite{RefDataset2} \\
				5.805 ${\textrm{GeV}}$ & 0.1042 & 0.0941 & 0.0918 & 0.0922 & 0.0954 & 0.0969 $\pm$ 0.07 \cite{RefDataset2} & 0.8958 & 0.9059 & 0.9082 & 0.9078 & 0.9046 & 0.9031 $\pm$ 0.07 \cite{RefDataset2} \\
				6.119 ${\textrm{GeV}}$ & 0.1032 & 0.0938 & 0.0878 & 0.0944 & 0.0973 & 0.10 $\pm$ 0.07 \cite{RefDataset2} & 0.8968 & 0.9062 & 0.9122 & 0.9056 & 0.9027 & 0.9 $\pm$ 0.07 \cite{RefDataset2} \\
				6.418 ${\textrm{GeV}}$ & 0.1026 & 0.0905 & 0.0879 & 0.0946 & 0.0944 & 0.0972 $\pm$ 0.06 \cite{RefDataset2} & 0.8974 & 0.9095 & 0.9121 & 0.9054 & 0.9056 & 0.9028 $\pm$ 0.06 \cite{RefDataset2} \\\
				6.704 ${\textrm{GeV}}$ & 0.1047 & 0.0860 & 0.0903 & 0.0951 & 0.0917 & - & 0.8953 & 0.914 & 0.9097 & 0.9049 & 0.9083 & - \\
				\midrule
				Dataset 3 &&&&&&&&&&&& \\
				3.466-4.409 ${\textrm{GeV}}$ & 0.1586 & 0.1254 & 0.1487 & 0.1479 & 0.1488 & 0.1268 $\pm$ 0.29 \cite{RefDataset3} & 0.8414 & 0.8746 & 0.8513 & 0.8521 & 0.8512 & 0.8732 $\pm$ 0.29 \cite{RefDataset3} \\
				4.409-5.187 ${\textrm{GeV}}$ & 0.1435 & 0.1158 & 0.1364 & 0.1355 & 0.1344 & 0.1223 $\pm$ 0.21 \cite{RefDataset3} & 0.8565 & 0.8842 & 0.8636 & 0.8645 & 0.8656 & 0.8777 $\pm$ 0.21 \cite{RefDataset3} \\
				5.187-5.863 ${\textrm{GeV}}$ & 0.1213 & 0.1024 & 0.1280 & 0.1078 & 0.1091 & 0.1171 $\pm$ 0.21 \cite{RefDataset3} & 0.8787 & 0.8976 & 0.872 & 0.8922 & 0.8909 & 0.8829 $\pm$ 0.21 \cite{RefDataset3} \\
				5.863-6.47 ${\textrm{GeV}}$ & 0.1092 & 0.0986 & 0.1083 & 0.1032 & 0.1030 & 0.1146 $\pm$ 0.20 \cite{RefDataset3} & 0.8908 & 0.9014 & 0.8917 & 0.8968 & 0.897 & 0.8854 $\pm$ 0.20 \cite{RefDataset3} \\
				6.47-7.024 ${\textrm{GeV}}$ & 0.0994 & 0.0933 & 0.0932 & 0.0982 & 0.0969 & 0.0961 $\pm$ 0.22 \cite{RefDataset3} & 0.9006 & 0.9067 & 0.9068 & 0.9018 & 0.9031 & 0.9039 $\pm$ 0.22 \cite{RefDataset3} \\
				7.024-7.538 ${\textrm{GeV}}$ & 0.0918 & 0.0907 & 0.0910 & 0.0959 & 0.0887 & 0.0945 $\pm$ 0.25 \cite{RefDataset3} & 0.9082 & 0.9093 & 0.909 & 0.9041 & 0.9113 & 0.9055 $\pm$ 0.25 \cite{RefDataset3} \\
				\midrule
				Dataset 4 &&&&&&&&&&&& \\
				13.748 ${\textrm{GeV}}$ & 0.0904 & 0.0881 & 0.0868 & 0.0872 & 0.0891 & 0.0794 \cite{RefDataset4} & 0.9096 & 0.9119 & 0.9132 & 0.9128 & 0.9109 & 0.9206 \cite{RefDataset4} \\
				16.823 ${\textrm{GeV}}$ & 0.0942 & 0.0846 & 0.0840 & 0.0862 & 0.0870 & 0.0772 \cite{RefDataset4} & 0.9058 & 0.9154 & 0.916 & 0.9138 & 0.913 & 0.9228 \cite{RefDataset4} \\
				19.416 ${\textrm{GeV}}$ & 0.0864 & 0.0820 & 0.0856 & 0.0832 & 0.0860 & 0.0763 \cite{RefDataset4} & 0.9136 & 0.918 & 0.9144 & 0.9168 & 0.914 & 0.9237 \cite{RefDataset4} \\
				21.471 ${\textrm{GeV}}$ & 0.0775 & 0.0827 & 0.0860 & 0.0838 & 0.0846 & 0.0752 \cite{RefDataset4} & 0.9225 & 0.9173 & 0.914 & 0.9162 & 0.9154 & 0.9248 \cite{RefDataset4} \\
				22.956 ${\textrm{GeV}}$ & 0.0767 & 0.0818 & 0.0840 & 0.0816 & 0.0835 & 0.0735 \cite{RefDataset4} & 0.9233 & 0.9182 & 0.916 & 0.9184 & 0.9165 & 0.9265 \cite{RefDataset4} \\
				24.536 ${\textrm{GeV}}$ & 0.0768 & 0.0810 & 0.0816 & 0.0816 & 0.0852 & 0.0732 \cite{RefDataset4} & 0.9232 & 0.919 & 0.9184 & 0.9184 & 0.9148 & 0.9268 \cite{RefDataset4} \\
				26.019 ${\textrm{GeV}}$ & 0.0753 & 0.0803 & 0.0808 & 0.0805 & 0.0842 & 0.07 \cite{RefDataset4} & 0.9247 & 0.9197 & 0.9192 & 0.9195 & 0.9158 & 0.93 \cite{RefDataset4} \\
				\bottomrule
			\end{tabular}
	\end{table}	
	\subsubsection{Model Discrepancy and Error Analysis}
\begin{figure}[h!]
		\centering
		\includegraphics[width=0.97\textwidth]{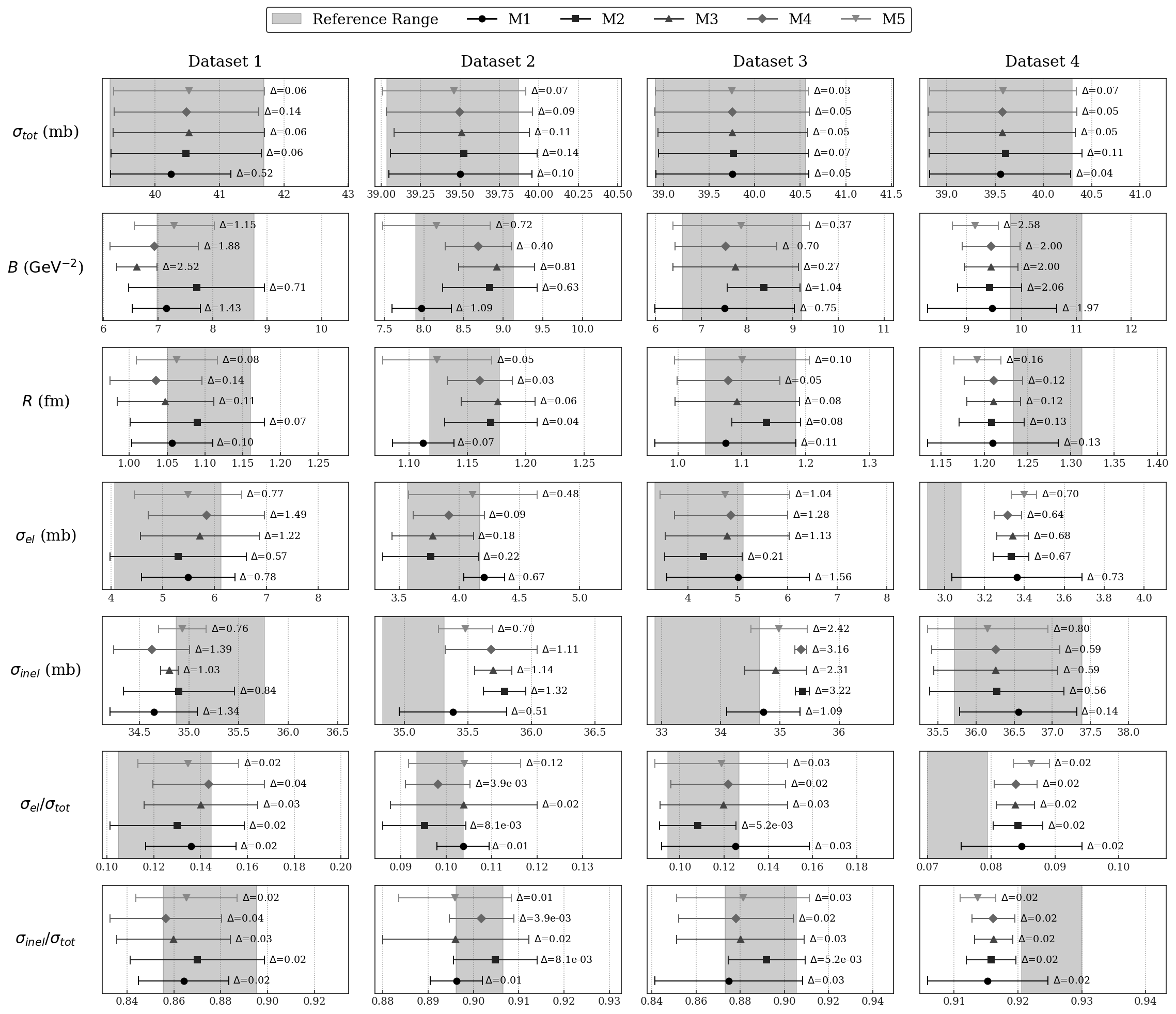}
		\caption{A comparative plot of calculated intervals against reference ranges ($\text{solid\ gray\ blocks}$) for all quantities. The localized boundary deviation ($\Delta$) for each of the five models ($M_{1}$–$M_{5}$) is annotated explicitly alongside each interval}
		\label{fig:ComparativeRangesPlot}
	\end{figure}

	To evaluate discrepancies in terms of calculated ranges (or intervals) of all the derived quantities, the least differences between the calculated ranges and the reference ranges are estimated across each dataset by:
	\begin{equation}
		\Delta_i = |Ref_{min}-(Model_{min})_i|+|Ref_{max}-(Model_{max})_i|,
	\end{equation}
	where i = 1, 2,...5 is the model number. For the total cross section in Dataset 1, the reference values range from 41.69 mb to 39.30 mb (ref. \cite{RefDataset1}. For Dataset 2, the reference value is a fixed normalization value of 39.0 mb, as in ref. \cite{RefDataset2}. But if the resulting values from fits 1, 2, and 3 are considered, their values show a decrease in $\sigma_{\rm tot}$ from 4.74 GeV to 6.704 GeV in the ranges 39.8246 mb - 39.0134 mb, 39.8423 mb - 39.0199 mb, and 39.9492 mb - 39.0726 mb, respectively. An average reference range of 39.0353 mb - 39.8720 mb is used across this dataset. For Dataset 3, the reference value is a fixed normalization value of 38.0 mb as in ref. \cite{RefDataset3}. The resulting values from fits 1, 2, and 3 show a decrease in $\sigma_{\rm tot}$ values from 3.466-4.409 GeV to 7.024-7.538 GeV in the ranges 40.5275 mb - 38.9018 mb, 40.5396 mb - 38.9037 mb, and 40.6055 mb - 38.931 mb, respectively. The average value of the ranges of the three fits is 40.5575-38.9122 mb. For Dataset 4, the reference values are in the range 38.8 mb to 40.3 mb, which show a gradual increase with c.m. energy as in ref. \cite{RefDataset4}. For the results of the other six quantities, reference ranges are estimated from the reference values mentioned in their corresponding tables. The results of the least differences are written in Table \ref{tab:ranges_table} for all the derived quantities across each dataset, and a comparative plot of these ranges is shown in Fig. \ref{fig:ComparativeRangesPlot}. Models 2 and 4 show a highest number (8) of the least difference values among all the five models.
	
\begin{figure}[h!]
		\centering
		\includegraphics[width=0.63\textwidth]{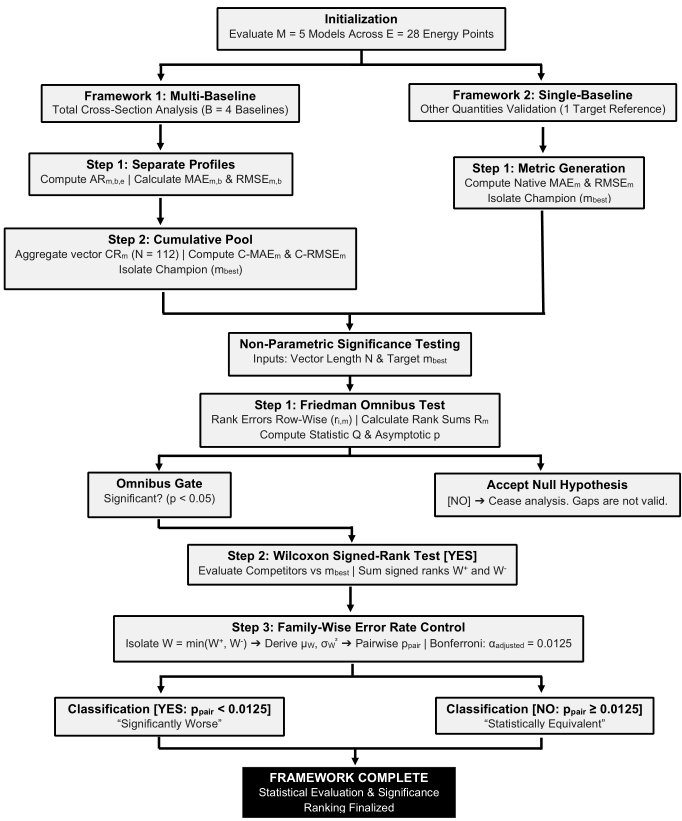}
		\caption{Evaluation workflow scheme displaying Framework 1 and Framework 2 converging into a non-parametric significance testing sequence with family-wise error control}
		\label{fig:Errorflowchart}
	\end{figure}

	An error analysis is performed in two distinct evaluation frameworks that are schematically shown in Fig. \ref{fig:Errorflowchart}) to determine the best models by computing the mean absolute error (MAE) and root mean square error (RMSE):
	\begin{equation}
		\text{MAE}_{m, b} = \frac{1}{E}\sum_{e=1}^{E} |Y_{m, e} - T_{b, e}|,
	\end{equation}
	\begin{equation}
		\text{RMSE}_{m, b} = \sqrt{\frac{1}{E}\sum_{e=1}^{E} (Y_{m, e} - T_{b, e})^2},
	\end{equation}
	where $m$ is the model, $Y_{m, e}$ is the model prediction, and $T_{b, e}$ is the baseline target (reference value) at energy level $e$. The first framework is applied across the results of the total cross section, which establishes independent discrepancy profiles by computing the MAE and RMSE for each model separately against four baseline values (the experimental reference values and three empirical fits). These absolute residuals are pooled into a cumulative data stream ($\mathbf{CR}_m$) of length $N = 112$ tracking points per model to calculate the following global integrated metrics:
	\begin{equation}
		\text{C-MAE}_m = \frac{1}{N}\sum_{i=1}^{N} \mathbf{CR}_{m, i},
	\end{equation}
	\begin{equation}
		\text{C-RMSE}_m = \sqrt{\frac{1}{N}\sum_{i=1}^{N} (\mathbf{CR}_{m, i})^2},
	\end{equation}
	These metrics allow us to identify the overall baseline best model by selecting the candidate that minimized the global $\text{C-MAE}$ score. In the case of the other quantities ($B$, $R$, $\sigma_{\mathrm{el}}$, $\sigma_{\mathrm{inel}}$, $\sigma_{\mathrm{el}}$/$\sigma_{\mathrm{tot}}$, and $\sigma_{\mathrm{inel}}$/$\sigma_{\mathrm{tot}}$), the second evaluation framework simplifies this approach by validating the results of the five models directly against reference values. Across both configurations, the definitive best models are identified by minimizing the absolute error parameters, and their mathematical superiority is rigorously validated using a non-parametric Friedman omnibus test \cite{friedman1937} to analyze the structural rank sums ($R_m$) through the test statistic ($Q$): 
	\begin{equation}
		Q = \frac{12}{N \cdot M(M + 1)} \sum_{m=1}^{M} R_m^2 - 3N(M + 1),
	\end{equation}
	where $M=5$. Upon confirming global significance ($p < 0.05$), post-hoc pairwise Wilcoxon signed-rank tests \cite{wilcoxon1945} are executed under a family-wise Bonferroni correction \cite{dunn1961} ($\alpha_{\text{adjusted}} = \frac{0.05}{M-1} = 0.0125$) to benchmark all remaining competitors directly against the leading models. This integrated model-validation procedure based on the predicted results of the scattering observables by the models, adapted from Dem\v{s}ar \cite{demsar2006}, allows us to identify whether top-performing models possess a statistically significant structural advantage across the entire energy domain.
	\renewcommand{\arraystretch}{0.550} 
	\setlength{\tabcolsep}{4.20pt} 
	\begin{table}[tbp] 
		\centering 
		\caption{The comparison table detailing the calculated ranges, reference ranges, and localized least differences ($\Delta$) across all models ($M_1$ to $M_5$) and quantities for Datasets 1 to 4} 
		\label{tab:ranges_table} 
		\footnotesize 
		\resizebox{\textwidth}{!}{ 
			\begin{tabular}{cccccccc} 
				\toprule
				\textbf{Quantity} & \textbf{Dataset} & $\boldsymbol{M_1}$ & $\boldsymbol{M_2}$ & $\boldsymbol{M_3}$ & $\boldsymbol{M_4}$ & $\boldsymbol{M_5}$ & \textbf{Ref. Range} \\
				\midrule
				\multirow{8}{*}{\shortstack[l]{$\sigma_{\mathrm{tot}}$ \\ (mb)}} & \multirow{2}{*}{Dataset 1} & 39.311--41.179 & 39.316--41.645 & 39.350--41.698 & 39.360--41.606 & 39.359--41.694 & \multirow{2}{*}{39.300--41.690} \\
				&                    & ($\Delta$: 0.5223) & ($\Delta$: 0.0609) & ($\Delta$: \textbf{0.0580}) & ($\Delta$: 0.1442) & ($\Delta$: 0.0625) & \\
				\cmidrule(lr){2-8}
				& \multirow{2}{*}{Dataset 2} & 39.051--39.955 & 39.060--39.988 & 39.083--39.939 & 39.033--39.960 & 39.009--39.916 & \multirow{2}{*}{39.035--39.872} \\
				&                    & ($\Delta$: 0.0977) & ($\Delta$: 0.1410) & ($\Delta$: 0.1141) & ($\Delta$: 0.0904) & ($\Delta$: \textbf{0.0705}) & \\
				\cmidrule(lr){2-8}
				& \multirow{2}{*}{Dataset 3} & 38.920--40.596 & 38.948--40.590 & 38.938--40.579 & 38.907--40.600 & 38.911--40.586 & \multirow{2}{*}{38.912--40.558} \\
				&                    & ($\Delta$: 0.0461) & ($\Delta$: 0.0688) & ($\Delta$: 0.0479) & ($\Delta$: 0.0471) & ($\Delta$: \textbf{0.0297}) & \\
				\cmidrule(lr){2-8}
				& \multirow{2}{*}{Dataset 4} & 38.823--40.284 & 38.817--40.398 & 38.818--40.333 & 38.805--40.344 & 38.823--40.343 & \multirow{2}{*}{38.800--40.300} \\
				&                    & ($\Delta$: \textbf{0.0393}) & ($\Delta$: 0.1143) & ($\Delta$: 0.0506) & ($\Delta$: 0.0496) & ($\Delta$: 0.0654) & \\
				\midrule
				\multirow{8}{*}{\shortstack[l]{$B$ \\ ($\text{GeV}^{-2}$)}} & \multirow{2}{*}{Dataset 1} & 6.5306-7.7745 & 6.4596-8.9536 & 6.2392-6.9844 & 6.1201-7.7414 & 6.5641-8.0293 & \multirow{2}{*}{6.98-8.76} \\
				&                    & ($\Delta$: 1.4349 ) & ($\Delta$: \textbf{0.714}) & ($\Delta$: 2.5164) & ($\Delta$: 1.8785) & ($\Delta$: 1.1466) & \\
				\cmidrule(lr){2-8}
				& \multirow{2}{*}{Dataset 2} & 7.5948-8.3493 & 8.2343-9.4263 & 8.4396-9.3986 & 8.2703-9.1009 & 7.4777-8.8344 & \multirow{2}{*}{7.90-9.13} \\
				&                    & ($\Delta$: 1.0859) & ($\Delta$: 0.6306) & ($\Delta$: 0.8082) & ($\Delta$: \textbf{0.3994}) & ($\Delta$: 0.7179) & \\
				\cmidrule(lr){2-8}
				& \multirow{2}{*}{Dataset 3} & 5.9875-9.0428 & 7.5736-9.1583 & 6.3829-9.1268 & 6.4230-8.6575 & 6.3796-9.3671 & \multirow{2}{*}{6.58-9.2} \\
				&                    & ($\Delta$: 0.7497) & ($\Delta$: 1.0353) & ($\Delta$: \textbf{0.2703}) & ($\Delta$: 0.6995) & ($\Delta$: 0.3675) & \\
				\cmidrule(lr){2-8}
				& \multirow{2}{*}{Dataset 4} & 8.2903-10.6444 & 8.8351-10.0078 & 8.9671-9.9338 & 8.9260-9.9778 & 8.7418-9.5778 & \multirow{2}{*}{9.8-11.1} \\
				&                    & ($\Delta$: \textbf{1.9653}) & ($\Delta$: 2.0571) & ($\Delta$: 1.9991) & ($\Delta$: 1.9962) & ($\Delta$: 2.5804) & \\
				\midrule
				\multirow{8}{*}{\shortstack[l]{$R$ \\ (fm)}} & \multirow{2}{*}{Dataset 1} & 1.003-1.1102 & 1.0013-1.1788 & 0.9840-1.1116 & 0.9746-1.0961 & 1.0093-1.1163 & \multirow{2}{*}{1.05-1.16} \\
				&                    & ($\Delta$: 0.0968) & ($\Delta$: \textbf{0.0675}) & ($\Delta$: 0.1144) & ($\Delta$: 0.1393) & ($\Delta$: 0.0844) & \\
				\cmidrule(lr){2-8}
				& \multirow{2}{*}{Dataset 2} & 1.0857-1.1383 & 1.1304-1.2095 & 1.1445-1.2078 & 1.1329-1.1885 & 1.0773-1.1709 & \multirow{2}{*}{1.1178-1.1773} \\
				&                    & ($\Delta$: 0.0711) & ($\Delta$: 0.0448) & ($\Delta$: 0.0572) & ($\Delta$: \textbf{0.0263}) & ($\Delta$: 0.0469) & \\
				\cmidrule(lr){2-8}
				& \multirow{2}{*}{Dataset 3} & 0.9640-1.1847 & 1.0841-1.1922 & 0.9953-1.1902 & 0.9984-1.1592 & 0.9950-1.2057 & \multirow{2}{*}{1.0431-1.1847} \\
				&                    & ($\Delta$: 0.1097) & ($\Delta$: 0.0791) & ($\Delta$: 0.0839) & ($\Delta$: \textbf{0.0498}) & ($\Delta$: 0.0997) & \\
				\cmidrule(lr){2-8}
				& \multirow{2}{*}{Dataset 4} & 1.1343-1.2853 & 1.1710-1.2463 & 1.1797-1.2417 & 1.1770-1.2444 & 1.1648-1.2192 & \multirow{2}{*}{1.2334-1.3126} \\
				&                    & ($\Delta$: 0.1264) & ($\Delta$: 0.1287) & ($\Delta$: 0.1246) & ($\Delta$: \textbf{0.1246}) & ($\Delta$: 0.1620) & \\
				\midrule
				\multirow{8}{*}{\shortstack[l]{$\sigma_{\mathrm{el}}$ \\ (mb)}} & \multirow{2}{*}{Dataset 1} & 4.5872-6.3933 & 3.9842-6.6108 & 4.5661-6.8617 & 4.7243-6.9643 & 4.4524-6.5207 & \multirow{2}{*}{4.0758-6.1275} \\
				&                    & ($\Delta$: 0.7772) & ($\Delta$: \textbf{0.5749}) & ($\Delta$: 1.2245) & ($\Delta$: 1.4853) & ($\Delta$: 0.7698) & \\
				\cmidrule(lr){2-8}
				& \multirow{2}{*}{Dataset 2} & 4.0364-4.3784 & 3.3620-4.1616 & 3.4393-4.1215 & 3.6155-4.2103 & 3.5778-4.6463 & \multirow{2}{*}{3.5685-4.172} \\
				&                    & ($\Delta$: 0.6743) & ($\Delta$: 0.2169) & ($\Delta$: 0.1797) & ($\Delta$: \textbf{0.0853}) & ($\Delta$: 0.4836) & \\
				\cmidrule(lr){2-8}
				& \multirow{2}{*}{Dataset 3} & 3.5754-6.4418 & 3.5355-5.0912 & 3.5459-6.0377 & 3.7321-6.0061 & 3.4500-6.0428 & \multirow{2}{*}{3.3423-5.111} \\
				&                    & ($\Delta$: 1.5639) & ($\Delta$: \textbf{0.213}) & ($\Delta$: 1.1303) & ($\Delta$: 1.2849) & ($\Delta$: 1.0395) & \\
				\cmidrule(lr){2-8}
				& \multirow{2}{*}{Dataset 4} & 3.0353-3.6882 & 3.2445-3.4227 & 3.2605-3.4206 & 3.2479-3.3859 & 3.3353-3.4604 & \multirow{2}{*}{2.9131-3.0831} \\
				&                    & ($\Delta$: 0.7273) & ($\Delta$: 0.671) & ($\Delta$: 0.6849) & ($\Delta$: \textbf{0.6376}) & ($\Delta$: 0.7) & \\
				\midrule
				\multirow{8}{*}{\shortstack[l]{$\sigma_{\mathrm{inel}}$ \\ (mb)}} & \multirow{2}{*}{Dataset 1} & 34.2041-35.0853 & 34.3367-35.4589 & 34.7166-34.8894 & 34.241-35.0052 & 34.6956-35.1729 & \multirow{2}{*}{34.8706-35.7625} \\
				&                    & ($\Delta$: 1.3437) & ($\Delta$: 0.8375) & ($\Delta$: 1.0271) & ($\Delta$: 1.3869) & ($\Delta$: \textbf{0.7646}) & \\
				\cmidrule(lr){2-8}
				& \multirow{2}{*}{Dataset 2} & 34.96-35.8068 & 35.6217-35.9573 & 35.5552-35.8465 & 35.3209-36.044 & 35.2695-35.6942 & \multirow{2}{*}{34.828-35.315} \\
				&                    & ($\Delta$: \textbf{0.5073}) & ($\Delta$: 1.3195) & ($\Delta$: 1.1422) & ($\Delta$: 1.1054) & ($\Delta$: 0.7042) & \\
				\cmidrule(lr){2-8}
				& \multirow{2}{*}{Dataset 3} & 34.1049-35.3441 & 35.265-35.4989 & 34.4065-35.4511 & 35.2545-35.454 & 34.5099-35.4606 & \multirow{2}{*}{32.889-34.6577} \\
				&                    & ($\Delta$: \textbf{1.0923}) & ($\Delta$: 3.2172) & ($\Delta$: 2.3109) & ($\Delta$: 3.1618) & ($\Delta$: 2.4238) & \\
				\cmidrule(lr){2-8}
				& \multirow{2}{*}{Dataset 4} & 35.7875-37.3214 & 35.3939-37.1512 & 35.445-37.0726 & 35.4193-37.0965 & 35.3621-36.9426 & \multirow{2}{*}{35.7169-37.3869} \\
				&                    & ($\Delta$: \textbf{0.1361}) & ($\Delta$: 0.5587) & ($\Delta$: 0.5862) & ($\Delta$: 0.588) & ($\Delta$: 0.7991) & \\
				\midrule
				\multirow{8}{*}{\shortstack[l]{$\sigma_{\mathrm{el}}/ \sigma_{\mathrm{tot}}$}} & \multirow{2}{*}{Dataset 1} & 0.1165-0.1552 & 0.1013-0.1587 & 0.1159-0.1645 & 0.1197-0.1673 & 0.1131-0.1563 & \multirow{2}{*}{0.1047-0.1445} \\
				&                    & ($\Delta$: 0.0225) & ($\Delta$: \textbf{0.0176}) & ($\Delta$: 0.0312) & ($\Delta$: 0.0378) & ($\Delta$: 0.0202) & \\
				\cmidrule(lr){2-8}
				& \multirow{2}{*}{Dataset 2} & 0.098-0.1095 & 0.086-0.1044 & 0.0878-0.12 & 0.0911-0.1053 & 0.0917-0.1164 & \multirow{2}{*}{0.0935-0.1038} \\
				&                    & ($\Delta$: 0.0102) & ($\Delta$: 0.0081) & ($\Delta$: 0.0219) & ($\Delta$: \textbf{0.0039}) & ($\Delta$: 0.0144) & \\
				\cmidrule(lr){2-8}
				& \multirow{2}{*}{Dataset 3} & 0.0918-0.1586 & 0.0907-0.1254 & 0.091-0.1487 & 0.0959-0.1479 & 0.0887-0.1488 & \multirow{2}{*}{0.0945-0.1268} \\
				&                    & ($\Delta$: 0.0345) & ($\Delta$: \textbf{0.0052}) & ($\Delta$: 0.0254) & ($\Delta$: 0.0225) & ($\Delta$: 0.0278) & \\
				\cmidrule(lr){2-8}
				& \multirow{2}{*}{Dataset 4} & 0.0753-0.0942 & 0.0803-0.0881 & 0.0808-0.0868 & 0.0805-0.0872 & 0.0835-0.0891 & \multirow{2}{*}{0.07-0.0794} \\
				&                    & ($\Delta$: 0.0201) & ($\Delta$: 0.019) & ($\Delta$: \textbf{0.0182}) & ($\Delta$: 0.0183) & ($\Delta$: 0.0232) & \\
				\midrule
				\multirow{8}{*}{\shortstack[l]{$\sigma_{\mathrm{inel}}/ \sigma_{\mathrm{tot}}$}} & \multirow{2}{*}{Dataset 1} & 0.8448-0.8835 & 0.8413-0.8987 & 0.8355-0.8841 & 0.8327-0.8803 & 0.8437-0.8869 & \multirow{2}{*}{0.8555-0.8953} \\
				&                    & ($\Delta$: 0.0225) & ($\Delta$: \textbf{0.0176}) & ($\Delta$: 0.0312) & ($\Delta$: 0.0378) & ($\Delta$: 0.0202) & \\
				\cmidrule(lr){2-8}
				& \multirow{2}{*}{Dataset 2} & 0.8905-0.902 & 0.8956-0.914 & 0.88-0.9122 & 0.8947-0.9089 & 0.8836-0.9083 & \multirow{2}{*}{0.8962-0.9065} \\
				&                    & ($\Delta$: 0.0102) & ($\Delta$: 0.0081) & ($\Delta$: 0.0219) & ($\Delta$: \textbf{0.0039}) & ($\Delta$: 0.0144) & \\
				\cmidrule(lr){2-8}
				& \multirow{2}{*}{Dataset 3} & 0.8414-0.9082 & 0.8746-0.9093 & 0.8513-0.909 & 0.8521-0.9041 & 0.8512-0.9113 & \multirow{2}{*}{0.8732-0.9055} \\
				&                    & ($\Delta$: 0.0348) & ($\Delta$: \textbf{0.0052}) & ($\Delta$: 0.0254) & ($\Delta$: 0.0225) & ($\Delta$: 0.0278) & \\
				\cmidrule(lr){2-8}
				& \multirow{2}{*}{Dataset 4} & 0.9058-0.9247 & 0.9119-0.9197 & 0.9132-0.9192 & 0.9128-0.9195 & 0.9109-0.9165 & \multirow{2}{*}{0.9206-0.93} \\
				&                    & ($\Delta$: 0.0201) & ($\Delta$: 0.019) & ($\Delta$: \textbf{0.0182}) & ($\Delta$: 0.0183) & ($\Delta$: 0.0232) & \\
				\bottomrule 
			\end{tabular} 
		} 
	\end{table}
	
\begin{figure}[h!]
		\centering
		\includegraphics[width=0.8\textwidth]{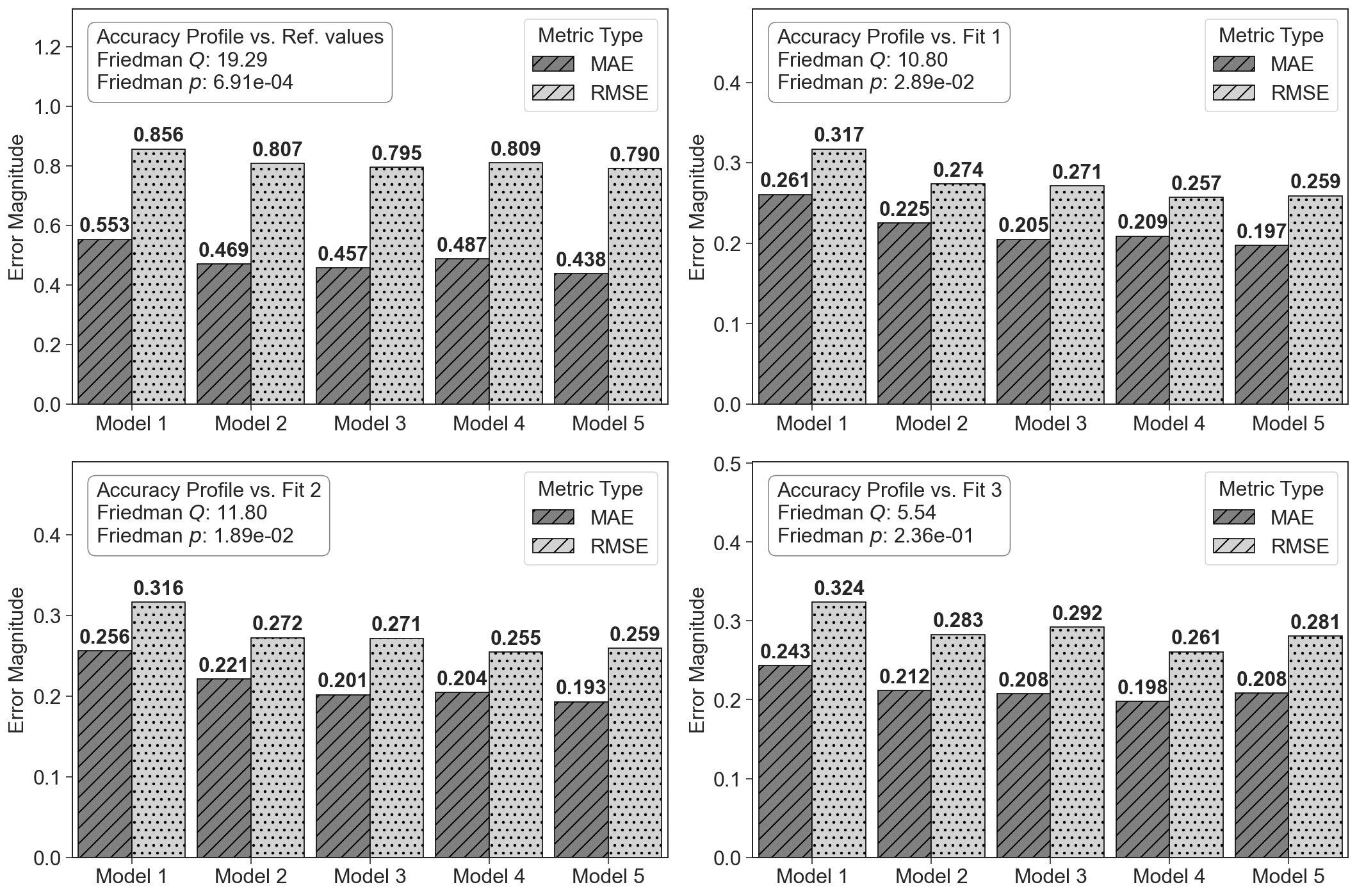}
		\caption{Comparative MAE and RMSE metric profiles of the calculated total cross sections across the five models in each of the four cases (Ref. values, fits 1-3)
		}
		\label{fig:Totalerrorplot}
	\end{figure}

\begin{figure}[h!]
		\centering
		\includegraphics[width=0.5\textwidth]{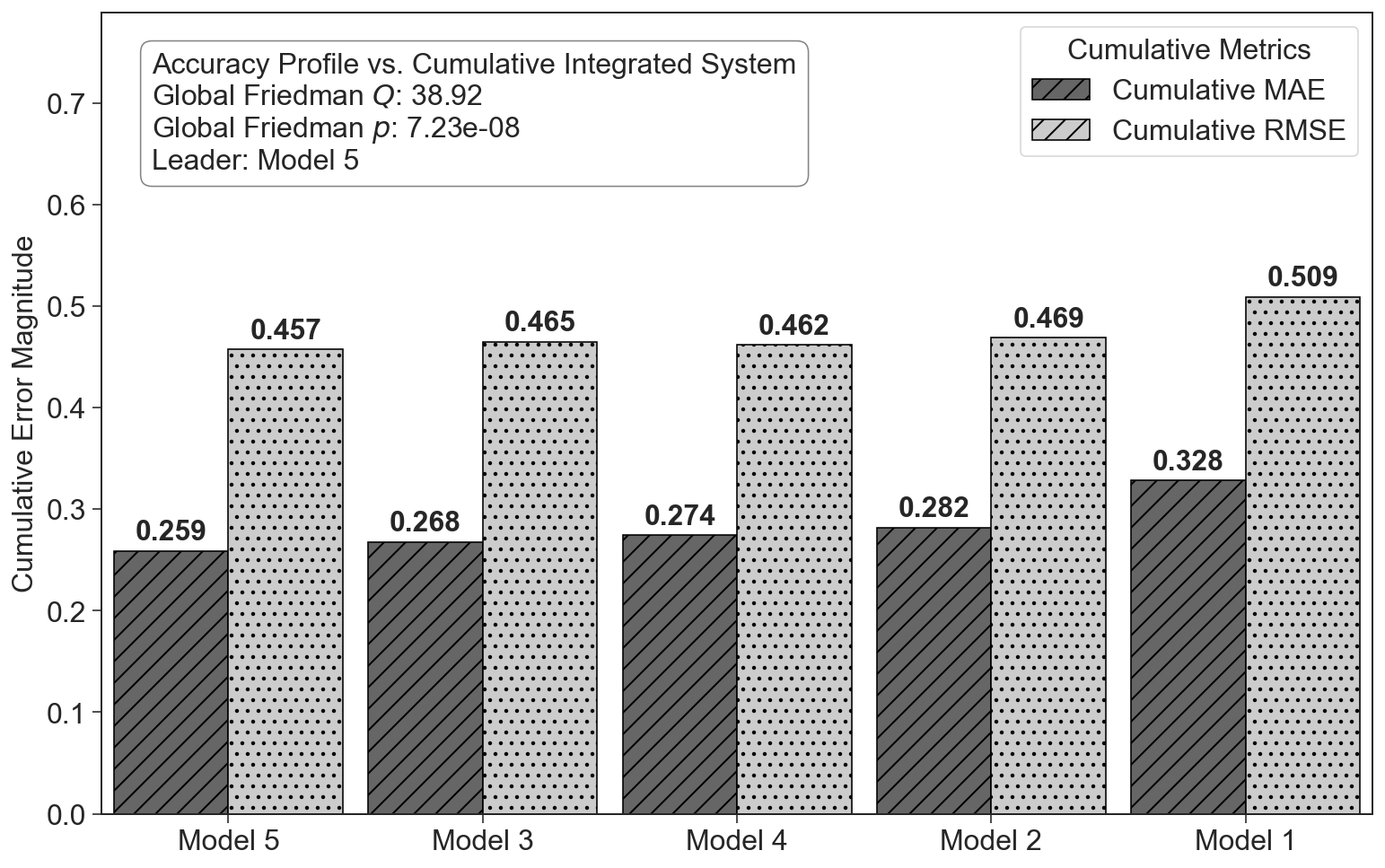}
		\caption{C-MAE and C-RMSE metric profiles of the calculated total cross sections (combined Ref. values and fits 1-3)
		}
		\label{fig:CTotalerrorplot}
	\end{figure}

	The aforementioned two complementary levels allow evaluation of the model performance, first by comparing the calculated and reference ranges (mentioned in Table \ref{tab:ranges_table}), and second by the MAE- and RMSE-based evaluations (mentioned in Tables \ref{tab:total_cross_section_summary} and \ref{tab:master_summary_1}), where the differences in estimated accuracy are assessed to determine their statistical significance. In the range-based comparisons, the direct descriptive picture of the model behavior is provided across each quantity. While the mean of the absolute deviations is estimated by MAE, and the larger errors acquire more weight in estimations of RMSE. Consequently, a model which shows the smallest deviation can be described as descriptively closest without being statistically superior. This distinction is particularly important for $\sigma_{\mathrm{tot}}$, where the agreement indicates that the models effectively approximate the effective forward absorptive strength. It is found in the two-phase statistical validation of $\sigma_{\mathrm{tot}}$ (shown in the Figs. \ref{fig:Totalerrorplot} and \ref{fig:CTotalerrorplot}) that in Phase 1, Model 5 gives the smallest discrepancies against the experimental reference data and the first two fitted curves, with $\mathrm{MAE}=0.4376$ and $\mathrm{RMSE}=0.7900$, and the corresponding residual-distribution comparisons indicate statistically significant differences relative to the competing models ($p<0.05$). Although a localized descriptive advantage for Fit 3 ($\mathrm{MAE}=0.1976$) is shown by Model 4, the omnibus result ($p=0.23$) does not establish statistically significant superiority. In the integrated Phase 2 analysis, based on $N=112$ observations, Model 5 is further identified as having the lowest cumulative discrepancy, with $\mathrm{C-MAE}=0.2590$ and $\mathrm{C-RMSE}=0.4574$; Model 3 is found to remain statistically equivalent ($\mathrm{C-MAE}=0.2677$), whereas Models 1, 2, and 4 exhibit significantly larger discrepancies ($p<0.0125$). Thus, these results show that for calculations of $\sigma_{\mathrm{tot}}$, Model 5 is supported by the statistical evidence beyond a merely descriptive preference and establishes it as an efficient model among the five models.
	
	The results across the remaining observables reveal why model performance cannot be reduced to a single universal ranking. The forward logarithmic slope ($B$) provides information that is absent from the forward normalization because $B$ is connected to the transverse momentum scale by $t\simeq-q_\perp^2$, and through an impact parameter interpretation with spatial distribution of the interaction. A larger $B$ corresponds qualitatively to a broader transverse interaction profile and a narrower distribution in $q_\perp$. The increase of $B$ with energy is reflected in the shrinkage of the diffraction cone and, in Regge phenomenology, contains a contribution proportional to $2\alpha'\ln s$. Our results for $B$ indicate that the general evolution of $B$ over Datasets 1-3 is shown by all the models, but all five models underestimate the experimental $9.8$--$11.1 (\mathrm{GeV}/c)^{-2}$ range in Dataset 4, despite their reasonable description of $\sigma_{\mathrm{tot}}$. It indicates that forward interaction strength is produced, while its transverse distribution does not broaden sufficiently at highest energies. Model 1 gives the lowest MAE for both $B$ and $R$, $0.5868$ and $0.0400$, respectively, but the associated Friedman tests give $p>0.05$, showing no statistically significant differences among the models. In the range-based and error analysis, Model 2 is found to give the lowest descriptive errors for $\sigma_{\mathrm{el}}$, $R_{\mathrm{el}}$, and $R_{\mathrm{inel}}$, with MAEs of $0.2698$, $0.0073$, and $0.0073$, respectively. But the corresponding Omnibus tests do not reach statistical significance, so these results are considered only as localized descriptive advantage. A significant model-to-model variation is found by the Friedman test with $Q=13.87$ and $p=7.73\times10^{-3}$, where Model 1 gives the lowest errors, $\mathrm{MAE}=0.5009$ and $\mathrm{RMSE}=0.6577$, and the post-hoc comparison identifies Model 2 as significantly worse under the adjusted criterion ($p<0.0125$). 
	
	In case of the ratios $R_{\mathrm{el}}$ and $R_{\mathrm{inel}}$, the elastic fraction decreases from approximately $0.10$--$0.14$ in Dataset 1 to $0.07$--$0.08$ in Dataset 4, while the inelastic fraction increases from roughly $0.86$--$0.90$ to $0.92$--$0.93$. At highest energies, the correlated pattern $B_{\rm model}<B_{\rm exp}$, $R_{\mathrm{el,model}}>R_{\mathrm{el,exp}}$, and $R_{\mathrm{inel,model}}<R_{\mathrm{inel,exp}}$ follows from the $1/B$ dependence of $\sigma_{\mathrm{el}}$. This pattern indicates a common limitation in the high-energy momentum-transfer ($|t|$) dependence in the models rather than independent deviations of the three observables. The small values $R_{\mathrm{el}}\sim~0.07$--$0.16$ in present $np$ elastic scattering analysis are comparable to those of elastic $pp$ scattering at ISR energies, where they stabilize at $\approx$ $0.178 \pm 0.003$ \cite{UAmaldi1980}, and are comparatively lower than recently measured values $0.25$--$0.28$ at 13 TeV by TOTEM \cite{GAnchtev2019} at LHC, far below than the black-disk limit $0.5$. Therefore, the $np$ interaction in the present analysis is not interpreted as having reached a completely absorptive black-disk regime.
	
	Dataset 1 \cite{RefDataset1} provides a broad $|t|$ distribution, reaching approximately $0.14\lesssim|t|\lesssim19~\mathrm{GeV}^2$ at the highest momenta, thereby testing the transition from the forward diffraction cone to large-angle scattering. Dataset 2 \cite{RefDataset2} covers approximately $0.06\lesssim|t|\lesssim3~\mathrm{GeV}^2$, bridging the forward and intermediate-$|t|$ regions and demonstrating substantial $np$ and $pp$ similarity. Dataset 3 \cite{RefDataset3} shows the energy evolution and shrinkage of the $np$ diffraction region while showing that $np$ and $pp$ scattering are similar but not identical. Dataset 4 \cite{RefDataset4} the measurements are extended to Fermilab momenta of $70$--$400~\mathrm{GeV}/c$, where a compatibility of $pp$ parametrizations with the data in the forward slope is found but differences occur in the intermediate-$|t|$ region, which gives a phenomenological indication that the high energy-scattering amplitude cannot be characterized solely by its forward normalization and slope. The dip structure is a consequence arising from destructive interference among components of the scattering amplitude. From a Regge phenomenological viewpoint, the slowly varying forward strength can be associated with the interplay among a leading Pomeron and secondary Reggeon exchanges with lower intercepts, whose relative importance decreases with energy. According to this view, specifically isospin-dependent contributions (isovector $\rho$- and $a_2$-Reggeon) in the amplitude can consequently generate bigger differences in $np$ and $pp$ scattering at lower energies, while these contributions become less important in the forward region at higher energies, without requiring their contribution to vanish at intermediate or large $|t|$. Therefore, the present models can be considered as effective parametrizations of competing energy dependences and transverse structures, showing good fitting with the data rather than explicit Regge-pole or eikonal amplitudes. The results show that the main success of the models can be attributed to the slowly varying forward interaction strength and soft-energy evolution, where $\sigma _{\mathrm{tot}}$ slowly varies, $B$ and $R$ increase, $R_{el}$ decreases, and $R_{inel}$ increases. Meanwhile, their common limitation is the simultaneous description of high-energy transverse broadening ($B_{\rm model}<B_{\rm exp}$) and the partitioning of elastic and inelastic cross sections. These limitations identify the momentum-transfer dependence in the models as the most important direction for further refinement.
	\begin{figure}[h!]
		\centering
		\includegraphics[width=0.85\textwidth]{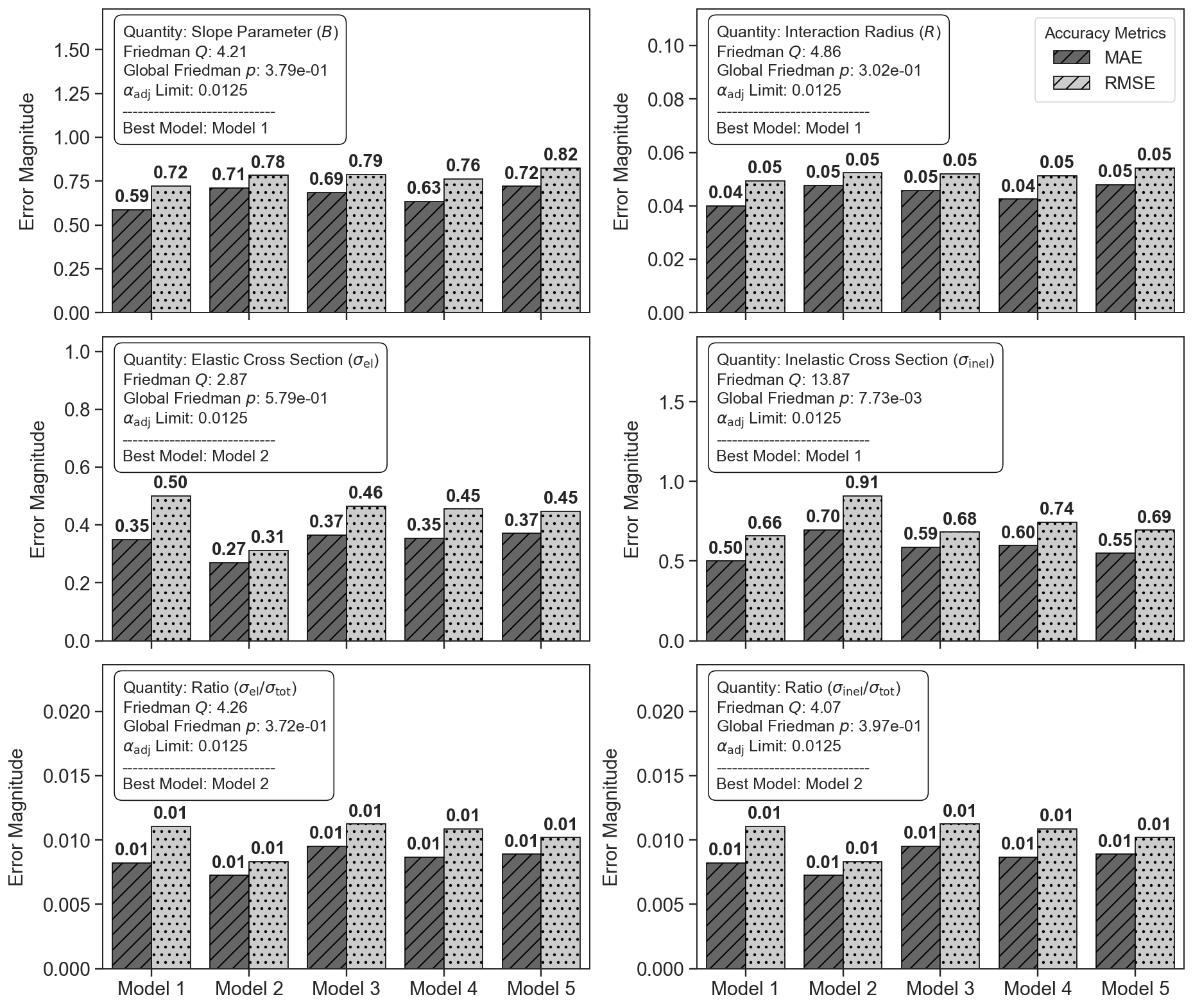}
		\caption{MAE and RMSE metric profiles of the other six calculated quantities across the five models
		}
		\label{fig:CombinedPlotError}
	\end{figure}
	\renewcommand{\arraystretch}{0.70}
	\setlength{\tabcolsep}{1.5pt}
	\begin{table}[!htbp]
		\centering
		\caption{Comprehensive Two-Phase Statistical Evaluation of Total Cross-Section Results(Framework 1)}
		\label{tab:total_cross_section_summary}
		\footnotesize
			\begin{tabular}{llcccc}
				\toprule
				\textbf{Benchmark / Profile} & \textbf{Model Name} & \textbf{MAE} & \textbf{RMSE} & \textbf{Friedman $p$-value} & \textbf{Significance Status} \\
				\midrule
				\multicolumn{6}{c}{PHASE 1: SEPARATE STATISTICAL BENCHMARK EVALUATIONS} \\
				\midrule
				\multirow{5}{*}{Ref. values} 
				& Model 5 & 0.4376 & 0.7900 & \multirow{5}{*}{$6.91 \times 10^{-4}$} & \textbf{Best Candidate} \\
				& Model 3 & 0.4568 & 0.7952 & & Statistically Equal \\
				& Model 2 & 0.4695 & 0.8069 & & Significantly Worse \\
				& Model 4 & 0.4874 & 0.8094 & & Significantly Worse \\
				& Model 1 & 0.5530 & 0.8558 & & Significantly Worse \\
				\midrule
				\multirow{5}{*}{Fit 1} 
				& Model 5 & 0.1971 & 0.2590 & \multirow{5}{*}{$2.89 \times 10^{-2}$} & \textbf{Best Candidate} \\
				& Model 3 & 0.2050 & 0.2714 & & Statistically Equal \\
				& Model 4 & 0.2088 & 0.2572 & & Statistically Equal \\
				& Model 2 & 0.2254 & 0.2739 & & Statistically Equal \\
				& Model 1 & 0.2606 & 0.3172 & & Statistically Equal \\
				\midrule
				\multirow{5}{*}{Fit 2} 
				& Model 5 & 0.1929 & 0.2589 & \multirow{5}{*}{$1.89 \times 10^{-2}$} & \textbf{Best Candidate} \\
				& Model 3 & 0.2014 & 0.2713 & & Statistically Equal \\
				& Model 4 & 0.2041 & 0.2545 & & Statistically Equal \\
				& Model 2 & 0.2207 & 0.2721 & & Statistically Equal \\
				& Model 1 & 0.2557 & 0.3160 & & Statistically Equal \\
				\midrule
				\multirow{5}{*}{Fit 3} 
				& Model 4 & 0.1976 & 0.2609 & \multirow{5}{*}{$2.36 \times 10^{-1}$} & \textbf{Best Candidate} \\
				& Model 3 & 0.2076 & 0.2920 & & Statistically Equal \\
				& Model 5 & 0.2084 & 0.2807 & & Statistically Equal \\
				& Model 2 & 0.2119 & 0.2827 & & Statistically Equal \\
				& Model 1 & 0.2434 & 0.3236 & & Statistically Equal \\
				\midrule
				\multicolumn{6}{c}{PHASE 2: INTEGRATED CUMULATIVE ACCURACY PROFILE} \\
				\midrule
				\multirow{5}{*}{\shortstack[l]{Integrated Global\\Discrepancy ($N=112$)}} 
				& Model 5 & 0.2590 & 0.4574 & \multirow{5}{*}{$1.42 \times 10^{-12}$} & GLOBAL CHAMPION \\
				& Model 3 & 0.2677 & 0.4650 & & Statistically Equivalent \\
				& Model 4 & 0.2745 & 0.4621 & & Significantly Worse \\
				& Model 2 & 0.2819 & 0.4691 & & Significantly Worse \\
				& Model 1 & 0.3282 & 0.5093 & & Significantly Worse \\
				\bottomrule
			\end{tabular}
	\end{table}
	\renewcommand{\arraystretch}{0.70}
	\setlength{\tabcolsep}{1.5pt}
	\begin{table}[!htbp]
		\centering
		\caption{Statistical Summary Table across the Six Secondary Physical Quantities (Framework 2)}
		\label{tab:master_summary_1}
		\footnotesize
			\begin{tabular}{llcccccc}
				\toprule
				\textbf{Quantity} & \textbf{Symbol} & \textbf{Model} & \textbf{MAE} & \textbf{RMSE} & \textbf{Friedman $Q$} & \textbf{Global $p$-value} & \textbf{Statistical Status} \\
				\midrule
				\multirow{5}{*}{Slope Parameter} & \multirow{5}{*}{$B$} 
				& Model 1 & 0.5868 & 0.7218 & \multirow{5}{*}{4.21} & \multirow{5}{*}{$3.79 \times 10^{-1}$} & \textbf{BEST CANDIDATE} \\
				& & Model 2 & 0.7115 & 0.7830 & & & Statistically Equal \\
				& & Model 3 & 0.6857 & 0.7880 & & & Statistically Equal \\
				& & Model 4 & 0.6349 & 0.7639 & & & Statistically Equal \\
				& & Model 5 & 0.7205 & 0.8243 & & & Statistically Equal \\
				\midrule
				\multirow{5}{*}{Interaction Radius} & \multirow{5}{*}{$R$} 
				& Model 1 & 0.0400 & 0.0493 & \multirow{5}{*}{4.86} & \multirow{5}{*}{$3.02 \times 10^{-1}$} & \textbf{BEST CANDIDATE} \\
				& & Model 2 & 0.0477 & 0.0524 & & & Statistically Equal \\
				& & Model 3 & 0.0458 & 0.0521 & & & Statistically Equal \\
				& & Model 4 & 0.0425 & 0.0513 & & & Statistically Equal \\
				& & Model 5 & 0.0480 & 0.0541 & & & Statistically Equal \\
				\midrule
				\multirow{5}{*}{Elastic Cross Section} & \multirow{5}{*}{$\sigma_{\text{el}}$} 
				& Model 1 & 0.3503 & 0.4998 & \multirow{5}{*}{2.87} & \multirow{5}{*}{$5.79 \times 10^{-1}$} & Statistically Equal \\
				& & Model 2 & 0.2698 & 0.3118 & & & \textbf{BEST CANDIDATE} \\
				& & Model 3 & 0.3659 & 0.4644 & & & Statistically Equal \\
				& & Model 4 & 0.3540 & 0.4550 & & & Statistically Equal \\
				& & Model 5 & 0.3705 & 0.4470 & & & Statistically Equal \\
				\midrule
				\multirow{5}{*}{Inelastic Cross Section} & \multirow{5}{*}{$\sigma_{\text{inel}}$} 
				& Model 1 & 0.5009 & 0.6577 & \multirow{5}{*}{13.87} & \multirow{5}{*}{$7.73 \times 10^{-3}$} & \textbf{BEST CANDIDATE} \\
				& & Model 2 & 0.6969 & 0.9086 & & & Significantly Worse \\
				& & Model 3 & 0.5866 & 0.6815 & & & Statistically Equal \\
				& & Model 4 & 0.5976 & 0.7434 & & & Statistically Equal \\
				& & Model 5 & 0.5489 & 0.6935 & & & Statistically Equal \\
				\midrule
				\multirow{5}{*}{Ratio} & \multirow{5}{*}{$\sigma_{\text{el}}/ \sigma_{\text{tot}}$} 
				& Model 1 & 0.0082 & 0.0111 & \multirow{5}{*}{4.26} & \multirow{5}{*}{$3.72 \times 10^{-1}$} & Statistically Equal \\
				& & Model 2 & 0.0073 & 0.0083 & & & \textbf{BEST CANDIDATE} \\
				& & Model 3 & 0.0095 & 0.0113 & & & Statistically Equal \\
				& & Model 4 & 0.0087 & 0.0109 & & & Statistically Equal \\
				& & Model 5 & 0.0089 & 0.0102 & & & Statistically Equal \\
				\midrule
				\multirow{5}{*}{Ratio} & \multirow{5}{*}{$\sigma_{\text{inel}}/ \sigma_{\text{tot}}$} 
				& Model 1 & 0.0082 & 0.0111 & \multirow{5}{*}{4.07} & \multirow{5}{*}{$3.97 \times 10^{-1}$} & Statistically Equal \\
				& & Model 2 & 0.0073 & 0.0083 & & & \textbf{BEST CANDIDATE} \\
				& & Model 3 & 0.0095 & 0.0113 & & & Statistically Equal \\
				& & Model 4 & 0.0087 & 0.0109 & & & Statistically Equal \\
				& & Model 5 & 0.0089 & 0.0102 & & & Statistically Equal \\
				\bottomrule
			\end{tabular}%
		\begin{flushleft}
		\end{flushleft}
	\end{table}	
	\subsection{Implications of the Models}
	A precise parametrization and fitting of the differential cross-section in ${np}$ elastic scattering at high energies has important implications for hadronic interactions, QCD dynamics, and hadronic structure. Such models can be potentially linked or embedded with QCD-based approaches and modern effective field theory (EFT) through several correspondences. First, the exponential $t$-dependence $\exp(B(s)t)$ notably follows from the impact-parameter representation of the scattering amplitude, where a Gaussian or quasi-Gaussian profile of the eikonal function in transverse space leads through a Fourier-Bessel transform to an approximately exponential behavior in momentum transfer. The logarithmic energy dependence $B(s)\propto \ln s$ in the leading effective slopes arises from Gribov diffusion in Reggeon Field theory, reflecting the growth of the interaction radius with energy, while deviations from a pure Gaussian profile account for observed departures from a simple exponential form \cite{Gribov1968,Ferreira1998}. Second, the multi-component exponential decomposition of the models allows us to effectively accommodate the separation of Pomeron, Odderon, and secondary Reggeon exchanges that are investigated in Regge-based effective approaches, where each term corresponds to distinct trajectories $\alpha_i(t)$ with signature factors, consistent with modern treatments of soft scattering such as the tensor-Pomeron framework \cite{Ewerz2014} and recent phenomenological implementations \cite{Lebiedowicz2014}. Third, the nontrivial features introduced in our models (e.g., Gaussian modulation terms) can be mapped to interference between multiple exchange channels or to unitarity corrections within the eikonal formalism, where zeros in the amplitude arise from destructive interference in impact-parameter space, which connects to the Good-Walker diffractive eigenstate formalism \cite{Good1960} and its modern realizations in multi-channel eikonal models \cite{Gotsman1999}. The energy evolution encoded in our models can be related to QCD-motivated evolution equations; the slow logarithmic growth of slopes and amplitudes reflects Gribov diffusion and is compatible with Balitsky-Fadin-Kuraev-Lipatov (BFKL)-type dynamics in the nonperturbative regime \cite{Lipatov1986}, while saturation-motivated approaches (e.g., Color Glass Condensate) provide a framework in which the effective exponential $t$-dependence arises after Fourier transformation from saturated gluon distributions \cite{Iancu2004}. Importantly, embedding our parametrizations into these frameworks would require promoting the energy-dependent amplitude coefficients $A_i(s)$ and slopes $B_i(s)$ to quantities derived from eikonal functions or Wilson-line correlators, thereby enabling a direct link between fitted parameters and QCD degrees of freedom. While the models do not explicitly enforce analyticity, crossing symmetry, or unitarity at the amplitude level, they can be systematically treated for these refinements by incorporating dispersion-relation constraints \cite{Block1986} and by matching to EFT-based amplitudes in appropriate kinematic limits.
	
	The proposed models may serve as fast, analytic inputs for nuclear data evaluations in transport simulations such as \textsc{Geant4} and \textsc{PHITS}, where extension to fully differential exclusive channels and consistent unitarization to interface with cascade generators are required in their current smooth analytic forms \cite{Agostinelli2003,Sato2018}. In this way, efficient generation of differential observables including $d\sigma/dt$, integrated elastic and total cross sections, angular distributions, and event-level scattering kernels for Monte Carlo propagation can be enabled. In applications to Accelerator-Driven Systems (ADS) and spallation neutron sources, the models provide energy-dependent cross sections, scattering amplitudes, and slope parameters that determine neutron yields, multiplicity spectra, and shielding-relevant observables, subject to inclusion of inelastic channels, medium corrections, and normalization constraints \cite{Bowman1992,Bauer2001}. Within Glauber theory, these parametrizations can be embedded into impact-parameter-dependent eikonal formalisms with fluctuations and inhomogeneity in nucleon structure \cite{Miller2007} to compute impact-parameter profiles, nuclear overlap functions $T(b)$, elastic and inelastic nucleus-nucleus cross sections, and initial-state geometrical quantities (e.g., eccentricities and participant distributions) relevant for heavy-ion collisions. Cosmic-ray air-shower simulations can be constrained through these models by their extrapolation to ultra-high energies, yielding proton-air cross sections, interaction lengths, and shower development observables such as $X_{\max}$ distributions, provided asymptotic unitarity constraints and matching to hadronic interaction models \cite{Engel2011} are incorporated. 
	
	The analytic structure of the models can be useful in Bayesian and machine-learning frameworks to achieve global fits, yielding parameter posteriors, correlations, uncertainty bands, and predictive extrapolations across energy regimes \cite{Sato2020,Neal2012}. The models can also be utilized for detector luminosity calibration and beam optics studies through precise determination of forward scattering rates and normalization-sensitive observables \cite{Antchev2013}. In modeling of neutrino-nucleus interaction, the models can be used to provide inputs to hadronic final-state interaction cross sections and angular distributions within intranuclear cascade frameworks \cite{Gallagher2011}. The models may also enable studies of factorization breaking and universality in soft QCD via scaling behavior, energy dependence of slopes, and correlations between $s$ and $t$ \cite{Donnachie:2002en,Collins:1977jy}. The explicit separation of crossing-even and crossing-odd components can potentially allow investigation of Odderon effects through differences in $pp$ and $p\bar{p}$ observables, including dip positions, differential cross-section ratios, and interference patterns \cite{Martynov2018}. Finally, in dispersion-relation analyses, these parametrizations provide smooth inputs for analytic continuation to the forward limit, enabling consistent determination of real-to-imaginary amplitude ratios, slope extrapolations, and high-energy bounds under analyticity and crossing constraints \cite{Block1986}.
	\section{Conclusions}
	\label{sec:IV}
	The primary motivation for this investigation into the inverse problem as a source of model building to establish potential linkages with QCD and a source of experimental information is driven by the absence of a pure QCD description of $np$ elastic scattering and the specific physical contexts underlying various phenomenological models. For this purpose, we have investigated the inverse problem for model development as a mechanism to bridge high-energy $np$ scattering with QCD by analyzing neutron-proton ($np$) elastic scattering using non-linear fitting by five distinct purely phenomenological models of the differential cross section over four experimental datasets spanning the few-GeV to Fermilab energy range. Several key features of the models include analyticity, differentiability, and integrability in both $s$ and $t$, which ensure their mathematical consistency to describe the data in the GeV and TeV energy domains. The proposed models effectively describe the empirical features of the experimental data and yield estimates of the derived quantities in elastic $np$ scattering. Important features in the plots of the data, such as the steep exponential decline at low $|t|$, the moderate deviations from a purely exponential shape at intermediate $|t|$ leading to the dip-bump structure, shrinkage of the forward peak, and shift of the diffractive minimum (a dip) towards lower $|t|$-values with increasing c.m. energies are reproduced in fits of the models, showing quite low values of the least $\chi^{2}$.
	
	The proposed models can potentially yield good fitting agreement with $pp$, $p\bar{p}$, and $n\bar{p}$ elastic scattering data not only in the high energy limit ($s\gg t$), but also in the intermediate-$\mid t\mid$ region, where a dip is observed in $pp$ data at ISR, as well as at higher GeV energy domains \cite{NAGY1979221,Pdesgrolard2000,Watanabe2023}. These models may also provide a good fit to the dip-bump structure, which is prominently present in the recent $pp$ data at LHC energies of refs. \cite{TOTEM_2018psk,Antchev_2013,Antchev_2011,TOTEM:2021imi,Antchev_2019}. The analysis of the correlation matrices reveal different behavior of Models 3 and 4 as compared to Models 1, 2, and 5. The number of fixed and varying parameters varied among the models upon confronting the models with data in non-linear-fitting and least redundancies (0) and dependencies (1) are found across the Models 3 and 4, which also showed the least penalty scores of 0.3324 and 0.3387, respectively. These correlation differences of the model parameters in fitting are associated to the differences in functional forms of the five models in terms of the variables s and t. The Residual Sum of Squares (RSS) values, obtained through the least $\chi^{2}$ from fitting across all the 28 c.m. energies, are incorporated into information criteria (AIC and BIC). Across Model 2, lowest AIC and BIC scores of values 193.62 and 246.89 are found. The lowest AIC and BIC scores reflect efficient fitting performance of Model 2 due to better $\chi^{2}$ minimization, while possessing a smaller number of total parameters than the other models.
	
	The fitted models are employed to derive important quantities ($\sigma_{\mathrm{tot}}$, $B$, $R$, $\sigma_{\mathrm{el}}$, $\sigma_{\mathrm{inel}}$, $R_{\mathrm{el}}$, and $R_{\mathrm{inel}}$) of $np$ elastic scattering, thereby constraining the forward interaction strength, transverse interaction scale, and elastic-inelastic partition. The results of $\sigma_{\mathrm{tot}}$ show that the models reproduce the slowly varying energy dependence reasonably well. The combined error analysis based on the MAE, RMSE, Friedman, and post-hoc analyses identifies Model 5 results in closest agreement with $\sigma_{\mathrm{tot}}$ reference values, while quantity-dependent model preferences are shown across the remaining observables. The lowest descriptive errors are found for $B$, $R$, and $\sigma_{\mathrm{inel}}$. While the lowest descriptive errors are found for $\sigma_{\mathrm{el}}$, $R_{\mathrm{el}}$, and $R_{\mathrm{inel}}$. However, most of these differences in the descriptive errors are not statistically significant. Thus, it is difficult to conclude any single parameterization as uniformly superior across all the seven observables. In this study, for the highest energies of the available experimental data, all the models underestimate the experimental diffraction slope $B$. This indicates an increase of the characteristic transverse interaction with energy, which is not adequately captured by their parameterizations. Through $1/B$ dependence of $\sigma_{\mathrm{el}}$, the low estimates are accompanied by an overestimation of the elastic fraction and underestimation of the inelastic fraction. Therefore, the models provide better estimation of the forward interaction strength than their momentum transfer dependence.
	
	Important advantages of the proposed models include their high accuracy, flexibility, extrapolation capabilities, analytic structure, and computational efficiency. The parameters of the models provide compact empirical descriptions of the measured scattering behaviour. The utility of the models can be increased for theoretical applications by transforming or promoting their parameters to connect with other theoretical and phenomenological approaches. For continuous advancements, it is crucial to consider the limitations as opportunities for development and improvement. The present analysis concludes that Model 2 is the most effective parametrization, containing a smaller number of parameters, while showing quite low error in fitting and the lowest AIC and BIC scores. The results of the quantities derived across Model 2 are mostly in good agreement in terms of the least differences between its calculated ranges and the reference ranges. In the MAE- and RMSE-based error analysis, its results are most accurate across $\sigma_{\mathrm{el}}$, $R_{el}$, and $R_{inel}$. This model can potentially provide an economical description of the elastic scattering due to fewer parameters in its considered form compared to the rest of the models in this study. The present analysis indicates that an efficient modeling strategy requires detailed investigation of different analyses and results. Generally, the good fitting agreement of our models with previous significant works (discussed in Section \ref{sec:III}), including the Regge pole model and related Regge frameworks, the Phillip Barger (PB) model, and the impact picture framework at GeV to TeV energy domains and intermediate to large $\mid t\mid$ across $pp$ and $p\bar{p}$ elastic scattering, as well as the holographic QCD approach at GeV energy domains in the Regge limit for $np$ and $n\bar{p}$ elastic scattering, emphasizes their physical relevance. In this study, an understanding is reinforced that the internal structure of the nucleons and the dynamics of elastic scattering can be consistently described by modeling the differential cross sections. Consequently, this study provides a basis for future refinement of phenomenological models and offers predictive strength for upcoming intermediate- and high-energy experiments.
	\section{Recommendations and Future Work}
	\label{sec:V}
	The proposed models of this study can be employed to describe elastic scattering data of $pp$, $p\bar{p}$, and $n\bar{p}$ at a broad range of c.m. energies. These models can be extended or modified by developing different parameterizations in terms of the Mandelstam variables s and t, especially those that have more effective theoretical roots. Modified approximate energy-dependent parameterizations \cite{Okorokov2008,Okorokov2015} in the effective slope parameters of the models can be used to investigate energy dependence in elastic hadronic scattering. The sub-leading terms can be parameterized depending on specific physical constraints, such as several meson-exchange contributions for elastic $np$, $n\bar{p}$, $pp$, and $p\bar{p}$ scattering \cite{Watanabe2023,Pdesgrolard2000}. These models can be considered for extensions with other phenomenological models \cite{Jenkovszky2011,Cudell2002,Block2005.1}. The parameters used in the models can also be modified or promoted to investigate polarization observables, comparisons with modern EFT-driven and QCD-driven approaches. These models can be considered for further validation with the availability of new data from the LHC and future colliders like FCC, particularly at broad $\mid t \mid$ ranges. The use of machine learning techniques may also offer improved predictive capabilities. We are hopeful that extending this study along the highlighted directions will yield deeper insights into hadronic physics.\\ \\
	\textbf{Data availability statement} Associated data is available in the manuscript.
	\section*{References}
	\renewcommand{\bibsection}{}
	
\end{document}